\documentclass[preprint, 10pt]{elsarticle}
\pdfoutput=1

\newif\ifInputs
\Inputstrue

\usepackage{palatino}
\usepackage[usenames]{color}
\usepackage{epsfig}
\usepackage[fleqn,reqno]{amsmath}
\usepackage{amsfonts,amsthm,bm}
\usepackage{amssymb}
\usepackage{mathrsfs}
\usepackage{stmaryrd}
\usepackage{subfigure}
\usepackage[titletoc]{appendix}
\usepackage{tabularx,booktabs}
\usepackage{graphics,caption}
\usepackage{fancybox}
\usepackage[top=1.2in,bottom=1.2in,left=1in, right=1in]{geometry}
\usepackage{array}
\usepackage{multirow}
\usepackage{wrapfig}
\usepackage{algorithmic,algorithm}
\usepackage{paralist}
\usepackage{ifthen}
\usepackage{enumitem}
\usepackage{tikz,pgfplots,filecontents}
\usepackage{float}
\usepackage{mdframed}
\usepackage{todonotes}

\graphicspath{{./figs/}}

\usepackage[pagebackref=false,bookmarks=false]{hyperref} 

\definecolor{dblue}{rgb}{0.03,0.3,0.62}
\definecolor{dorange}{rgb}{1,0.55,0}

\hypersetup{
  bookmarksnumbered=true,
  bookmarksopen=false,
  hypertexnames=false,     
  breaklinks=true,          
  unicode=false,
  pdffitwindow=true,        
  pdfnewwindow=true,        
  colorlinks=true,         
  linkcolor=dblue,
  anchorcolor=red,
  citecolor=dorange,
  filecolor=magenta,
  urlcolor=dblue,
  pdfstartview = FitH,
  pdfkeywords = {},
  pdfcreator = {LaTeX with hyperref package}
}

\definecolor{sblue}{cmyk}{0.98,0.13,0,0.43} 
\definecolor{sblue}{cmyk}{0.98,0.13,0,0.43} 

\newcommand{\mcaption}[2]{\caption{\em #1}\label{#2}}

\newcommand{\figref}[1]{Figure~\ref{#1}}
\newcommand{\tabref}[1]{Table~\ref{#1}}
\newcommand{\secref}[1]{\S\ref{#1}}

\newcounter{magicrownumbers}

\newcommand{\pec}{\text{Pe}} 

\definecolor{darkblue}{rgb}{0,0,0.8}
\definecolor{darkgreen}{rgb}{0,0.5,0}

\long\def\symbolfootnote[#1]#2{\begingroup \def\thefootnote{\fnsymbol{footnote}}\footnote[#1]{#2} \endgroup}

\newcommand{\calA}{\ensuremath{ \mathcal{A} }}

\newcommand{\calK}{\ensuremath{ \mathcal{K} }}
\newcommand{\calL}{\ensuremath{ \mathcal{L} }}

\newcommand{\hatphi       }{\ensuremath{ \widehat{\phi} }}

\newcommand{\del}{\ensuremath{\partial}}

\newcommand{\parder}[2]{\ensuremath{ \frac{\del #1}{\del #2} }}
\newcommand{\pparder}[2]{\ensuremath{ \frac{\del^2 #1}{\del {#2}^2} }}
\newcommand{\dder}[2]{\ensuremath{ \frac{\text{d} #1}{\text{d} #2} }}

\renewcommand{\cos}[1]{ \text{cos}\hspace{0.0cm}\left( {#1} \right) }

\newcolumntype{C}{>{\centering\arraybackslash} m{2.5cm}}

\usepackage{lineno}

\begin{document}

\title{Quantification of mixing in vesicle suspensions \\[1mm] using
numerical simulations in two dimensions}

\author[ut]{G\"{o}kberk Kabacao\u{g}lu} \ead{gokberk@ices.utexas.edu}
\author[fsu]{Bryan Quaife} \ead{bquaife@fsu.edu}
\author[ut]{George Biros}\ead{gbiros@acm.org}
\address[ut]{Institute for Computational Engineering and Sciences,\\
The University of Texas at Austin, Austin, TX, 78712, United States}
\address[fsu]{Department of Scientific Computing, \\ Florida State
University, Tallahassee, FL 32306, United States}

\begin{abstract} 
We study mixing in Stokesian vesicle suspensions  in two dimensions on
a cylindrical Couette apparatus using numerical simulations.  The
vesicle flow simulation is done using a boundary integral method and
the advection-diffusion equation for the mixing of the  solute  is
solved using a pseudo-spectral scheme.
We study the effect of the area fraction, the viscosity contrast
between  the inside (the vesicles) and the outside (the bulk) fluid,
the initial condition of the solute, and the mixing metric.  We compare
mixing in the suspension with mixing in the Couette apparatus without
vesicles.

On the one hand, the presence of vesicles  in most cases, slightly
suppresses mixing. This is because the solute can be only diffused
across the vesicle interface and not advected.  On the other hand,
there exist spatial distributions of the solute for which the
unperturbed Couette flow completely fails to mix whereas the presence
of vesicles enables mixing. We derive a simple condition that relates
the velocity and solute and can be used to characterize the cases in
which the presence of vesicles promotes mixing.
\end{abstract}

\maketitle

\section{Introduction} \label{s:intro}
Vesicles are closed phospholipid membranes suspended in a viscous
solution.  They are found in biological systems, and play an important
role in intracellular and intercellular transport.  Artificial vesicles
are used in a variety of drug-delivery systems and in the study of
biomembrane mechanics.  Vesicle-inspired mechanical models can be used
to approximate red blood cell mechanics, and non-local hydrodynamic
interactions.  Most vesicle suspension flows take place in vanishing
Reynolds number regime.  Although there has been a lot of work in
characterizing the dynamics of vesicles, there has been very little
work in characterizing mixing in vesicle flows.

\begin{figure}[!htb]
\begin{minipage}{\textwidth}
\centering   
   \setcounter{subfigure}{0}
   \renewcommand*{\thesubfigure}{(a)}
   \hspace{0cm}\subfigure[A vesicle suspension in a Couette apparatus]{\scalebox{0.65}{{\includegraphics{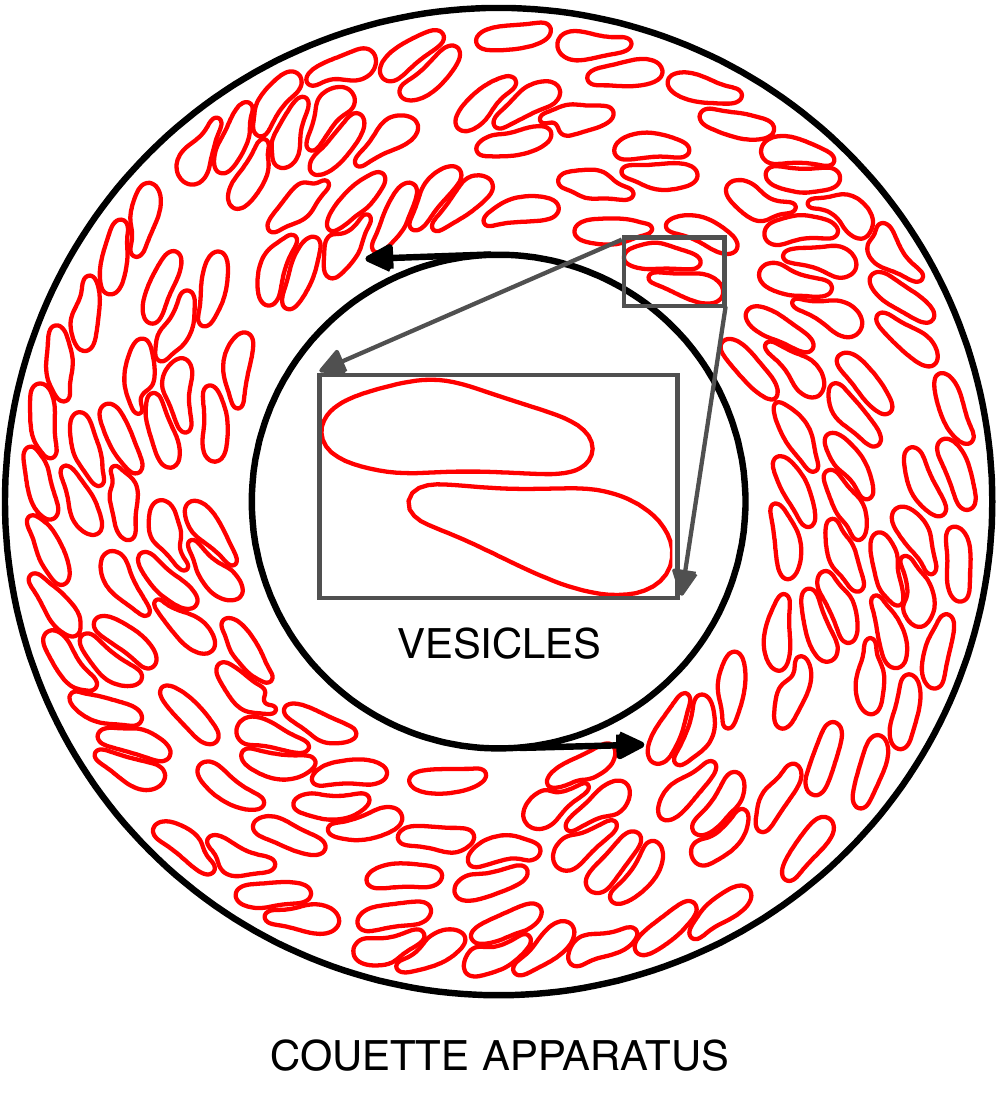}}}
   \label{f:geometry}}
  \setcounter{subfigure}{0}
  \renewcommand*{\thesubfigure}{(b)} 
  \hspace{0cm}\subfigure[Initial conditions for mixing simulations]{\scalebox{0.65}{{\includegraphics{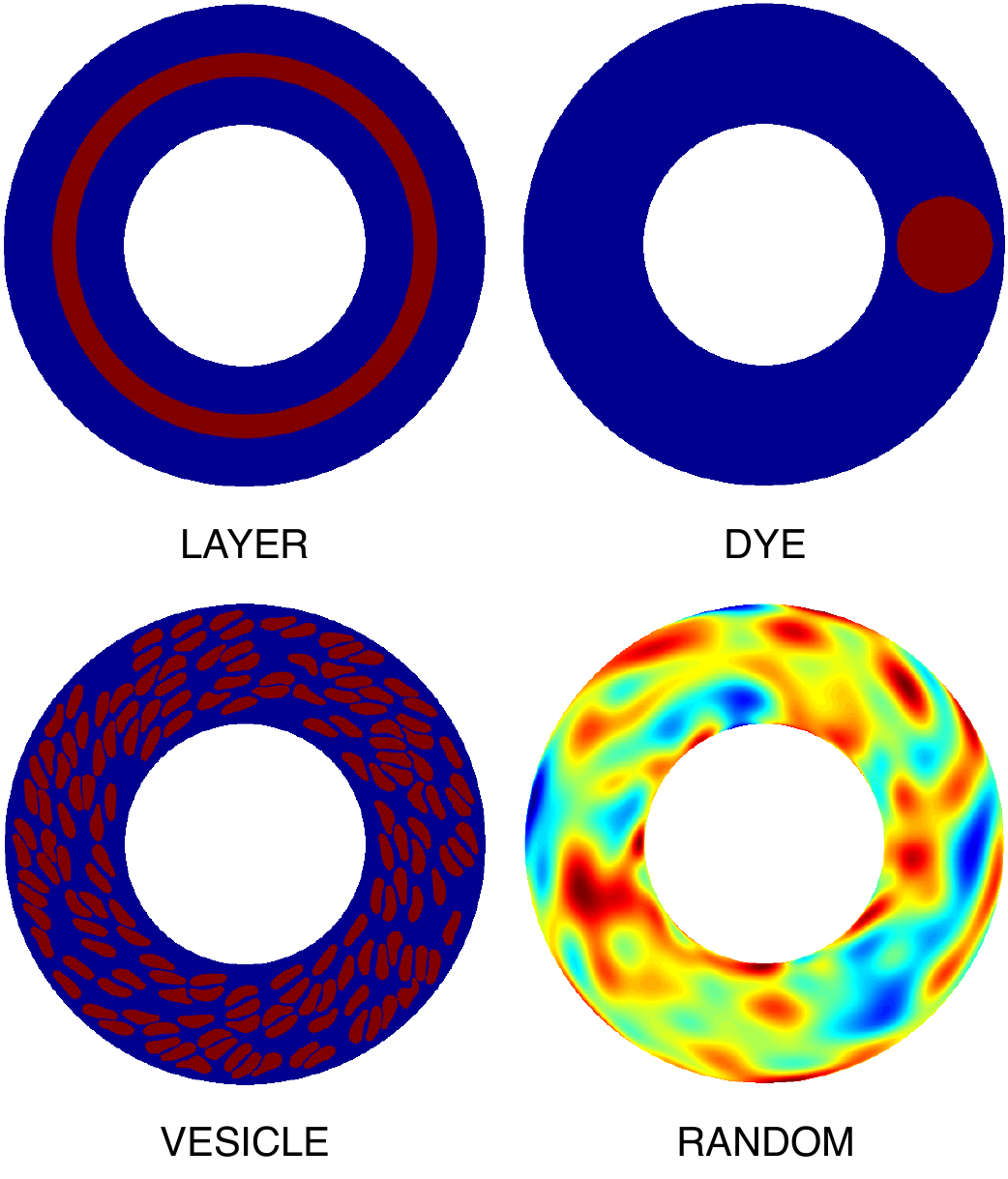}}} 
   \label{f:initConds}}
  \end{minipage}
\mcaption{{The} geometry used in this work is a cylindrical Couette
apparatus.  It consists of two concentric cylinders, where the inner
cylinder is rotating and the outer cylinder is stationary.  We also
show the vesicle suspension.  First we perform a vesicle-flow numerical
simulation to compute the velocity field of the suspension.  This step
is independent of the advection-diffusion equation since the
transported quantity is passive.  Once the velocity field is computed,
we solve the transport problem for a number of different initial
conditions depicted in \figref{f:initConds}. Red colors correspond to
high concentration (maximum value is 1) and blue colors correspond to
low concentration (minimum value is 0).}{f:method}
\end{figure}

\paragraph*{\textit{Contributions:}}
To the best of our knowledge, this is one of the first papers studying the effects of vesicle suspensions on mixing.  We consider a simple setup: a two-dimensional cylindrical Couette apparatus in the zero Reynolds number regime.  The size of the vesicles is comparable to the size of the apparatus, so we study systems for which it is not clear how to use an upscaled model since there is no separation of scales (see \figref{f:introFigure}, for an example). We only consider no-slip boundary conditions where the inner cylinder rotates at a fixed rate and the outer cylinder is stationary.  We study the system numerically with an integro-differential equation formulation for the fluid dynamics and a pseudo-spectral scheme for a passive advection-diffusion equation.  We compare the mixing in the
absence and the presence of vesicles.  There is no unique way to define mixing but our results are based mostly on negative Sobolev norms of the concentration.  We study the effects of the Peclet number, the area fraction, and the viscosity contrast between the fluids inside and outside of the vesicles.  We also study several different initial conditions for the passively transported quantity (\emph{``the solute''}).  The membranes of the vesicles in our model are assumed to be impermeable for the background fluid (\emph{``the solvent''}) and permeable for the solute.  Since, in the model, Lagrangian trajectories do not cross the vesicle membrane, this has the effect of reducing advective mixing.  Overall, we find that for the same average Peclet number, the presence of
vesicles \emph{slightly reduces mixing}.  Interestingly, however, this is not always the case.  There exist certain rather special initial conditions for the passively transported quantity that this is not the case.  For these conditions in the absence of vesicles there is no advective mixing while the presence of vesicles \emph{increases} mixing.  One such initial condition is the
\emph{``LAYER''} initial condition in \figref{f:initConds}.
    
\begin{figure}[!htb]
 \begin{minipage}{\textwidth}
\setcounter{subfigure}{0}
\renewcommand*{\thesubfigure}{(a)} 
      \hspace{0cm}\subfigure[$t = 0$]{\scalebox{0.5}{{\includegraphics{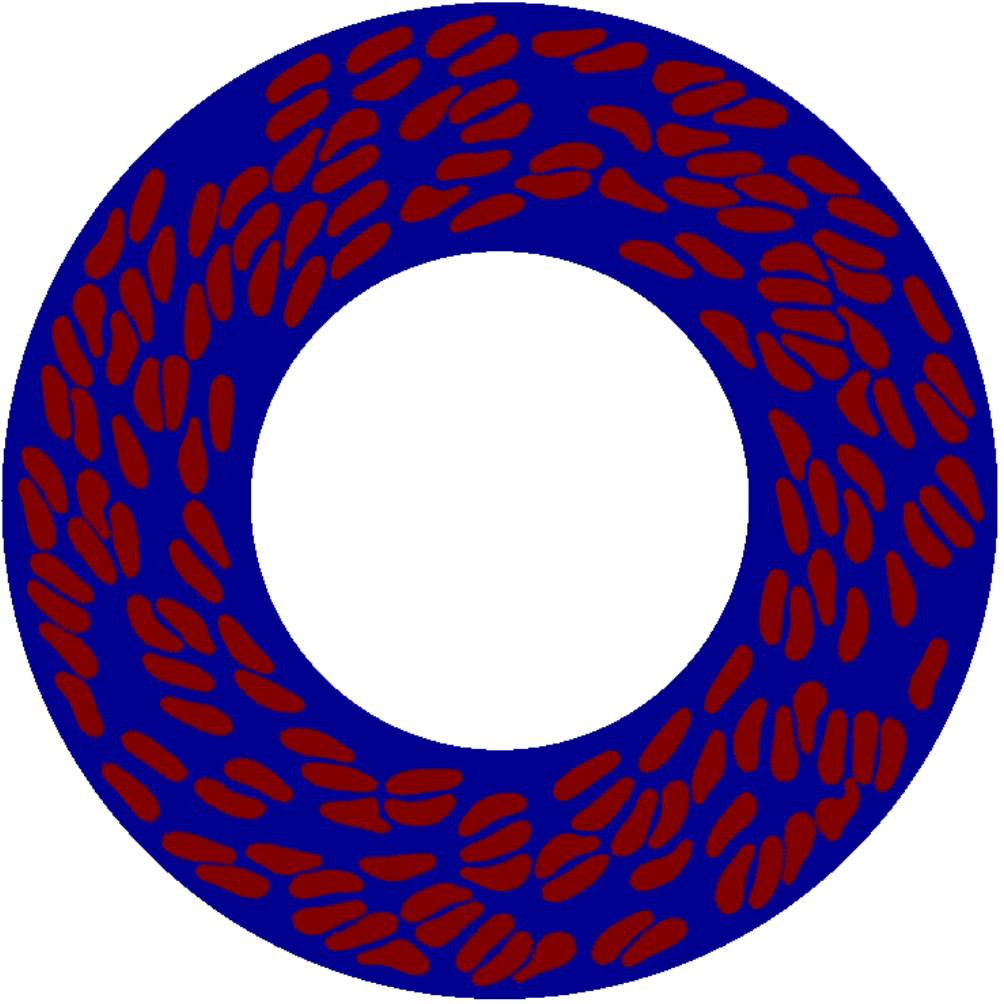}}}	
      \label{f:VF40S1}}
\setcounter{subfigure}{0}
\renewcommand*{\thesubfigure}{(b)} 
      \hspace{0cm}\subfigure[$t = 5$]{\scalebox{0.5}{{\includegraphics{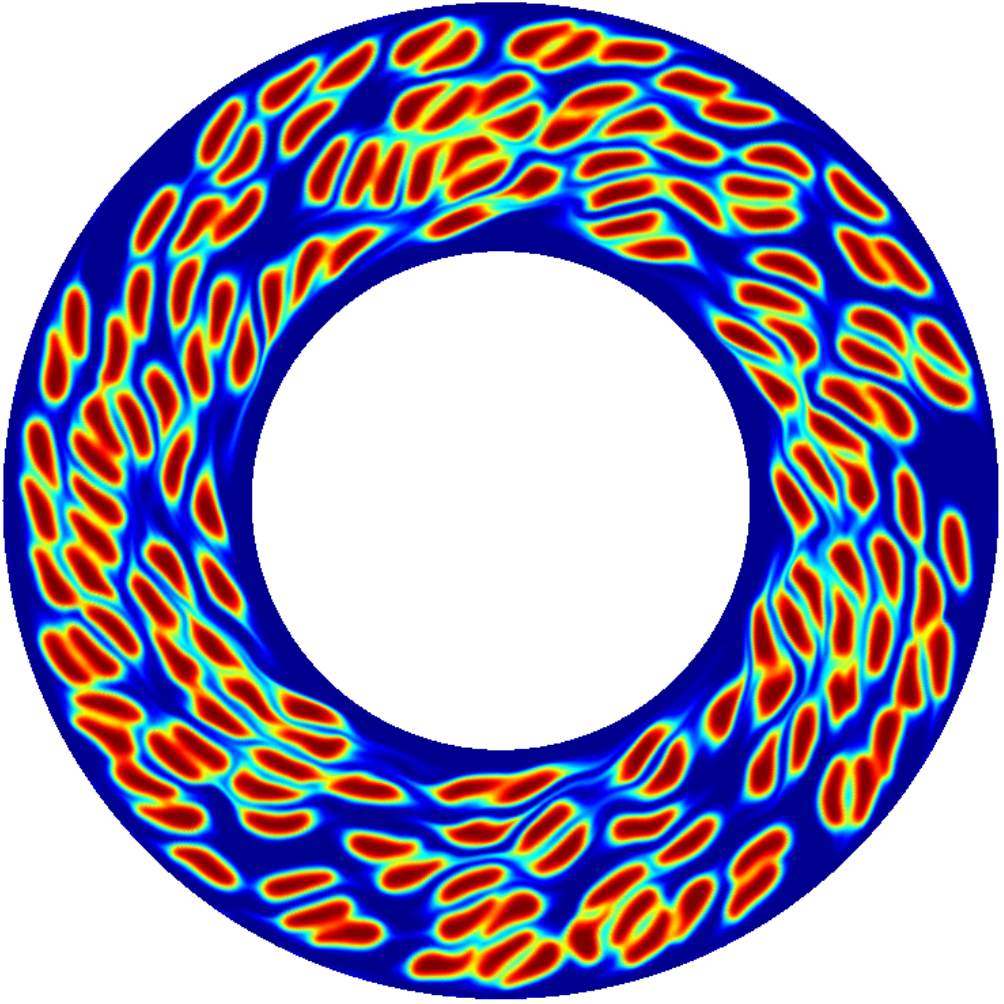}}}
      \label{f:VF40S2}}
\setcounter{subfigure}{0}
\renewcommand*{\thesubfigure}{(c)} 
      \hspace{0cm}\subfigure[$t = 10$]{\scalebox{0.5}{{\includegraphics{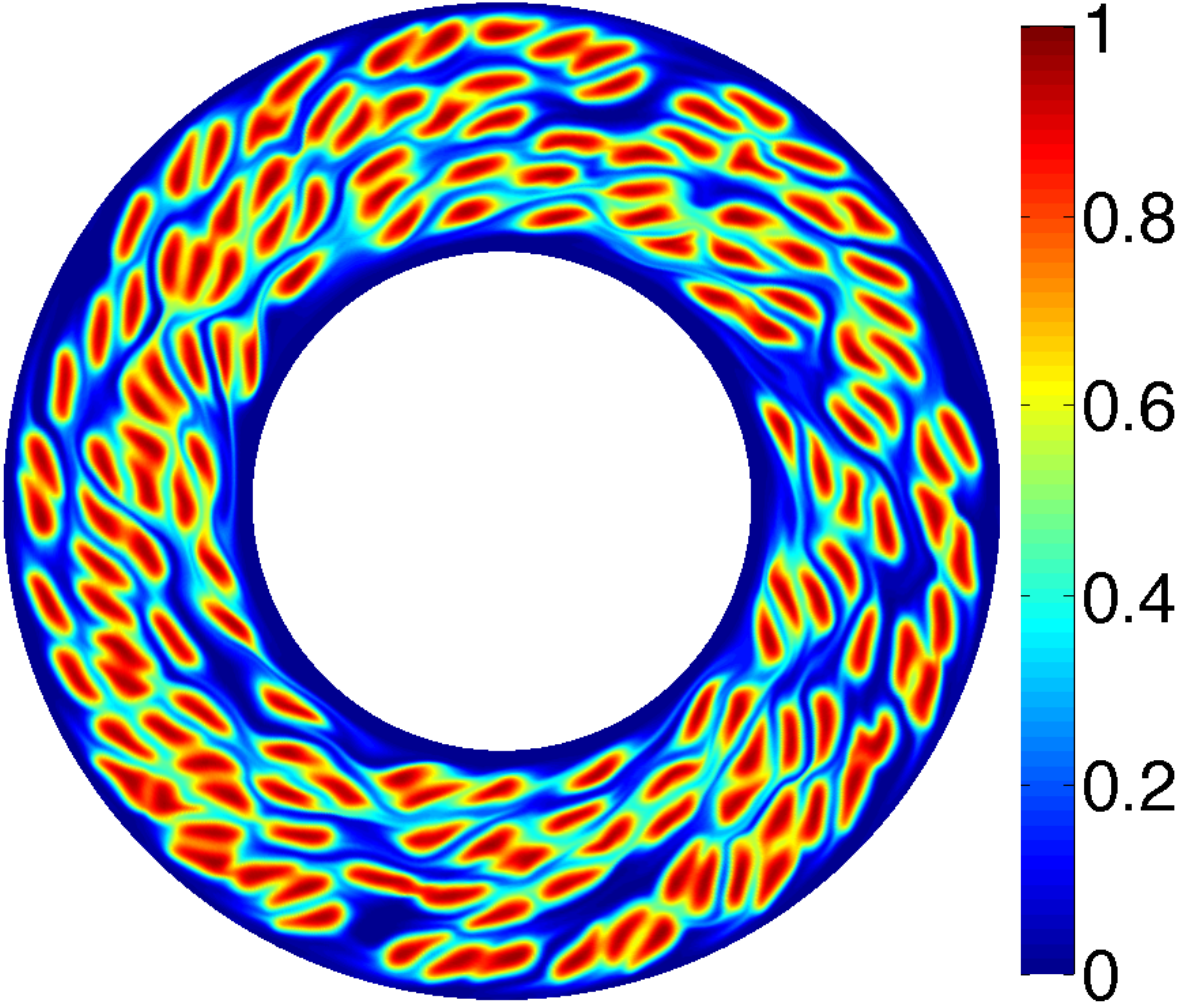}}}
      \label{f:VF40S3}}
  \end{minipage}    
  \begin{minipage}{\textwidth}    
\setcounter{subfigure}{0}
\renewcommand*{\thesubfigure}{(d)} 
      \hspace{0cm}\subfigure[$t = 25$]{\scalebox{0.5}{{\includegraphics{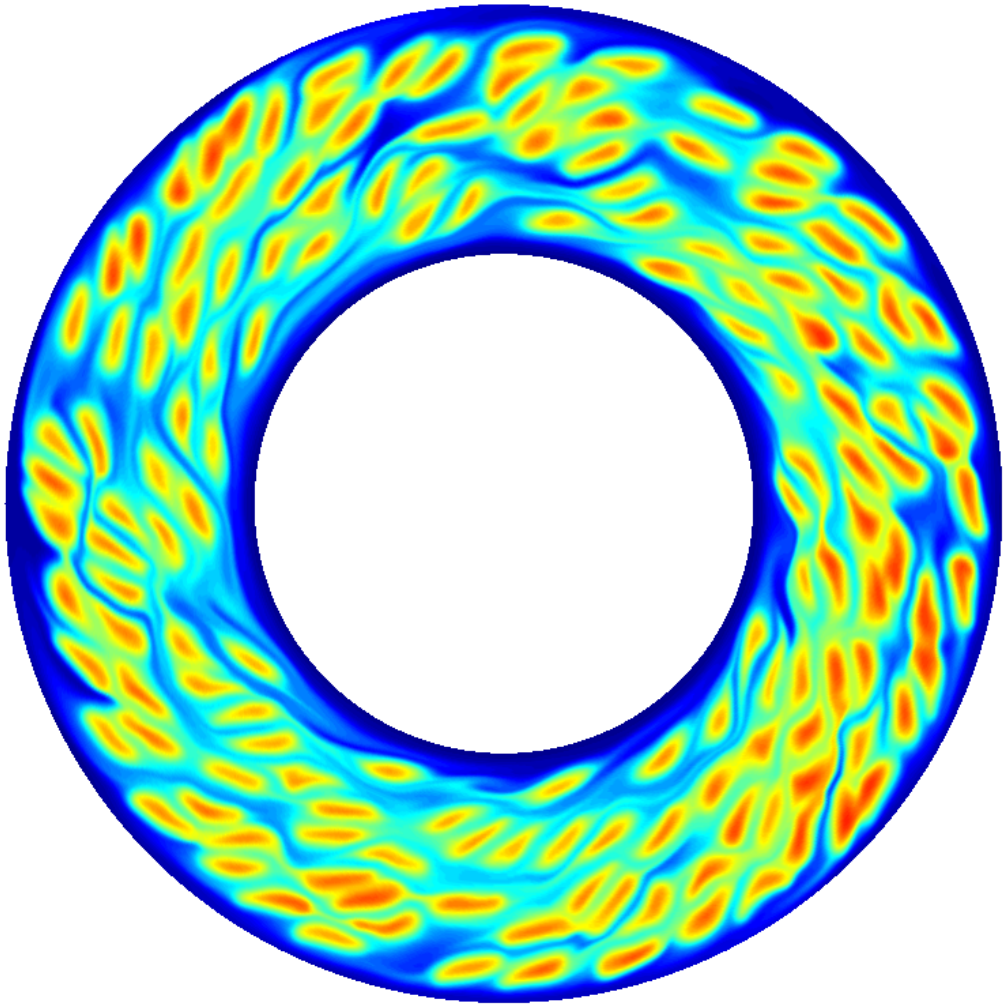}}}
      \label{f:VF40S4}}
\setcounter{subfigure}{0}
\renewcommand*{\thesubfigure}{(e)} 
      \hspace{0cm}\subfigure[$t = 75$]{\scalebox{0.5}{{\includegraphics{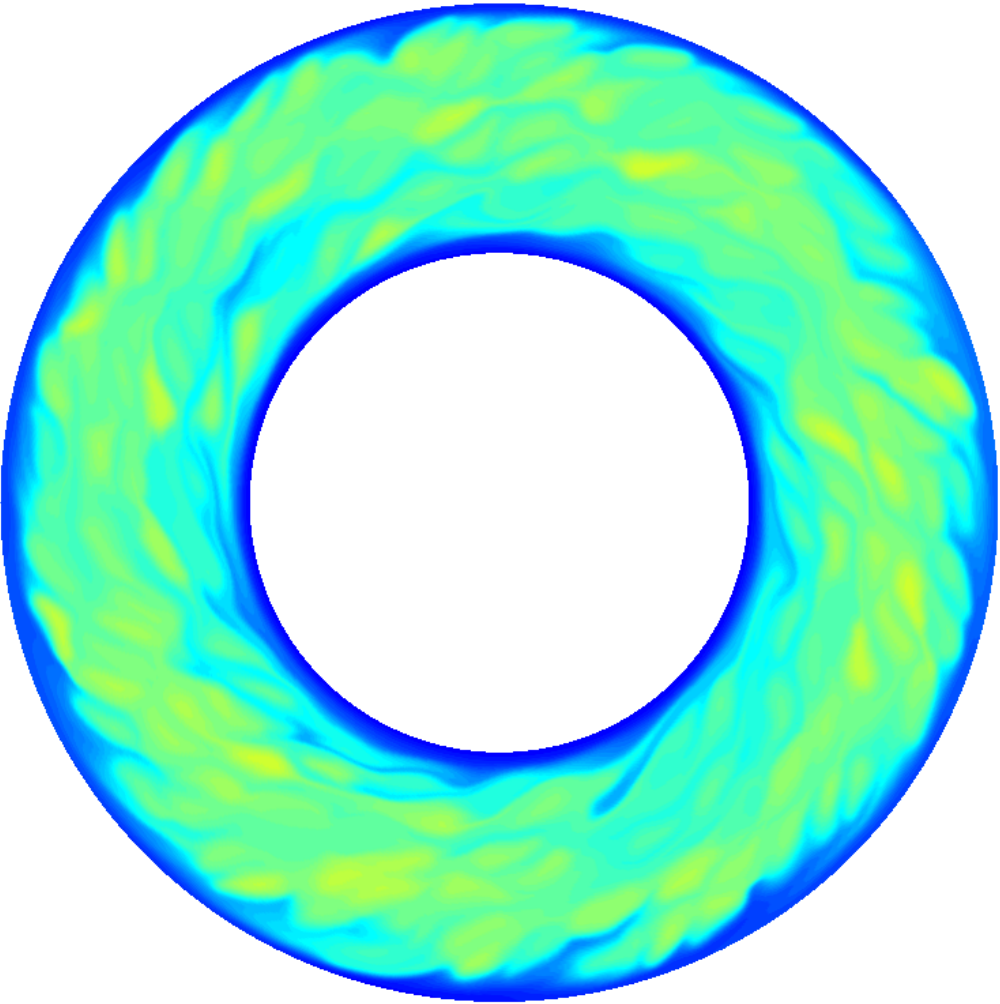}}}
      \label{f:VF40S5}}
\setcounter{subfigure}{0}
\renewcommand*{\thesubfigure}{(f)} 
      \hspace{0cm}\subfigure[$t = 150$]{\scalebox{0.5}{{\includegraphics{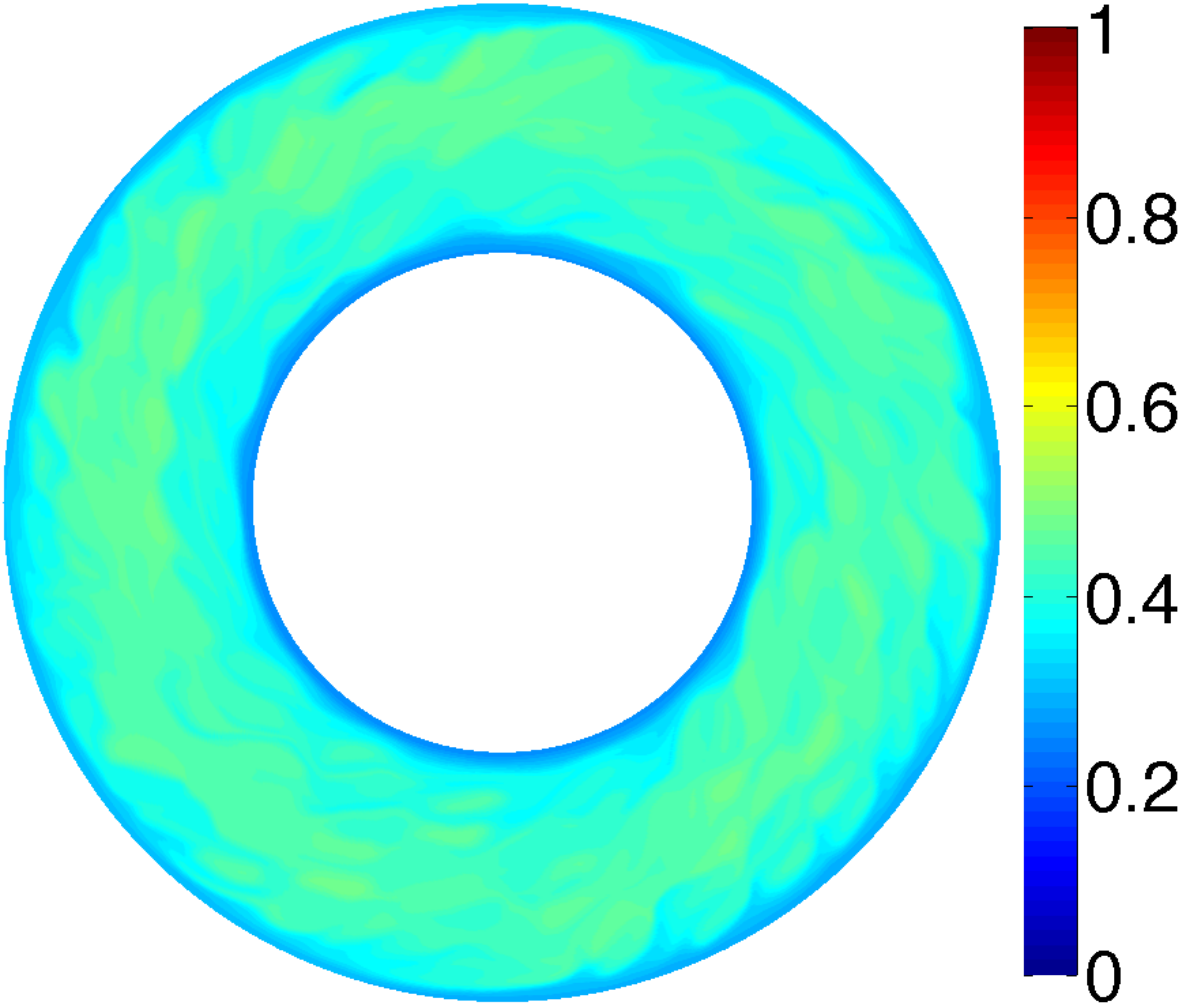}}}
      \label{f:VF40S6}} 
    \end{minipage}      
\mcaption{{Here} we show an example of a simulation of mixing in a
suspension of vesicles. The suspension has a $40\%$ area fraction of
vesicles and no viscosity contrast. The initial condition for the
solute is one inside the vesicles and zero outside the vesicles.  This
advection-dominated transport problem has $\pec = 1e+4$. In this
simulation, we use 256 and 1024 discretization points in the radial and
the azimuthal directions, respectively.  In the simulation of the
vesicle suspension, we discretize a vesicle with $N = 96$ points and a
wall with $N_{\text{wall}} = 256$ points. The time horizon ($T_h =
150$) corresponds to 24 revolutions of the inner cylinder in the
Couette apparatus. We take a total of 3,750 time steps in our advection
simulation.  All numerical simulations were performed using an in-house
MATLAB code.}{f:introFigure}

\end{figure}

\paragraph*{\textit{Limitations:}}
The main limitation is that we only consider a specific two-dimensional
flow.  So, generalizations to other type of flows are not immediate.
Also, we consider several metrics for mixing, but other metrics can be
considered.

\paragraph*{\textit{Related work:}}
Mixing has been studied extensively as it is important in many
scientific and industrial settings.  Classical works in
mixing~\cite{Annaswamy-Ghoniem95, Leiken-Marsden-e05} consider large
scale systems such as combustion in engines and pollution in seas.  We
are interested in mixing in microfluidic settings in which the flow
complexity is driven by moving boundaries or suspensions of deformable
particles.

Mixing in flows with moving and deformable boundaries have been
studied~\cite{Bottausci-Cardonne-e04, Ottino-Wiggins04,
Hessel-Schonfeld-e05}.  However, none of these works discuss mixing of
vesicle suspensions. Of course, there are many studies on the rheology
of vesicle flows such as tank-treading and tumbling
motions~\cite{Keller-Skalak82,Kraus-Lipowsky-e96, Nogouchi-Gompper05,
Misbah06, Ghigliotti-Misbah-e11}.  Vesicles also model red blood cells
and are used to study microcirculation~\cite{Popel-Johnson05}.
Therefore, mixing in vesicles can be related to mixing in capillaries
and arterioles.  Regarding the numerical method for vesicles, we use our
in-house algorithms for vesicle simulations~\cite{Veerapaneni-Biros-e09,
Rahimian-Biros-e10, Quaife-Biros14a, Quaife-Biros16} and we refer the
reader to~\cite{Quaife-Biros14a, Quaife-Biros16} for a review of the
related work on the two-dimensional vesicle simulations. 

Another important aspect in our work is the quantification of mixing.
Although there is extensive work on metrics for mixing, there is not a
universal measure~\cite{Khakhar-Ottino86, Ottino90, Mathew-Petzold-e05,
Lin-Doering-e11,Fourer-Schmid-e14}.  We review some of the metrics
specifically for advection.  Metrics derived from dynamical systems
consider the locations of tracer particles after a single period of a
periodic flow.  One example is the Poincar\'{e}
section~\cite{Aref-Balachandar86} which examines the position of
particles after multiple periods of the flow.  If the separation between
neighboring particles increases exponentially with each period, then we
say that the flow is chaotic, and the exponent, which is called the
stretching rate or Lyapunov exponent, quantifies the mixing.  In
particular, larger Lyapunov exponents correspond to better mixing, and
this approach is used in~\cite{Muzzio-Swanson91}.  Mixing can also be
measured statistically.  One measure is the mixing variance metric.  Another
 measure is the Kolmogorov-Sinai entropy that computes an
integral of Lyapunov exponents over a
domain~\cite{DAlessandro-Mezic-e99}.  Another set of metrics is based on
tracking the interface between the solute and the solvent.  When an
effective mixing takes place, this interface grows rapidly. The
exponential rate of the growth is called the interface
stretch~\cite{Meleshko-Aref96} and measures global stretching unlike
Lyapunov exponents~\cite{Adrover-Alvarez-e98,Alvarez-Giona-e98}.

The aforementioned metrics are appropriate to measure mixing for
advection dominated flows.  However, the introduction of diffusion
further enhances mixing.  We refer the reader to~\cite{Finn-Byrne-e04} for a more detailed discussion on  the different metrics for different Peclet numbers and initial conditions.  When quantifying mixing due to
diffusion, metrics that are based on the  the
solute are more informative.  For example, the Euclidean ($L^2$) norm and the maximum norm
($L^{\infty}$)~\cite{Rothstein-Gollub-e99, Ashwin-Kirkby-e02} can be
used.  However, $L^p$ norms do not decay in the absence of diffusion,
and therefore cannot quantify mixing due to advection. Thus, there is a
need for a metric that captures mixing due to both diffusion and
advection.  One metric that captures mixing due to diffusion and
advection is the negative index $H^{-1}$ Sobolev
norm~\cite{Mathew-Petzold-e05, Mathew-Petzold-e07}, which we will refer
to as the "{\em mix norm}".  Additionally,~\cite{Doering-Thiffeault06,
Thiffeault11} compare the $H^{-1}$ norm with $L^p$ norms.  In addition
to being able to capture mixing due to advection and diffusion, the
$H^{-1}$ norm depends on the initial concentration field.

Since all the numerical methods we adopt often appear in the literature,
we only briefly describe our numerical scheme.  We use a
Fourier-Chebyshev collocation~\cite{Trefethen00,Boyd13} method to
discretize the equation in space. We use a Strang operator splitting
time-stepping scheme~\cite{Strang68}, which we combine with a
semi-Lagrangian method for the advection~\cite{Robert81,
Xiu-Karniadakis01}. 

\paragraph*{\textit{Methodology:}}\label{s:method}
We consider the advective and the diffusive mixing of a passive scalar
in a Couette apparatus (see \figref{f:geometry}) that has a rotating
the inner cylinder and a stationary outer cylinder. The relevant
dimensionless number of the transport problem is the ratio of the
advective transport rate and the diffusive transport rate, or the
Peclet number $\pec$~\cite{Thiffeault-Gibbon-e04},
\begin{equation} \label{e:PeNumber}
  \pec = {\overline{V}L_c}/{D}. 
\end{equation}
Here $\overline{V}$ is the time average of the $L^2$ norm of the
velocity field $\mathbf{v}$, i.e.~$\overline{V} = \langle {\| \mathbf{v}
\|}_{L^2} \rangle$ and measures the kinetic energy, $L_c$ is the
characteristic length scale (the diameter of the apparatus), and $D$ is
the diffusivity of the transported quantity.
As an example, we discuss transport in microcirculation. The diameters of a
capillary and an arteriole are $\mathcal{O}(10 \mathrm{~\mu m})$ and the mean
velocities of blood flow in them are $\mathcal{O}(1)\mathrm{~mm/s}$ and
$\mathcal{O}(10)\mathrm{~mm/s}$, respectively~\cite{Goldsmith-Skalak75,
Wang-Popel93, Popel-Johnson05}.  The diffusivities of oxygen and an
iron-oxide nanoparticle are
$\mathcal{O}(10^{-3})\mathrm{~{mm}^2/s}$~\cite{Wang-Popel93} and
$\mathcal{O}(10^{-7})\mathrm{~{mm}^2/s}$~\cite{Nacev-Shapiro-e11}, respectively.
Therefore, the Peclet number for the transfer of oxygen or nanoparticles
ranges from $10$ to $10^4$.

Vesicle suspensions have several parameters such as the distribution of
sizes of the vesicles, the reduced volume, the bending resistance, the
"volume" fraction (in 2D the ratio between the area occupied by the
vesicles and the total area of the apparatus; we call it area fraction
throughout the paper), and the viscosity contrast between the fluid
inside the vesicle and the bulk fluid. All these parameters could
affect mixing. Here, however, we consider only two main vesicle parameters, the
area fraction and the viscosity contrast.  Of course, another parameter
is the imposed external velocity field.  In our case it is the velocity
generated by the rotating inner cylinder in the Couette apparatus and
we parameterize it by the Peclet number.  Again, taking an example from microcirculation,
the volume fraction of red blood cells
in human blood is typically around $45\%$ and their viscosity contrast
with plasma ranges from $5$ to $10$~\cite{Goldsmith-Skalak75}.

Our numerical simulations require two steps.  First, we
simulate the vesicle motion for various values of area fraction and
viscosity contrast and then compute the velocity field on a Fourier-Chebyshev
grid.  Second, using this velocity field, we simulate mixing in the
Couette apparatus for the initial conditions in \figref{f:initConds} and
denote the corresponding concentration of the solute with $\phi$. We remark that
numerical algorithms for the calculation of the velocity and $\phi$ are
very different. The suspension dynamics are computed using a boundary integral equation
while the advection-diffusion equations is computed using a pseudospectral method.
We also simulate mixing
of the same initial concentration in the apparatus with the same Peclet
number but without any vesicles, and we denote this concentration by
${\phi}_0$.  Using $\phi_0$, we define a mixing efficiency $\eta$ as
\begin{equation} \label{e:mixingEff}
  \eta = \frac{\| {\phi}_0 \|}{\| \phi \|},
\end{equation}
which compares the mixing efficiency of the Couette apparatus with
vesicles to that without vesicles (the default Couette flow). If $\eta$
is greater than one the vesicle flow mixes better.  When computing the
Peclet number~\eqref{e:PeNumber}, we use the spatio-temporal average of
the velocity field to quantify the advective transport rate. Since the
velocity field $\mathbf{v}$ depends on the volume fraction and the
viscosity contrast, it changes with the area fraction and viscosity
contrast of the suspension. Thus, we adjust the diffusivity $D$ to keep
the Peclet number the same in computing ${\phi}_0$ and $\phi$. In this
manner, we investigate the effects of area fraction and viscosity
contrast on the mixing efficiency.  Additionally, we look at these
effects under various Peclet numbers for the initial conditions in
\figref{f:initConds}.

\paragraph*{\textit{Organization of the paper:}} In \secref{s:vesicles}
we briefly summarize the formulation for the numerical simulation of
vesicle flows. We, then, present the temporal and the spatial
discretization methods for the advection-diffusion equation in
\secref{s:transport}. After we define the mixing metrics in
\secref{s:metrics}, we show the results of the numerical experiments and
discuss the effects of the area fraction, the viscosity contrast, and
the initial condition of the transported quantity on the mixing
efficiency in \secref{s:experiments}.

\paragraph*{\textit{Notation:}} We summarize the main notation used in
this paper in \tabref{t:notation}.
\begin{table}[htb]
\mcaption{List of frequently used notation.}{t:notation}
\centering
\begin{tabular}{l|l}
 \hline 
 Symbol & Definition \\ 
 \hline
  Pe 
  & Peclet number: ratio of the advective transport rate to the diffusive
  transport rate \\

  AF
  & Area fraction: ratio of the area occupied by vesicles to the area
  of the Couette apparatus \\ 

  VC &
  Viscosity contrast: ratio of the fluid viscosity inside a vesicle to
  the fluid viscosity in the bulk \\

  $N_r$ & 
  Number of collocation points in the radial direction $r$ \\

  $N_{\theta}$ & 
  Number of collocation points in the azimuthal direction $\theta$ \\

  ${\phi}_{0}$ & 
  Concentration in the absence of vesicles \\

  $\phi$ & 
  Concentration in the presence of vesicles \\

  $\eta$ & 
  Mixing efficiency: ratio of $\| {\phi}_0 \|$ to $\| {\phi} \|$ \\

  ${\mathbf{v}}_0$ & 
  Velocity field of a Couette flow without vesicles \\
  
  $\mathbf{v}$ & 
  Velocity field of a vesicle suspension \\
  
 \hline 
\end{tabular} 
\end{table} 

\section{Simulation of a vesicle suspension}\label{s:vesicles}
\begin{figure}
\begin{minipage}{\textwidth}   
   \setcounter{subfigure}{0}
   \renewcommand*{\thesubfigure}{(a)}
   \hspace{0.5cm}\subfigure[Domain of a vesicle simulation]{\scalebox{0.65}{{\includegraphics{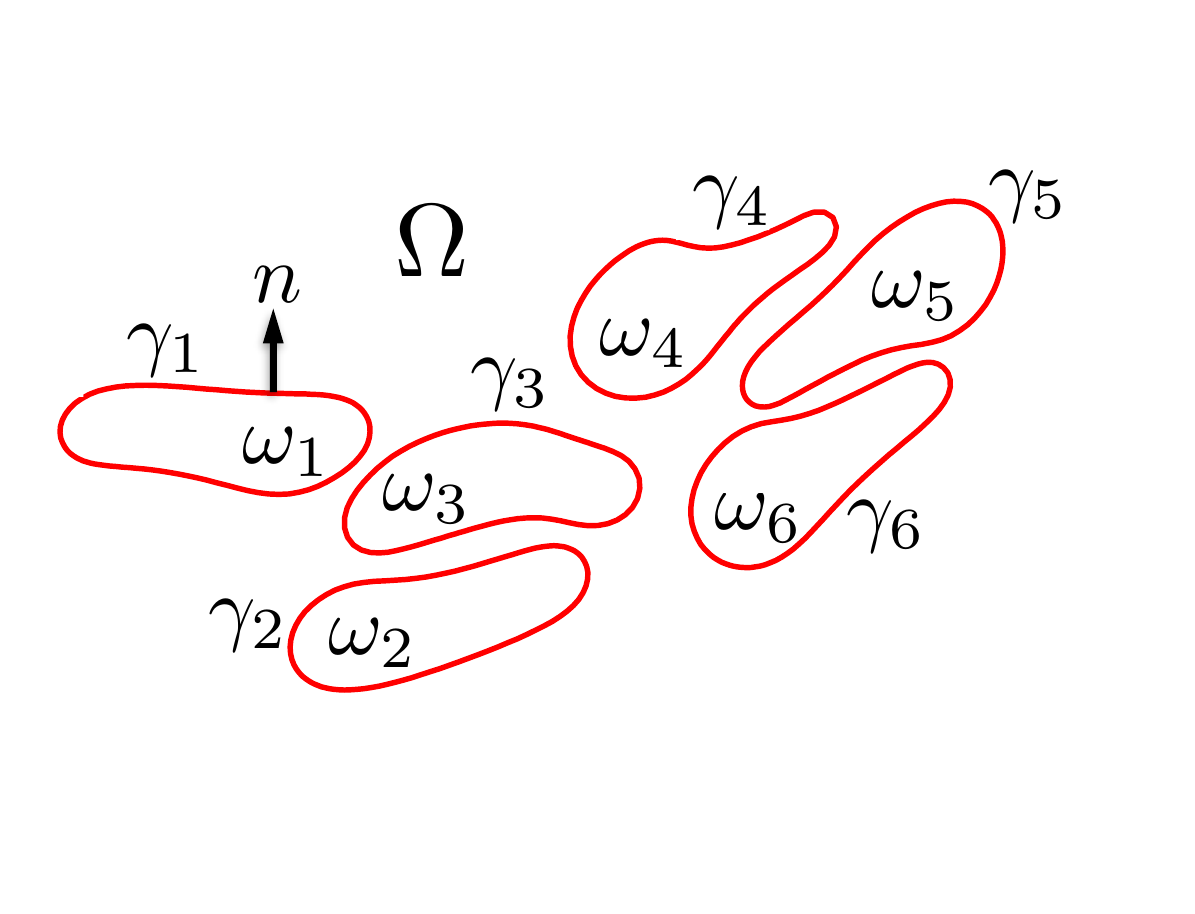}}}
   \label{f:vesicDomain}}
  \setcounter{subfigure}{0}
  \renewcommand*{\thesubfigure}{(b)} 
  \hspace{0.3cm}\subfigure[Grid points for a mixing simulation]{\scalebox{0.5}{{\includegraphics{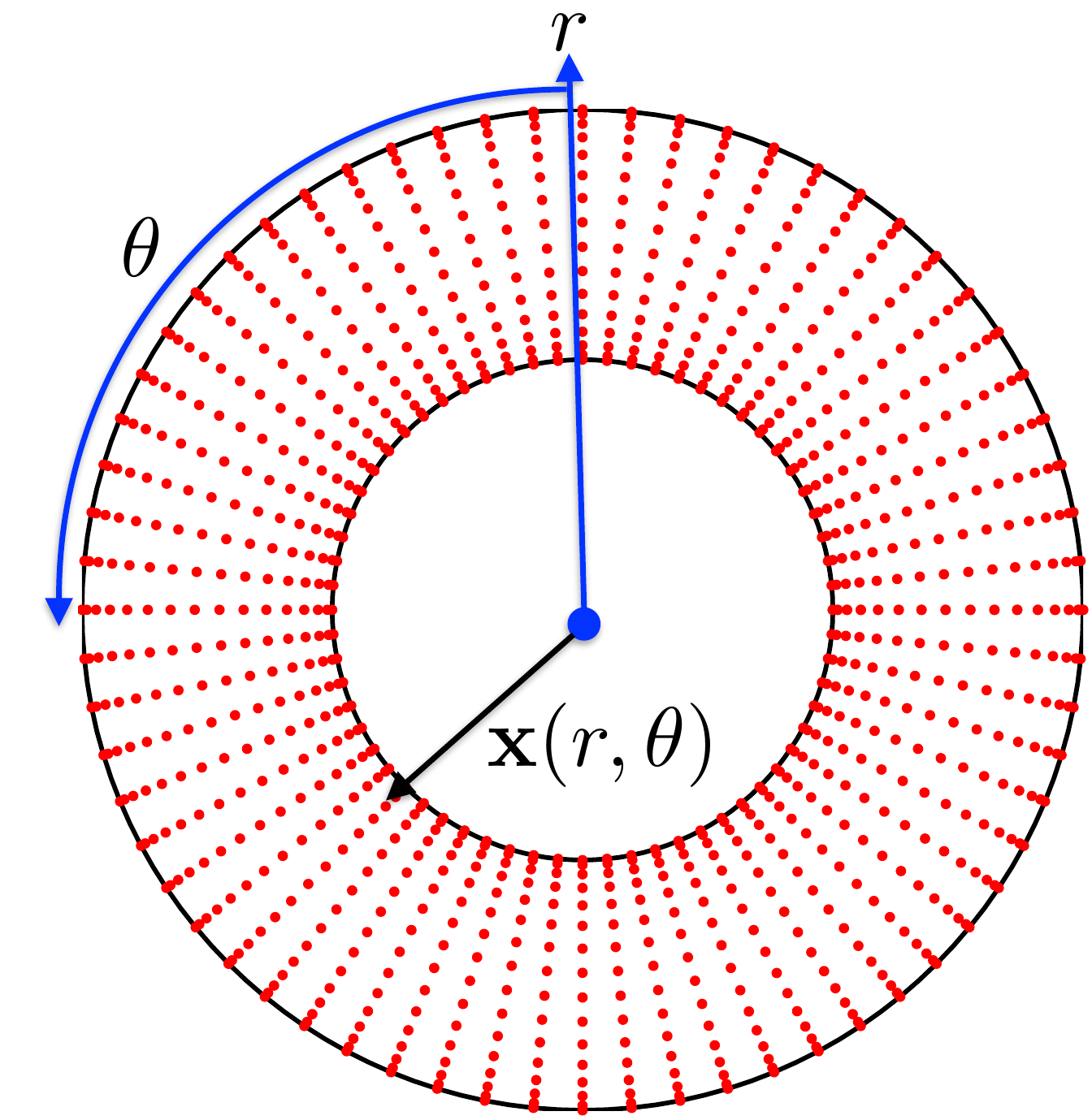}}} 
   \label{f:gridCoords}}
  \end{minipage}
 \mcaption{{We} illustrate the domain of a vesicle simulation in
 \figref{f:vesicDomain}. The interior and boundary of the $i^{th}$
 vesicle is denoted by ${\omega}_i$ and ${\gamma}_i$, respectively. The
 vesicle suspension formulation is described in \secref{s:vesicles}.
 \figref{f:gridCoords} shows the polar coordinate Fourier-Chebyshev
 grid where we discretize the transport equation.
 The transport equations are described in
 \secref{s:transport}.}{f:domainAndGrid}
\end{figure}

In this section, we briefly summarize the numerical scheme for the
vesicle dynamics.  We refer the reader to~\cite{Pozrikidis-92,
Quaife-Biros14a, Quaife-Biros16} for further details. The fluid domain
is denoted by $\Omega$, the boundary of the $i^{th}$ vesicle by
${\gamma}_i$, and the interior of the $i^{th}$ vesicle by
$\omega_{i}$.  We also write $\gamma = {\bigcup}_i {\gamma}_i$ and
$\omega = {\bigcup}_i {\omega}_i$.  The vesicle boundaries $\gamma_{i}$
are parameterized as ${\mathbf{x}}_{i}(s,t)$, where $s$ is the arc
length and $t$ is time. This notation is also described in
\figref{f:domainAndGrid}.

In Stokesian fluids, the inertial forces are negligible compared to the
viscous forces resulting in a small Reynolds number.  For $Re = 0$ the
momentum and continuity equations are
\begin{equation} \label{e:stokesCont}
  -\mu \Delta \mathbf{v} + \nabla p = 0, \quad 
  \text{div}(\mathbf{v}) = 0, \quad 
  \text{in}\: \Omega \setminus \gamma.
\end{equation}
Here $\mu$ is the fluid viscosity, $\mathbf{v}$ is the fluid velocity,
and $p$ is the pressure. We assume that the viscosities of the exterior
and interior fluids are constant. The no-slip boundary condition on the
interface of vesicles implies that 
\begin{equation} \label{e:noSlip}
  \frac{d{\mathbf{x}}_{i}}{dt} = 
  \mathbf{v}({\mathbf{x}}_{i}), \quad \text{on}\, \gamma_{i}. 
\end{equation}
In addition, we impose a no-slip velocity boundary condition on the
inner and outer boundaries of the Couette apparatus.  Next, we
assume that the surface of the vesicles is locally inextensible.
This constraint is to the
divergence of the velocity on $\gamma$ vanishing,
\begin{equation} \label{e:inExtensible}
  {\text{div}}_{\gamma_{i}} \mathbf{v}({\mathbf{x}}_{i}) = 0, 
  \quad \text{on}\, \gamma_{i}. 
\end{equation}
Another governing equation comes from the balance of momentum on the
interface of the vesicles. It enforces the jump in the surface traction
to be equal to the net force applied by the interface onto the fluid,
\begin{equation} \label{e:tracJump}
  \mathbf{f}({\mathbf{x}}_{i}) = [\![ T\mathbf{n} ]\!],  \quad
  \text{on}\, \gamma_{i},
\end{equation}
where $T = -pI + \mu \left(\nabla \mathbf{v} + {\nabla
\mathbf{v}}^T\right)$ is the Cauchy stress tensor, $\mathbf{n}$ is the
outward normal vector of $\gamma_{i}$, $[\![ \cdot ]\!]$ is the jump
across the interface, and $\mathbf{f}$ is the net force applied by the
interface onto the fluid. The net force $\mathbf{f}$ is the nonlinear
function of ${\mathbf{x}}_{i}(s,t)$
\begin{equation} 
  \label{e:formOfF}
  \mathbf{f}({\mathbf{x}}_{i}) = -{\calK}_b 
    \frac{{\partial}^4 {\mathbf{x}}_{i}}{\partial s^4} + 
    \frac{\partial}{\partial s}\left(\sigma({\mathbf{x}}_{i})
    \parder{{\mathbf{x}}_{i}}{s} \right), \quad \text{on}\, \gamma.
\end{equation}
The first term on the right hand side is the force due to bending
modulus ${\calK}_b$ and the second term is the force due to the tension
on the interface. We assume that there is no other force such as the
gravitational force on the interface. Finally, the position of the
boundaries of $M$ vesicles evolves as 
\begin{equation} 
\label{e:evolution}
  \frac{d{\mathbf{x}}_{i}}{dt} = {\mathbf{v}}_{\infty}({\mathbf{x}}_{i}) + 
  \sum_{j = 1}^M {\mathbf{v}}_{j}({\mathbf{x}}_{i}), 
  \quad i = 1,\ldots,M,
\end{equation}
where ${\mathbf{v}}_{\infty}({\mathbf{x}}_{i})$ is the background
velocity and ${\mathbf{v}}_{j}({\mathbf{x}}_{i})$ is the velocity
due to the $j^{th}$ vesicle acting on the $i^{th}$ vesicle.

The complete set of nonlinear
equations~\eqref{e:stokesCont}--\eqref{e:evolution} governs the
evolution of the vesicle interfaces. We use an integral equation method
since it can naturally handle the moving geometry, achieve high-order
accuracy, and resolve the viscosity contrast.  Because of the
high-order derivatives in \eqref{e:formOfF}, a semi-implicit method is
used. The tension and traction jump in \eqref{e:formOfF} for a given
${\mathbf{x}}_{i}(s,t)$ are computed spectrally with Fourier
differentiation. Finally, the boundary conditions \eqref{e:noSlip},
\eqref{e:inExtensible} result in a coupled system of equations with
integral, differential and algebraic components. Once the positions and
tensions of the vesicles are obtained, the integral equation
formulation allows us to compute the velocity $\mathbf{v}$ at any point
in the domain as a postprocessing step.  Then, this velocity field is
used in our advection-diffusion solver.

\section{Advection-diffusion equation} \label{s:transport}
The advection-diffusion equation governs mixing of a passive scalar.
Its nondimensional form with Neumann boundary conditions is
\begin{subequations} \label{e:governEqns}
\begin{alignat}{2}
\parder{\phi}{t} + \mathbf{v} \cdot \nabla \phi &= \frac{1}{\pec} \Delta \phi		& 	\quad \text{in}\: \Omega,  \label{e:transport} \\
\parder{\phi}{r} &= 0      											       & 	\quad \text{on}\: \Gamma.  \label{e:BCs}
\end{alignat}%
\end{subequations}
Here, $\phi$ is the concentration, $\mathbf{v}$ is the velocity, and
$t$ is time. Additionally, $\Omega$ is the Couette geometry
(\figref{f:geometry}) and $\Gamma$ is its boundary.  We now present the
numerical scheme we have adopted to solve~\eqref{e:governEqns}.

\subsection{Temporal discretization}\label{s:strang}
The Strang splitting method expresses the solution operator $\calL$
in~\eqref{e:transport} in terms of the advection (${\mathcal{L}}_A$) and
the diffusion (${\mathcal{L}_D}$) operators:
\begin{subequations} \label{e:strangSplit}
\begin{align}
\parder{\phi}{t} + {\calL}_A \phi & = 0, \label{e:advOperator}\\
\parder{\phi}{t} + {\calL}_D \phi & = 0. \label{e:diffOperator}
\end{align}
\end{subequations}
Here, ${\mathcal{L}}_A = \mathbf{v} \cdot \nabla$ and ${\mathcal{L}}_D
= -\frac{1}{\pec} \Delta$. Given some initial concentration field, the
Strang splitting updates the concentration in three steps: first it
solves the advection equation~\eqref{e:advOperator} in $[t^n,
t^{(n+1)/2}]$, second the diffusion equation~\eqref{e:diffOperator} in
$[t^n,t^{n+1}]$, and third the advection equation~\eqref{e:advOperator}
in $[t^{(n+1)/2}, t^{n+1}]$.  This splitting is second-order accurate,
but the methods used to solve each subproblem also determines the
accuracy of the complete scheme. In this study, we solve the advection
problem~\eqref{e:advOperator} and the diffusion
problem~\eqref{e:diffOperator} with a semi-Lagrangian method and a
Crank-Nicolson method, respectively.  This decoupling results in an
unconditionally stable scheme~\cite{Christlieb-Qiu-e14}.

\paragraph*{\textit{Semi-Lagrangian method for advection:}}
The advection equation~\eqref{e:advOperator} in Lagrangian form is
\begin{equation*}
  \label{e:lagrangAdv}
  \dder{\phi}{t} = \parder{\phi}{t} + \mathbf{v} \cdot \nabla \phi = 0, 
\end{equation*}%
which means that $\phi$ is constant along the characteristic path
$\mathbf{x}(t)$ which satisfies
\begin{align}
  \label{e:chars}
  \dder{\mathbf{x}}{t} & = \mathbf{v}\left(\mathbf{x},t\right).
\end{align}

In the semi-Lagrangian method, first we solve~\eqref{e:chars} backward
in time to find the Lagrangian point or ``{\em departure point}"
$\mathbf{x}_d$ that arrives at a point $\mathbf{x}_a$ that coincides
with the discretization points used for the diffusion solve (see
\figref{f:gridCoords}).  This trajectory is computed with the
second-order explicit midpoint rule
\begin{subequations} \label{e:midPoint}
\begin{align}
{\mathbf{x}}_m & = {\mathbf{x}}_a - \mathbf{v}\left({\mathbf{x}}_a,t^n\right)\frac{{\Delta t}_A}{2}, \label{e:midPoint1} \\
{\mathbf{x}}_{d} & = {\mathbf{x}}_a - \mathbf{v}\left({\mathbf{x}}_m,t^n + \frac{{\Delta t}_A}{2}\right){\Delta t}_A. \label{e:midPoint2}
\end{align}%
\end{subequations}
Here ${\Delta t}_A$ is the time step size for the advection problem
(usually ${\Delta t}_D \geq {\Delta t}_A$ where ${\Delta t}_D$ is the
time step size for the diffusion problem).  Since we have integrated
along the characteristic, the concentration at ${\mathbf{x}}_a$
satisfies $\phi\left({\mathbf{x}}_a,t^{n+1}\right) =
\phi\left({\mathbf{x}}_d,t^{n}\right)$. In general, the departure points
${\mathbf{x}}_d$ do not coincide with the grid points, thus
we interpolate the concentration at ${\mathbf{x}}_d$ using cubic
interpolation with $\phi({\mathbf{x}}_a,t)$. Additionally, we also
interpolate the velocity at the mid-point ${\mathbf{x}}_m$
in~\eqref{e:midPoint2} using the same method.  This particular
semi-Lagrangian scheme is second-order accurate in
time~\cite{Falcone-Ferretti98, Xiu-Karniadakis01}. 

\paragraph*{\textit{Crank-Nicolson for diffusion:}}

We use a Crank-Nicolson scheme to discretize the diffusion
equation~\eqref{e:diffOperator} in time 
\begin{equation} \label{e:crankNicols}
\frac{{\phi}^{n+1} - {\phi}^{n}}{{\Delta t}_D} = \frac{1}{\pec} {\nabla}^2 \left(\frac{{\phi}^{n+1} + {\phi}^{n}}{2}\right).
\end{equation}
Since~\eqref{e:crankNicols} is not $L$-stable, high frequency
components of $\phi$ can lead to spurious numerical
oscillations~\cite{Crank-Nicolson1947}.  Since we choose discrete
initial conditions (see \figref{f:initConds}), high frequencies
components will be present.~\cite{Leveque07}. Therefore, we require a
method that behaves as a numerical low-pass filter so that high
frequencies are suppressed. We apply the $L$-stable backward Euler
method initially to smooth the initial
condition~\cite{Zvan-Forsyth-e00, Wade-Deininger-e07}. Since backward
Euler is only first-order accurate, it is only applied for $t \in
[0,\Delta t_{D}]$ with a time step size ${\Delta t}^{BE} = {\Delta
t}_{D}^{2}$.  Then, to achieve second-order accuracy, Crank-Nicholson
is used for $t > \Delta t_{D}$.

\subsection{Spatial discretization}\label{s:collocation}
Taken advantage of symmetries in the geometry, We use polar coordinates
$(r,\theta)$ (see \figref{f:gridCoords}) and a pseudo-spectral
representation of $\phi$. Since $\phi$ is periodic in $\theta$, we use
a Fourier series in $\theta$
\begin{equation} \label{e:FourDisc}
\phi(r,\theta,t) = \sum_{k = -N_{\theta}/2 + 1}^{N_{\theta}/2} {\hatphi}_{k}(r,t) e^{ik\theta}.
\end{equation}
Then, we discretize the Fourier coefficients ${\hatphi}_{k}$ in $r$ using Chebyshev polynomials as 
\begin{equation*} \label{e:ChebDisc}
{\hatphi}_{k}(r,t) = \sum_{m = 0}^{N_r - 1} {\hatphi}_{k,m}(t) \cos{m\alpha}.
\end{equation*}
Here, $N_{\theta}$ is the number of uniformly distributed collocation
points in $\theta \in [0, 2\pi]$ and $N_r$ is the number of collocation
points in $r$. Additionally, $\alpha = \pi m/(N_r-1)$ $\in$ $[0,\pi]$
and we define the radial coordinate as $r =
\frac{1}{2}\left(1-\cos\alpha\right)\left(r_2 - r_1\right) + r_1$,
where the radii of the inner and the outer cylinders are $r_1$ and
$r_2$, respectively. The resulting grid points are illustrated in
\figref{f:gridCoords}.

After substituting~\eqref{e:FourDisc} into~\eqref{e:crankNicols} and
applying the operator ${\nabla}^2$, the resulting diagonal set of
linear equations is
\begin{equation} \label{e:systOfEqns}
{\calA}_k^{-}{\hatphi}_k^{n+1} (r) =  {\calA}_k^{+}{\hatphi}_k^{n} (r),
\end{equation}
where the operators ${\calA}_k$ are
\begin{equation*} \label{e:LinOperator}
{\calA}_k^{\mp} = \frac{I}{\Delta t} \mp \frac{1}{2\pec} \left( \frac{1}{r} \parder{}{r} + \pparder{}{r} + \frac{k^2}{r^2}\right),
\end{equation*}
and $I$ is the identity matrix.  Equation~\eqref{e:systOfEqns} is
efficiently solved using the fast cosine transform.

We have tested our numerical scheme on different initial conditions and
velocity fields.  For smooth velocity fields and initial conditions,
the method is second-order accurate in time and spectrally accurate in
space.  We have also tested our solver on velocity fields that are not
continuous, such as those for vesicle suspensions, and we achieve
similar convergence rates for smooth initial conditions.

\section{Metrics of mixing}\label{s:metrics}
To measure mixing in advection-dominated transport, some of the early
work~\cite{Mathew-Petzold-e05, Thiffeault11} suggests that the $H^{-1}$
norm is appropriate, and discusses the disadvantages of $L^p$ norms.
In this section, we define and compare the $L^1$, $L^2$,
and $H^{-1}$ norms on an example problem.

The $L^p$ norm of the concentration $\phi$ is 
\begin{equation*} \label{e:LpNorm}
{\| \phi \|}_{L^p} = {\left( \int_{\Omega} | \phi(\mathbf{x}) |^p
d\Omega \right)}^{1/p}.
\end{equation*}
We only use $p = 1, 2$.  The $H^{-1}$ norm is a negative Sobolev norm
and is defined as
\begin{equation*} \label{e:MixNorm}
{\| \phi \|}_{H^{-1}} = \left(\int_{\Omega} g(\mathbf{x}) \phi(\mathbf{x})
d\mathbf{x}\right)^{1/2}
\end{equation*}
where $g$ is the solution of the boundary value problem
\begin{alignat*}{2}
  \left(I - \Delta\right) g(\mathbf{x}) &= \phi(\mathbf{x})
  & \quad \mathbf{x} \in \Omega \\
  g(\mathbf{x}) &= 0 & \quad \mathbf{x} \in \Gamma
\end{alignat*}
In $L^{2}$ and $H^{-1}$, smaller norms of $\phi$ correspond to a more
mixed concentration field.

\paragraph*{Remark:} By integrating~\eqref{e:transport} over $\Omega$,
integrating by parts, and applying the Neumann boundary
condition~\eqref{e:BCs} and the incompressibility constraint, we have
\begin{equation*}
  \frac{\partial}{\partial t}\int_{\Omega}|\phi(\mathbf{x},t)|d\mathbf{x} = 0.
\end{equation*}
Here we have used the fact that the concentration field is positive.
Since the completely mixed state corresponds to a uniform concentration
$\overline{\phi}$, we have
\begin{equation*}
  \int_{\Omega}|\phi(\mathbf{x},t)|d\mathbf{x} = 
  \overline{\phi} \int_{\Omega} d\mathbf{x},
\end{equation*}
for all time.  Therefore the $L^1$ norm is not an appropriate norm to
measure mixing.

\begin{figure}[!htb]
 \begin{minipage}{\textwidth}
\setcounter{subfigure}{0}
\renewcommand*{\thesubfigure}{(a-1)} 
      \hspace{1.7cm}\subfigure[$\phi$ at $t = 0$]{\scalebox{0.4}{{\includegraphics{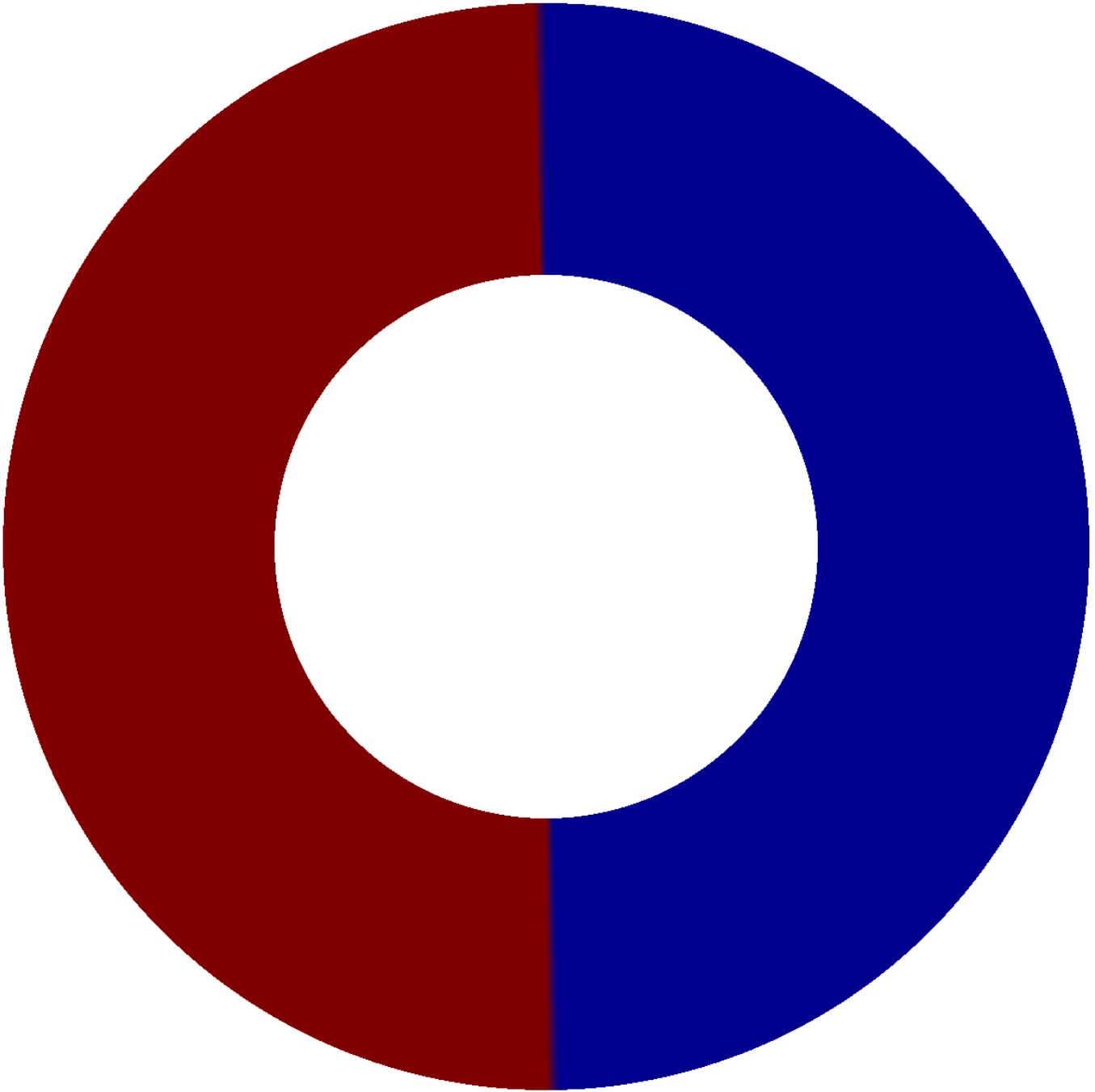}}}	
      \label{f:moonIC}}
\setcounter{subfigure}{0}
\renewcommand*{\thesubfigure}{(a-2)} 
      \hspace{1.6cm}\subfigure[$\phi$ at $t = 150$ with Pe = $5e+3$]{\scalebox{0.4}{{\includegraphics{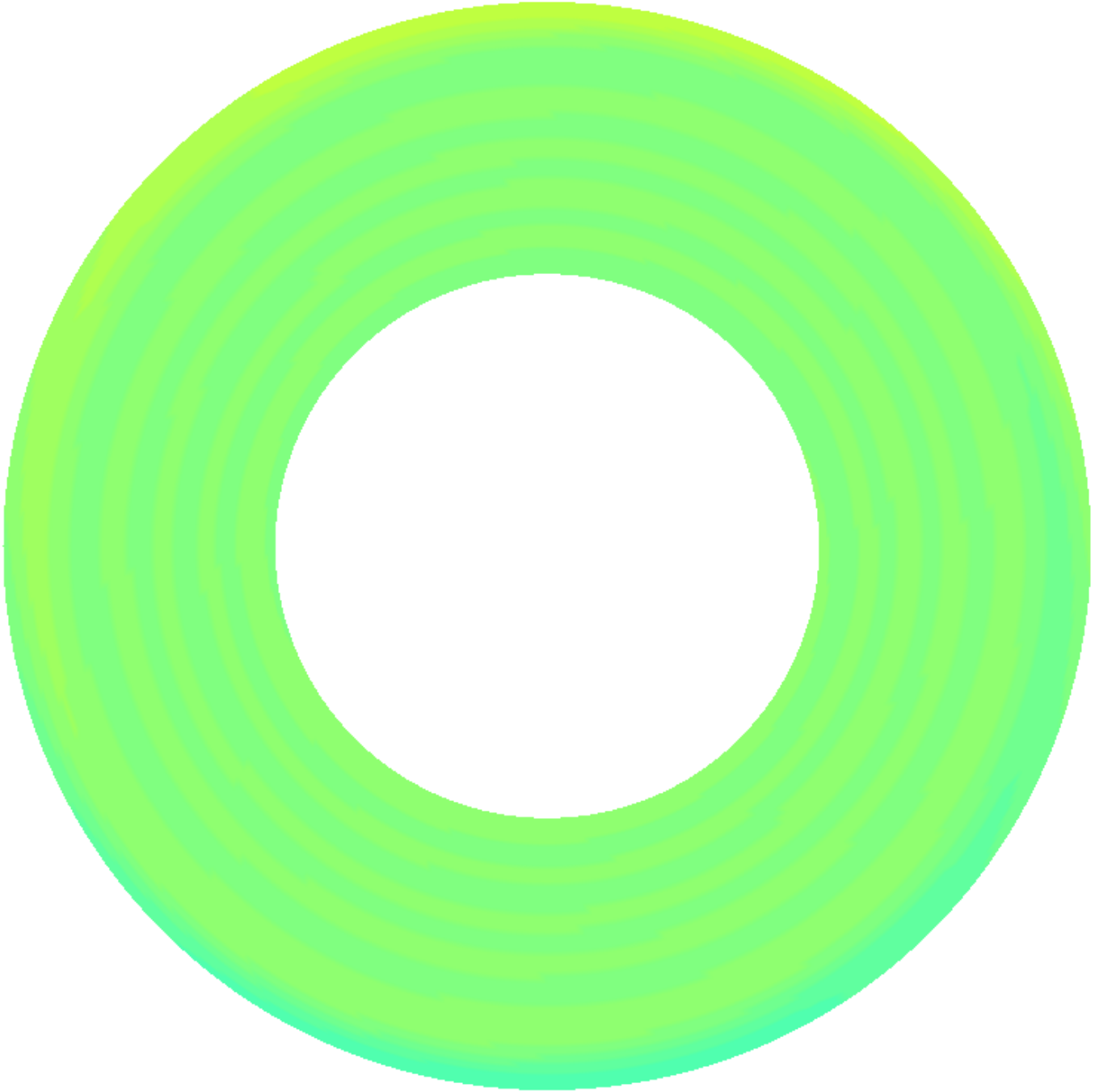}}}	
      \label{f:moonICHor}}
\end{minipage}
\begin{minipage}{\textwidth}
\centering           
\setcounter{subfigure}{0}
\renewcommand*{\thesubfigure}{(b-1)} 
      \hspace{-0.3cm}\subfigure[Pe = $\infty$]{\scalebox{0.4}{{\includegraphics{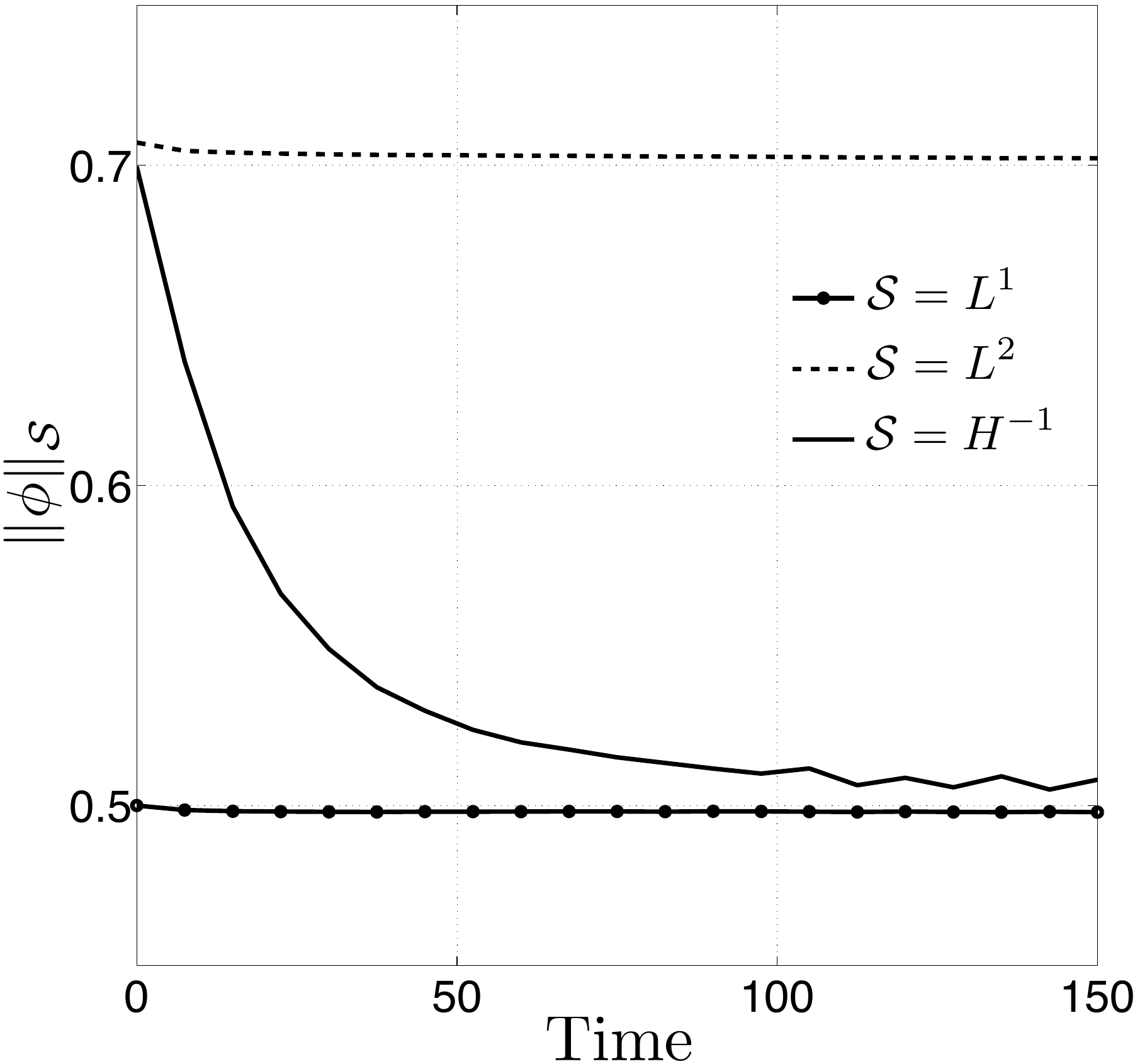}}}
      \label{f:normsPeInf}}
\setcounter{subfigure}{0}
\renewcommand*{\thesubfigure}{(b-2)} 
      \hspace{-0.2cm}\subfigure[Pe = $5e+3$]{\scalebox{0.4}{{\includegraphics{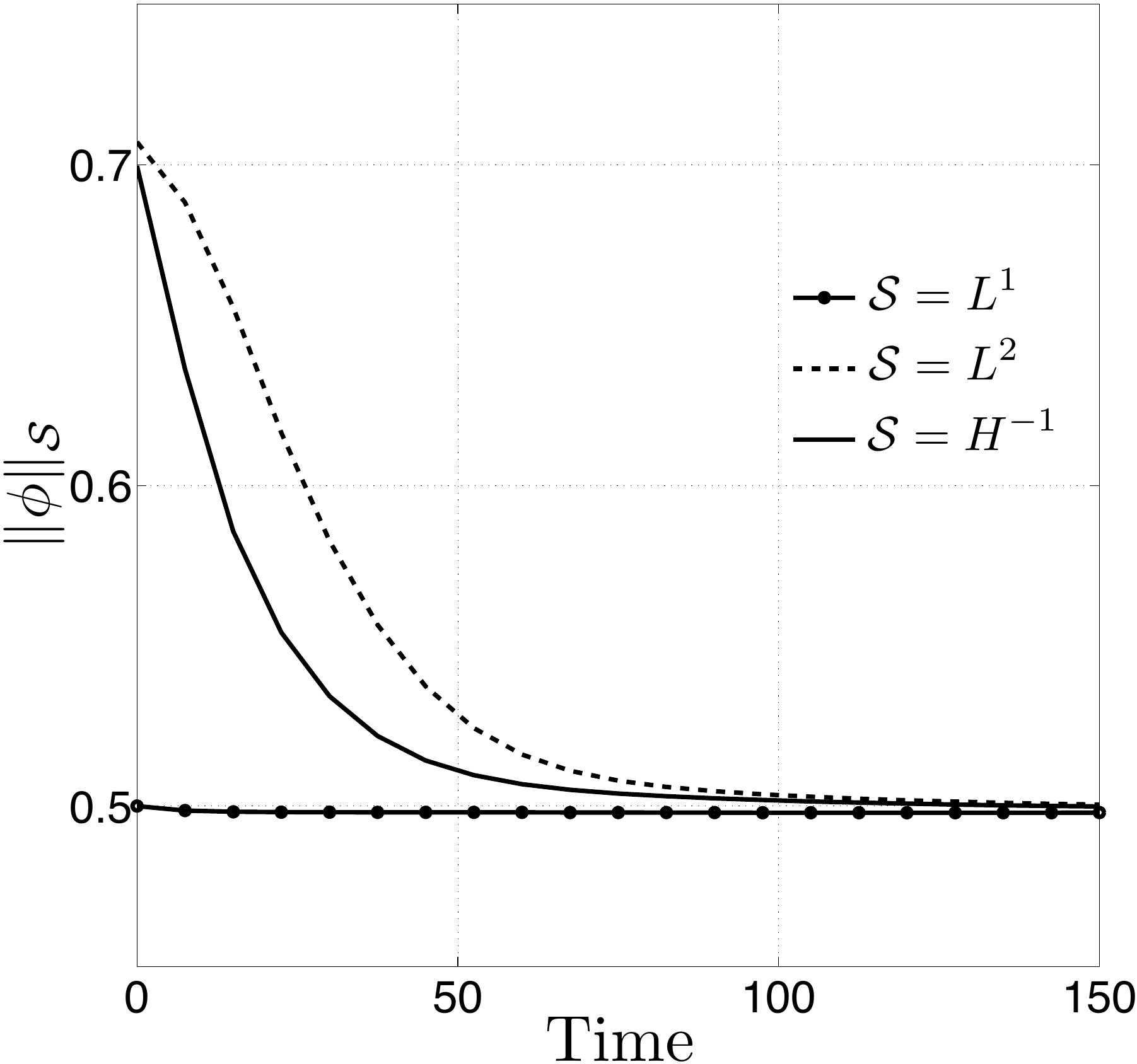}}}
      \label{f:normsPe}}
\end{minipage}
\mcaption{{We illustrate the differences between $L^{1}$,
$L^{2}$, and $H^{-1}$ norms when applied as mixing metrics of a
concentration field.  The advection velocity is a simple cylindrical
Couette flow without vesicles. We consider two Peclet numbers
$\pec = \infty$ and $\pec = 5e+3$.  Mixing measures are usually normalized by
their initial values in the literature, but we do not normalize here so
that we can demonstrate that the three norms converge to the same value
as $t \rightarrow \infty$. We measure the degree of mixing by $L^1$,
$L^2$ and $H^{-1}$ norms for both Peclet numbers in
\figref{f:normsPeInf} and \figref{f:normsPe}.  \figref{f:normsPeInf}
shows that the $L^1$ and $L^2$ norms do not decay without diffusion,
even though mixing is taking place by advection.  Additionally, we
observe in \figref{f:normsPe} that as the concentration becomes uniform
over the domain (i.e. $\phi (\mathbf{x}) = \overline{\phi}$), the $L^2$
and the $H^{-1}$ norms approach one another and ultimately converge to
the constant $L^1$ norm.}}{f:compareNorms}
\end{figure}

We illustrate the remark in \figref{f:compareNorms}.  The initial
condition is depicted in \figref{f:moonIC} and we consider a simple
Couette flow without any vesicles.  The number of collocation points
are $N_{r} = 128$ and $N_{\theta} = 512$.  The rest of the parameters
are in \tabref{t:MixingParams}. First, \figref{f:normsPeInf} shows that
neither the $L^1$ norm nor the $L^2$ are able to capture mixing due to
pure advection, while the $H^{-1}$ norm decays as the concentration is
mixed. Second, \figref{f:normsPe} demonstrates that in the presence of
diffusion, the $L^{1}$ norm is still independent of time, but the
$L^{2}$ and $H^{-1}$ norms decrease with mixing.  Additionally, in the
presence of diffusion, since the concentration becomes uniform, the
three norms approach the same value.

Since we are interested in advection-dominated transport, the $H^{-1}$
norm is preferable to the $L^{2}$ norm to quantify mixing and to define
the mixing efficiency~\eqref{e:mixingEff}.  If the efficiency $\eta>1$,
then ${\| {\phi}_0 \|} > {\| {\phi} \|}$ and thus the presence of
vesicles promotes mixing.  Similarly, if $\eta < 1$, then the presence
of vesicles suppresses mixing.

\section{Numerical experiments} \label{s:experiments}
In this section, we discuss the effects of the presence of vesicles on
mixing. We investigate the dependence of the mixing efficiency on the
area fraction (AF) and the viscosity contrast (VC).  The area
fraction of a vesicle suspension is the ratio of the area occupied by
the vesicles to the area of the fluid bulk. The viscosity contrast is a
ratio of viscosities of interior and exterior fluids.  Additionally,
since the $H^{-1}$ norm depends on an initial configuration, we run
tests on the different initial conditions in \figref{f:initConds}.  Let
us summarize the experiments we perform.
\begin{itemize}
\item \textbf{Effects of area fraction (\figref{f:VF10Snaps},
\figref{f:VF20Snaps} and \figref{f:VF40Snaps}):} Here we aim to
understand whether the presence of more vesicles in a suspension
promotes or suppresses mixing. For this purpose, we simulate mixing in
vesicle suspensions with area fractions $10\%$, $20\%$, and $40\%$, and
with the layer initial condition and Pe = $1e+4$. We visualize several
time snapshots of the concentration field, the vesicle positions, and
the velocity field (in fact, we visualize its {\em difference} from the
pure Couette flow).  The results indicate, for this setup, that
increasing the area fraction promotes mixing.

\item \textbf{Effects of the Peclet number and the initial condition
(\figref{f:VFMixing1} and \figref{f:VFMixing2}):} We simulate mixing in
the vesicle suspensions with the three area fractions of $10\%$,
$20\%$, and $40\%$ with various Peclet numbers and initial conditions
(\figref{f:initConds}).  In order to quantify the effect of the Peclet
number and initial condition, we plot the mixing efficiency $\eta$ with
respect to time. The results show that the presence of vesicles
manifest its effects on mixing at very high Peclet numbers. However,
depending on the initial condition, the vesicles promote (layer initial
condition), suppress (dye initial condition), and do not affect (random
initial condition) mixing.

\item \textbf{Effects of viscosity contrast (\figref{f:VCMixing}):} We
consider vesicle suspensions with a area fraction of $5\%$ and various
viscosity contrasts.  Since mixing will be shown to depend on the
initial condition, we perform the viscosity contrast tests only on the
layer initial condition, which is the one with the greatest change in
the mixing efficiency.  The results indicate that the viscosity
contrast does not significantly effect the mixing efficiency.
\end{itemize}

We consider vesicles of reduced area $0.65$.  The inner boundary
rotates at a constant angular velocity while the outer boundary is
fixed.  The inner boundary of the simulations with area fractions of
5\% and 40\% completes 32 rotations, while it completes 21 rotations in
all other simulations.  In all the runs, the vesicles are discretized
with 64 points, and in all but one run, the outer walls are discretized
with 128 points.  For the simulation with 40\% volume fraction, the
outer boundary is discretized with 256 points.  Additionally, the error
is controlled with an adaptive time integration
scheme~\cite{Quaife-Biros16}.

We list the physical parameters and values of the mixing simulation in
\tabref{t:MixingParams}. We discretize the transport equation with $N_r
= 256$ collocation points in the radial direction and with $N_{\theta}
= 1024$ collocation points in the azimuthal direction.  Crank-Nicholson
is used to solve the diffusion equation~\eqref{e:crankNicols} with the
time step size ${\Delta t}_D = 0.04$ and a Semi-Lagrangian method is
used to solve the advection equation~\eqref{e:midPoint} with the time
step size ${\Delta t}_A = 0.01$. Here, a unit time corresponds to $1.5
ms$ and a single rotation of the inner boundary requires $2\pi$ time
units.

\begin{table}[H]
\mcaption{{Physical} parameters for mixing simulations.}{t:MixingParams}
\centering
\begin{tabular}{cccc}
\hline
& Notation & Units & Value \\
 \hline 

 Angular velocity of the inner cylinder & $\omega$ & rad/unit time & 
 $1$ \\  
   
 Radius of inner cylinder &  $r_1$ & 
unit length & $10$ \\
  
 Radius of outer cylinder & $r_2$  & 
unit length  & $20$ \\
 
 unit length &  &
$\mu m$  & $3$ \\
 
 \hline 
\end{tabular} 
\end{table} 

For physically meaningful experiments, it is necessary that the dynamic
system of a vesicle suspension is statistically stationary. To address
this issue, let us discuss how we detect the statistical stationarity
in this study.
\subsection{Statistical analysis}\label{s:stats}
As we mentioned earlier, the velocity from the vesicle simulations are
used to drive the advective part of mixing. Here we describe the
procedure we use to ensure that the velocity is not polluted by effects
of the initial position and shape of the vesicles.  In all experiments
we assume that the vesicle suspensions eventually become statistically
stationary.  That is, we assume that all artifacts of the initial
condition vanish when the system reaches this statistical equilibrium,
and we use statistics of the velocity field to determine the onset of
this statistical equilibrium.  We start the mixing simulation once
these statistics, which we will define shortly, become time independent.

The presence of vesicles perturbs the velocity field of the default
Couette flow. We define the perturbation $\tilde{\mathbf{v}}$ as
$\mathbf{v} = {\mathbf{v}}_0 + \tilde{\mathbf{v}}$, where
$\mathbf{v}$ is the velocity field of the vesicle suspension and
${\mathbf{v}}_0$ is the velocity field in the absence of vesicles.  We
monitor stationarity of the time series of $\nu(t) = \|
\tilde{\mathbf{v}} \|_{L^2}$.  A stationary time series has statistical
properties that do not change over time, i.e.~its mean and variance over
any statistically representative window are unchanging. Our statistical
analysis involves, first, determining the statistically representative
window size $w'$ of the time series and, second, finding the time when
the statistical equilibrium is first reached. 

Given a time series, the window size $w'$ is chosen so that any sample of
the of this window size has the same mean and variance, independent of
the location of the window.  For example, suppose we are monitoring $
\nu(t) = {\| \tilde{\mathbf{v}} \|}_{L^2}$.  Then, $w'$ is chosen
with the following numerical decision scheme:
\begin{itemize}
\item We obtain a number of samples of a window width $w$ from the time
series $\nu(t)$ starting at randomly chosen times $t_i \, \in \, [0,
T_h-w]$ where $T_h$ is the time horizon. We denote these samples with
${\nu}_{w,i} = (\nu(t_{i}),\nu(t_{i}+w))$.

\item We compute the Fourier transform of the oscillations of the mean
${\tilde{\nu}}_{w,i} = {\nu}_{w,i} - \langle {\nu}_{{w},i} \rangle$,
where $\langle \cdot \rangle$ denotes the mean value $\langle \cdot
\rangle = \frac{1}{{w}} \int_{t_i}^{t_i + w} \cdot \,\, dt$. We then
sum the Fourier coefficients to form $\mu_{w,i}$ which represents a
property (in this case, the energy) of the particular window.

\item As the window size $w$ increases, the ensemble average ${\mu}_{w}
= \frac{1}{M}\Sigma_{i=1}^M  {\mu}_{w,i}$ converges to the mean of the
entire time series, and the standard deviation,
$\{{\mu}_{{w},i}{\}}_{i=1}^{M}$, decreases, where $M$ is the number of
samples. In \figref{f:statsAnalysis}, we plot the mean ${\mu}_{w}$ and
the standard deviation of the ensemble as a function of the window size
$w$ in the top row.  The statistically representative window size $W$
is the one which delivers a small standard deviation and a converged 
mean.

\end{itemize}
We plot the means of the samples as a function of window size $w$ for
the suspensions at AF = $10\%, 20\%, 40\%$ in the top row of
\figref{f:VFStatistics}. We choose the window size $w' = 25$ time units
for the suspensions at AF = $10\%$ and $20\%$, and $w' = 40$ time units
for those at AF = $5\%$ and $40\%$. Although we do not show the results
for the vesicle suspensions with VC = 5 and VC = 8, we have repeated
this analysis for them.   Alternatively, a similar decision scheme is
frequently used to find a statistically representative volume element
of a random microstructure (see~\cite{Torquato02} for details). Here,
the samples are ideally independent and identically distributed.
However, we cannot choose such samples since the time series we have is
too short.  the samples are supposed to be independent and identically
distributed.  However, we cannot choose such samples since the time
series we have is too short.

\begin{figure}[H]
 \begin{minipage}{\textwidth}
\setcounter{subfigure}{0}
\renewcommand*{\thesubfigure}{(a-1)} 
      \hspace{0cm}\subfigure[AF = $10\%$]{\scalebox{0.3}{{\includegraphics{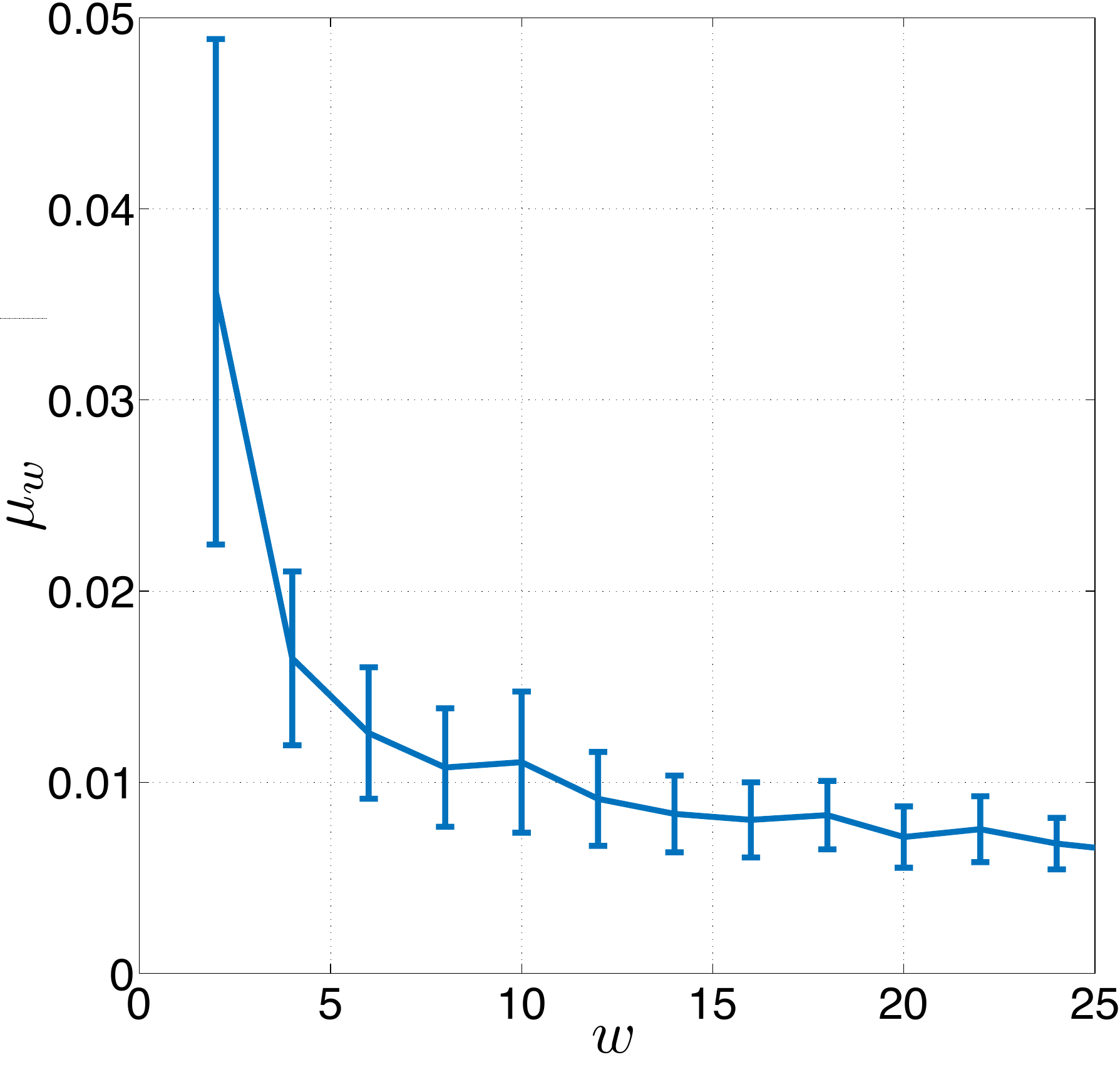}}}
      \label{f:windowVF10}} 
\setcounter{subfigure}{0}      
\renewcommand*{\thesubfigure}{(a-2)} 
      \hspace{-0.4cm}\subfigure[AF = $20\%$]{\scalebox{0.3}{{\includegraphics{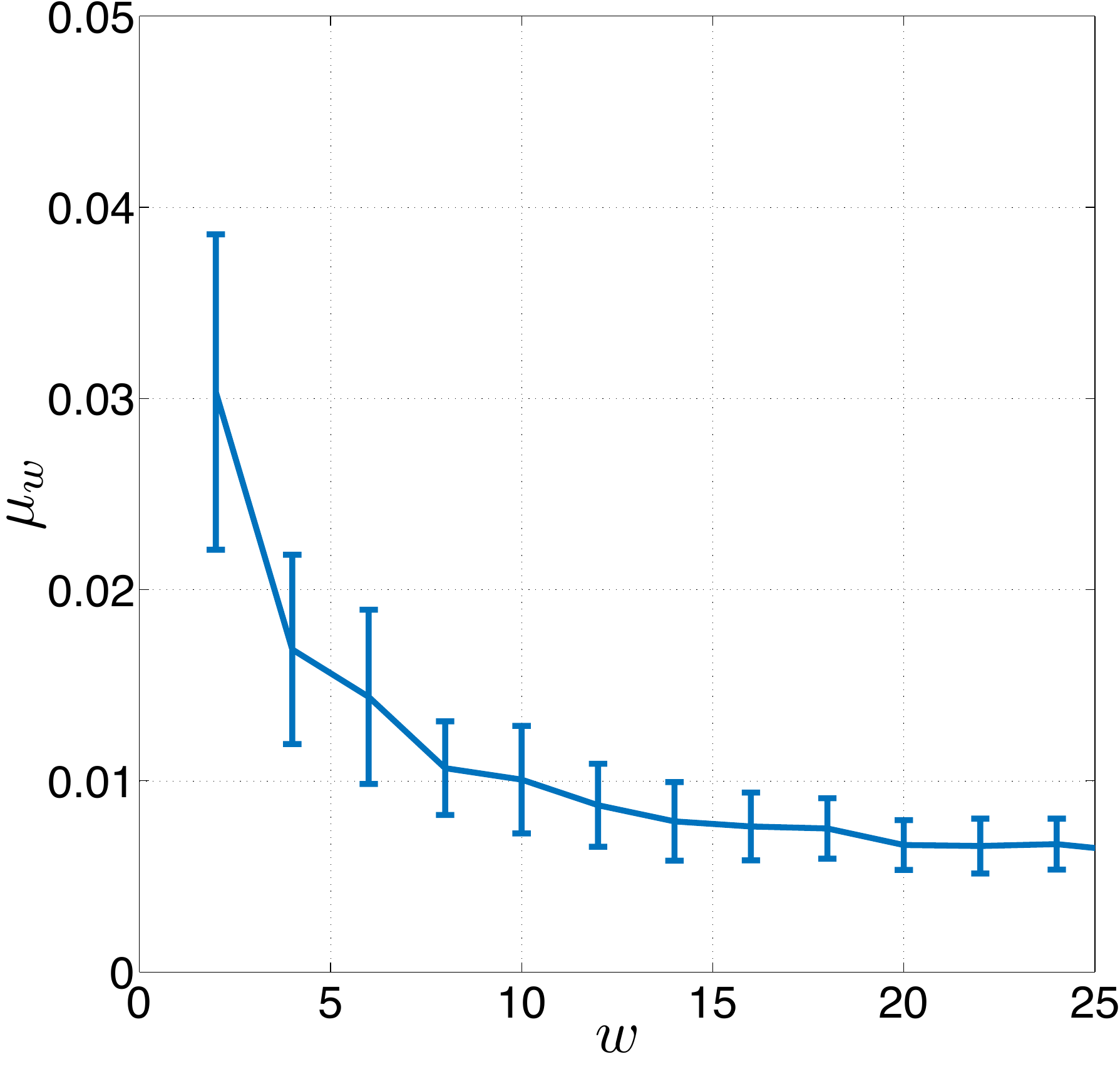}}}
      \label{f:windowVF20}}
\setcounter{subfigure}{0}
\renewcommand*{\thesubfigure}{(a-3)} 
      \hspace{-0.4cm}\subfigure[AF = $40\%$]{\scalebox{0.3}{{\includegraphics{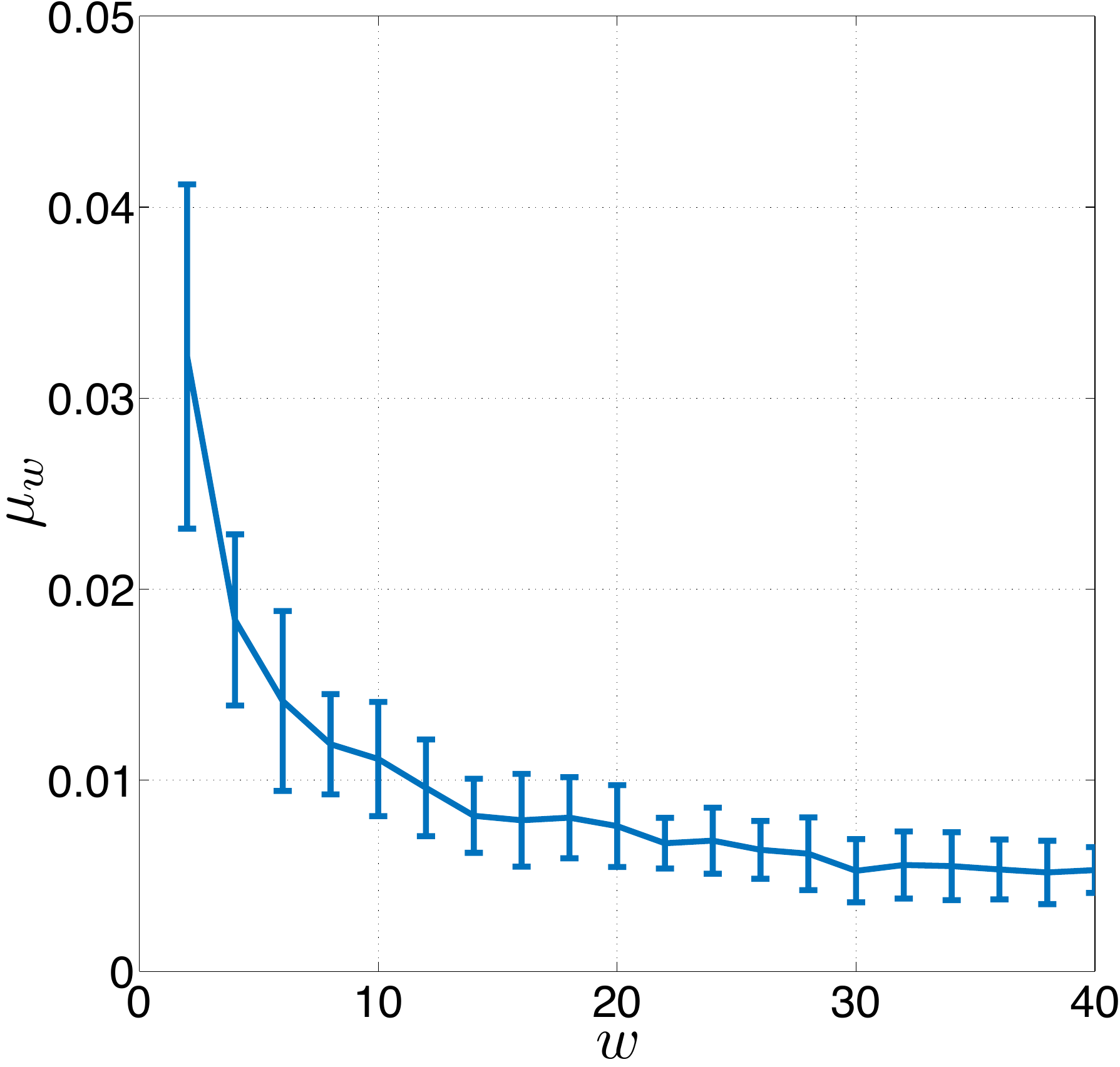}}}
      \label{f:windowVF40}}
\end{minipage}             
\begin{minipage}{\textwidth}
\setcounter{subfigure}{0}
\renewcommand*{\thesubfigure}{(b-1)} 
      \hspace{0cm}\subfigure[$\|\mathbf{v}\|_{L^{2}}$]{\scalebox{0.29}{{\includegraphics{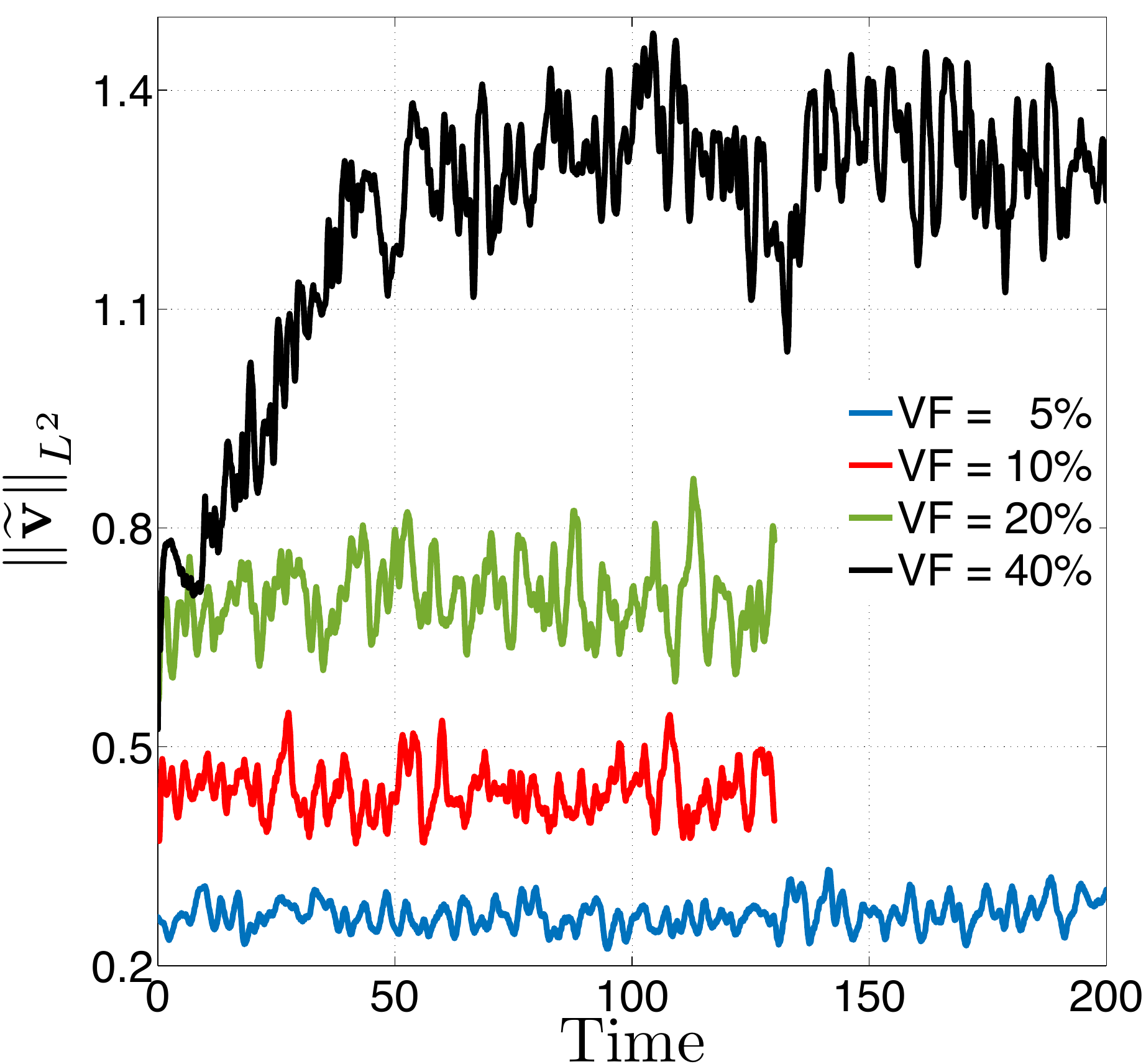}}}
      \label{f:spatRMS}}
\setcounter{subfigure}{0}
\renewcommand*{\thesubfigure}{(b-2)} 
      \hspace{-0.4cm}\subfigure[Mean in ($t_i, t_i + w'$)]{\scalebox{0.29}{{\includegraphics{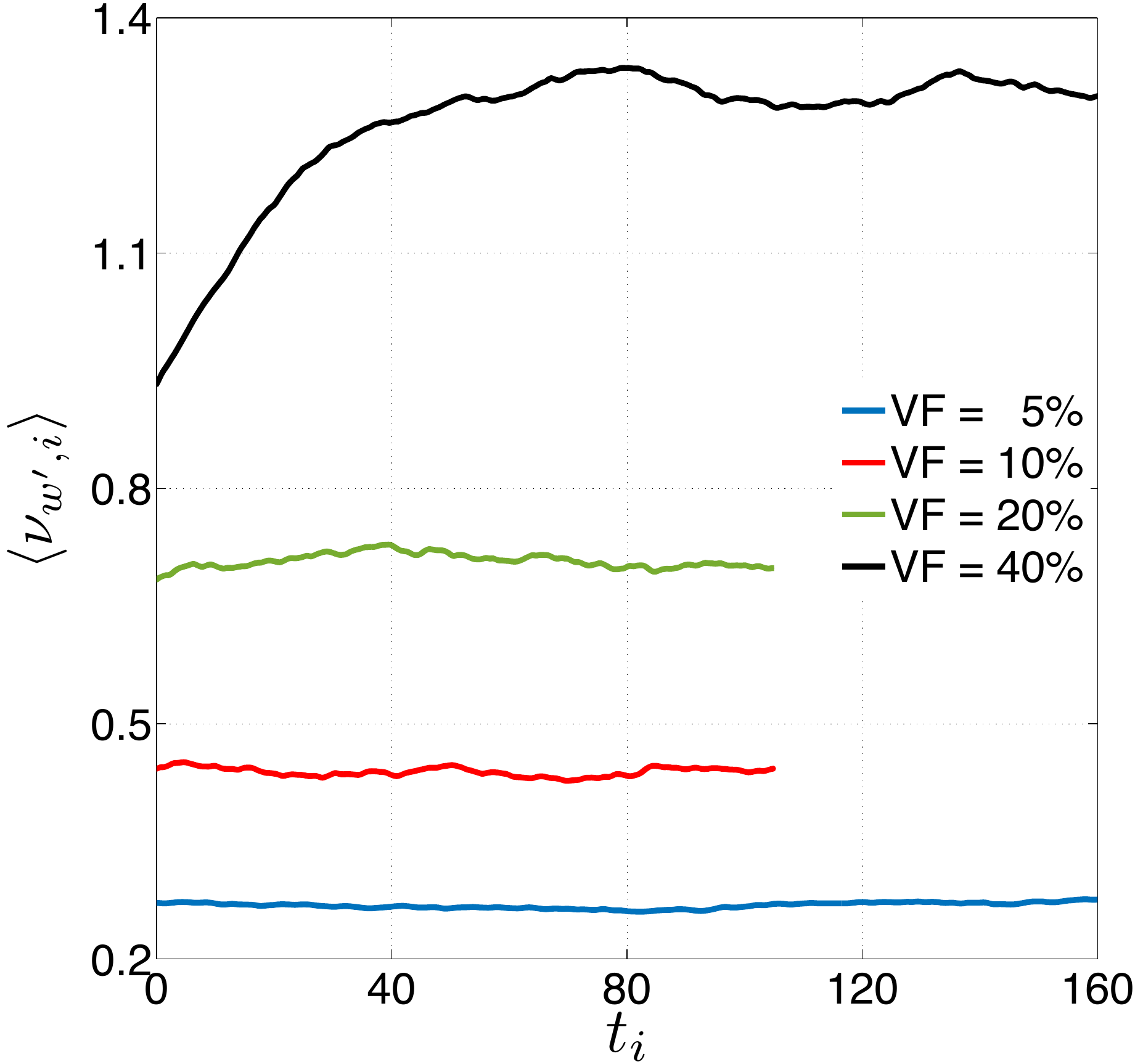}}}
      \label{f:EqAv}}
\setcounter{subfigure}{0}
\renewcommand*{\thesubfigure}{(b-3)} 
      \hspace{-0.4cm}\subfigure[Standard deviation in ($t_i, t_i + w'$)]{\scalebox{0.29}{{\includegraphics{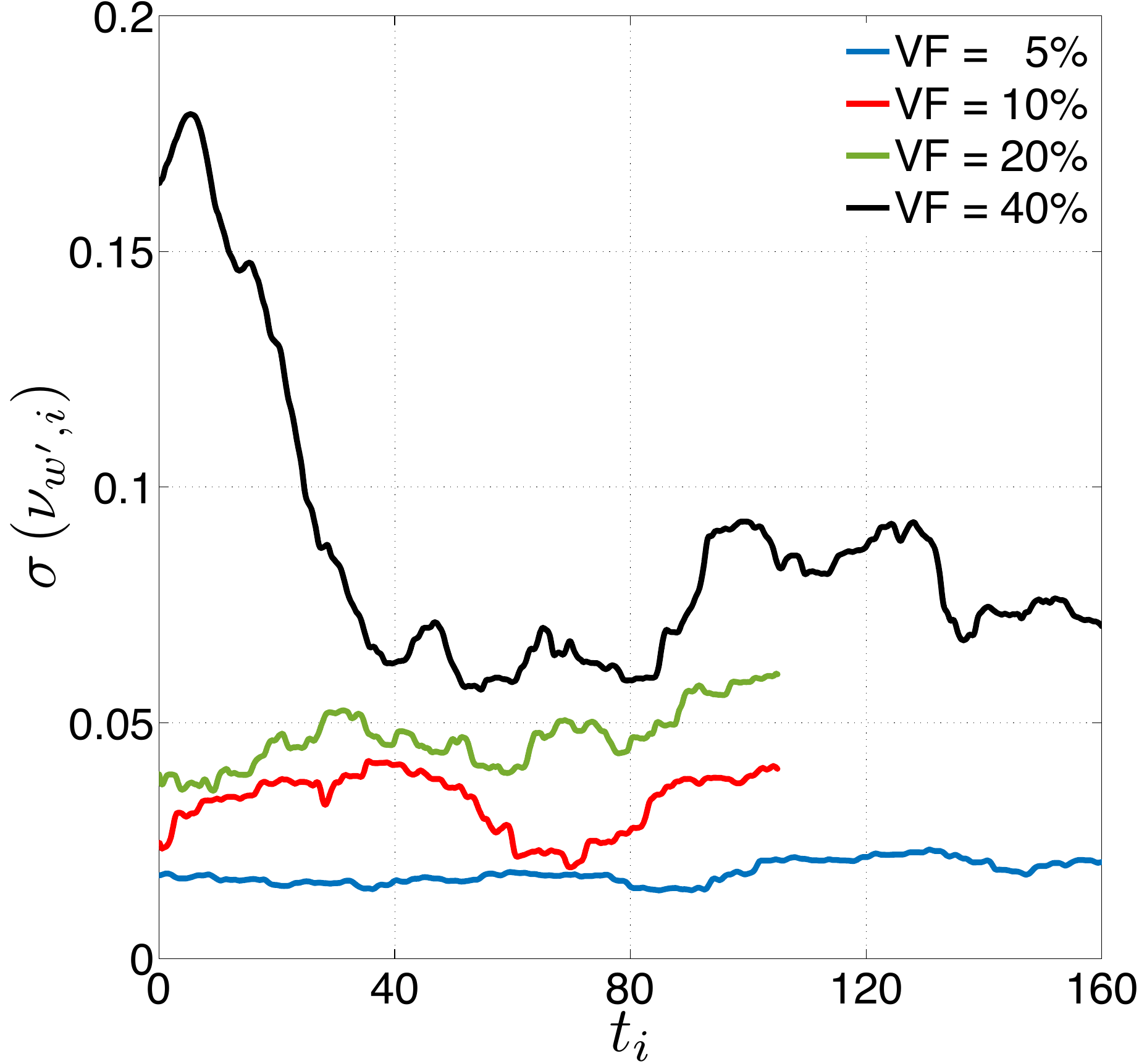}}}	
      \label{f:EqRMS}}      
 \end{minipage}              
\mcaption{{We} consider the $L^2$ norm of the perturbations
$\tilde{\mathbf{v}}$ (see \figref{f:spatRMS}) to examine the
statistical properties of the vesicle suspensions. The dynamics become
weakly stationary when the mean and the standard deviation of its
statistical properties become time independent.  Therefore, we compute
the mean and the standard deviation of ${\| \tilde{\mathbf{v}}
\|}_{L^2}$ int the window ($t_i, t_i + w'$), for all $i$ (see
\figref{f:EqAv} and \figref{f:EqRMS}).  It is necessary that the width
is statistically representative.  Therefore, we look at the energy of
randomly chosen samples from ${\| \tilde{\mathbf{v}} \|}_{L^2}$ of
various window widths $w$ for each area fraction in the top
row.}{f:VFStatistics}
\end{figure}

Once the representative window size $w'$ of the time series $\nu(t)$ is
chosen, we need to determine when the statistical equilibrium is first
reached. To do so, we choose samples $\nu_{w',i}$ of size $w'$ at every
discrete time $t_i \, \in \, [0,T_h-w']$ and compute their means
$\langle \nu_{w',i} \rangle$ and standard deviations
$\sigma(\nu_{w',i})$.  We determine the time when the statistical
equilibrium is attained by examining when the mean and standard
deviation plateau.  In particular, we require that $| \langle
\nu_{w',i+1}\rangle- \langle\nu_{w',i} \rangle|/|\langle
\nu_{w',i}\rangle|$ to be less than some tolerance.  We summarize this procedure in the bottom
row of \figref{f:statsAnalysis}, and the results for the different area
fractions are in the bottom row of \figref{f:VFStatistics}.
The mean (\figref{f:EqAv}) and the standard deviation
(\figref{f:EqRMS}) converge after $t_i = 100$ in AF = $5\%$, $t_i = 40$
in AF = $10\%$ and AF = $20\%$, and $t_i = 100$ in AF = $40\%$. 
\begin{figure}[H]
\includegraphics[width=\textwidth]{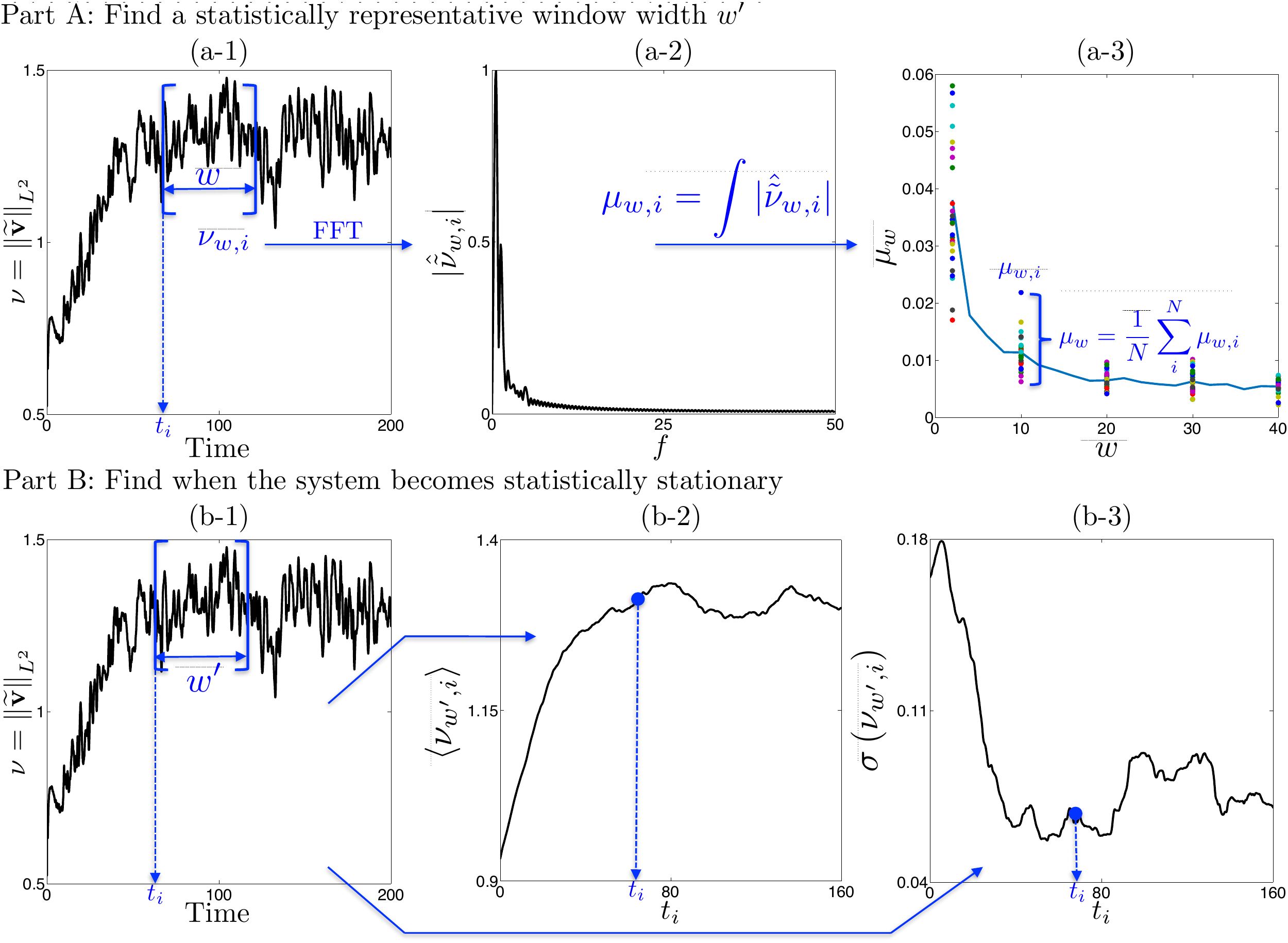}
\mcaption{{We} outline the procedures to find the statistically
representative window width $w'$ and to detect the statistical equilibrium
for the vesicle suspension of AF = $40\%$, in the first and second
rows, respectively.}{f:statsAnalysis}
\end{figure}

\subsection{Effects of area fraction}\label{s:volFrac}

We simulate mixing in vesicle suspensions with the area fractions of
$10 \%$, $20 \%$, and $40 \%$, and with no viscosity contrast.  We use
the layer initial condition. We fix the Peclet number to Pe=$10^{4}$
for all the area fractions by adjusting the diffusivity based on the
value of $\langle \| \mathbf{v} \|_{L^2} \rangle$. We show the vesicle
positions, the magnitude of the perturbation in the Couette flow due to
the vesicles $\| \tilde{\mathbf{v}} \|$ (see \secref{s:stats} for
its definition) and the concentration $\phi$ for the area fractions of
$10 \%$ in \figref{f:VF10Snaps}, $20 \%$ in \figref{f:VF20Snaps}, and
$40 \%$ in \figref{f:VF40Snaps}. The results show that as the area
fraction increases, the maximum value of $\| \tilde{\mathbf{v}} \|$
increases from approximately 1.5 to 4 wherein the maximum of the
magnitude of the velocity field is 10 (see the second columns in
\figref{f:VF10Snaps} and \figref{f:VF40Snaps}).  The corresponding
concentration fields observably differ as the area fraction of the
vesicles increases (see the third columns in \figref{f:VF10Snaps} and
\figref{f:VF40Snaps}).  In addition to the qualitative results in
\figref{f:VF10Snaps}, \figref{f:VF20Snaps}, and \figref{f:VF40Snaps},
the first row in \figref{f:VFMixing1} demonstrates the corresponding
mixing efficiencies.  We see that the presence of vesicles enhances
mixing for this particular initial condition and increasing the area
fraction increases the efficiency as high as $\eta \approx 1.35$ when
AF $ = 40\%$.

\begin{figure}[H]
\vspace{-0.5cm}
 \begin{minipage}{\textwidth}
\setcounter{subfigure}{0}
\renewcommand*{\thesubfigure}{(a-1)} 
      \hspace{0.5cm}\subfigure[Vesicles at $t = 0$]{\scalebox{0.42}{{\includegraphics{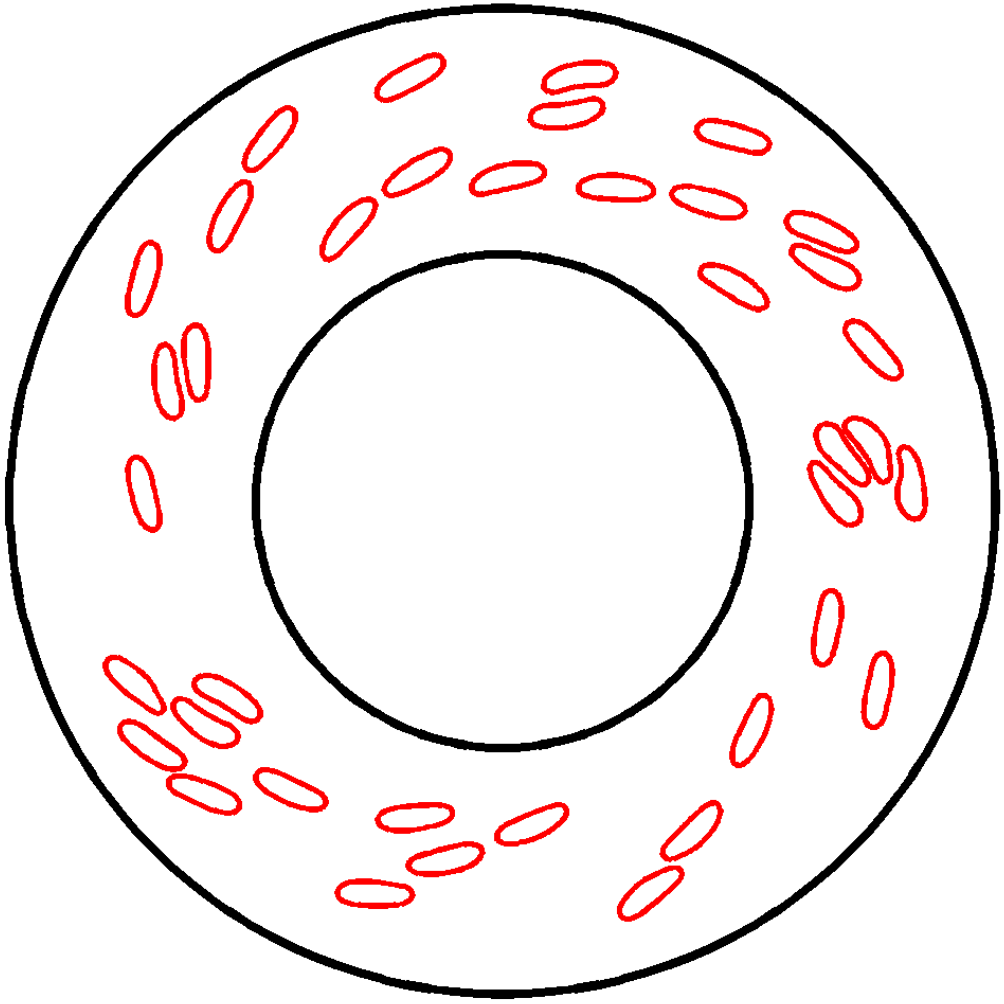}}}	
      \label{f:VF10VesS1}}
\setcounter{subfigure}{0}
\renewcommand*{\thesubfigure}{(a-2)} 
      \hspace{0.2cm}\subfigure[$\| \tilde{\mathbf{v}} \|$ at $t = 0$]{\scalebox{0.43}{{\includegraphics{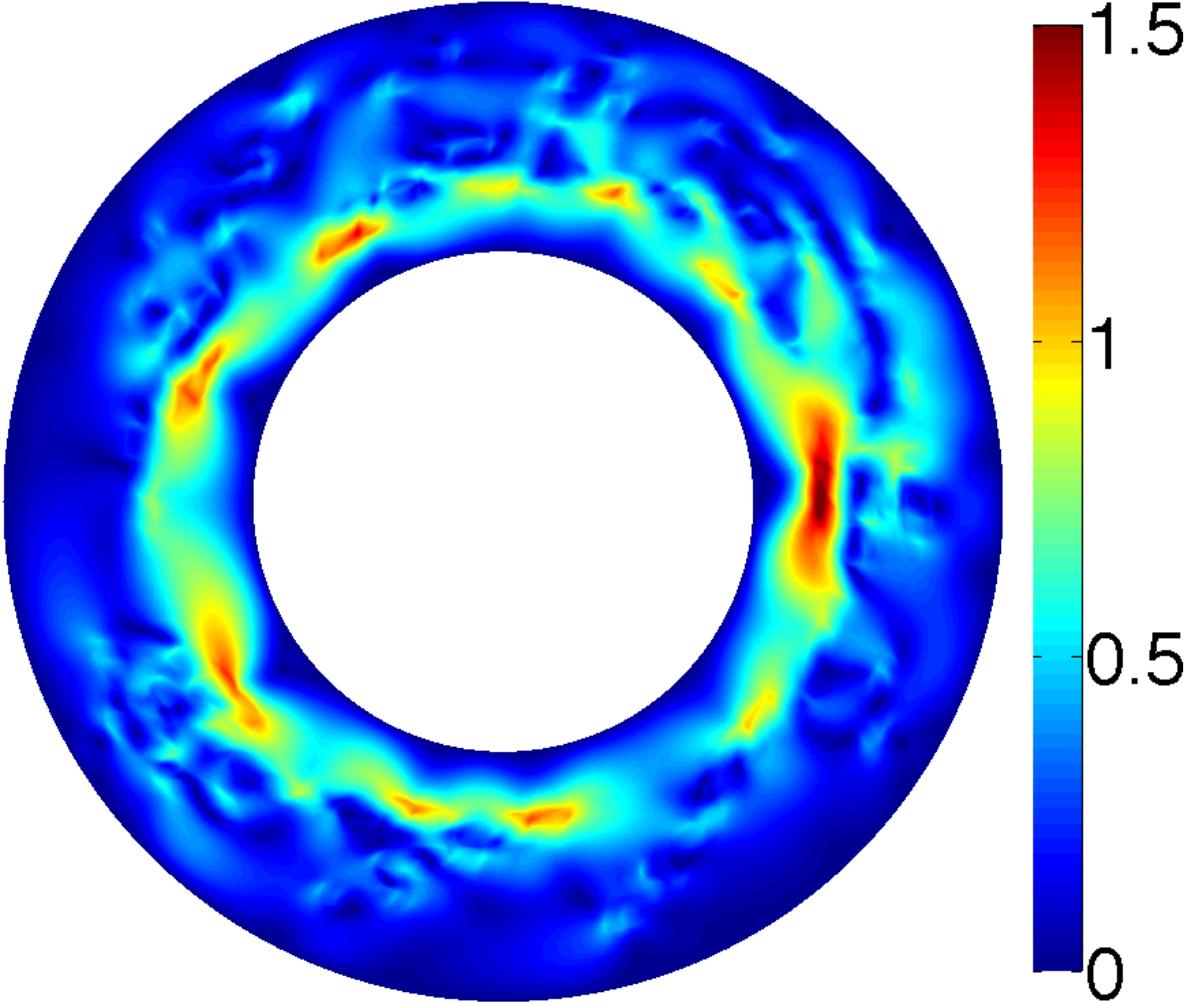}}}
      \label{f:VF10VelS1}}
\setcounter{subfigure}{0}
\renewcommand*{\thesubfigure}{(a-3)} 
      \hspace{0cm}\subfigure[$\phi$ at $t = 0$]{\scalebox{0.43}{{\includegraphics{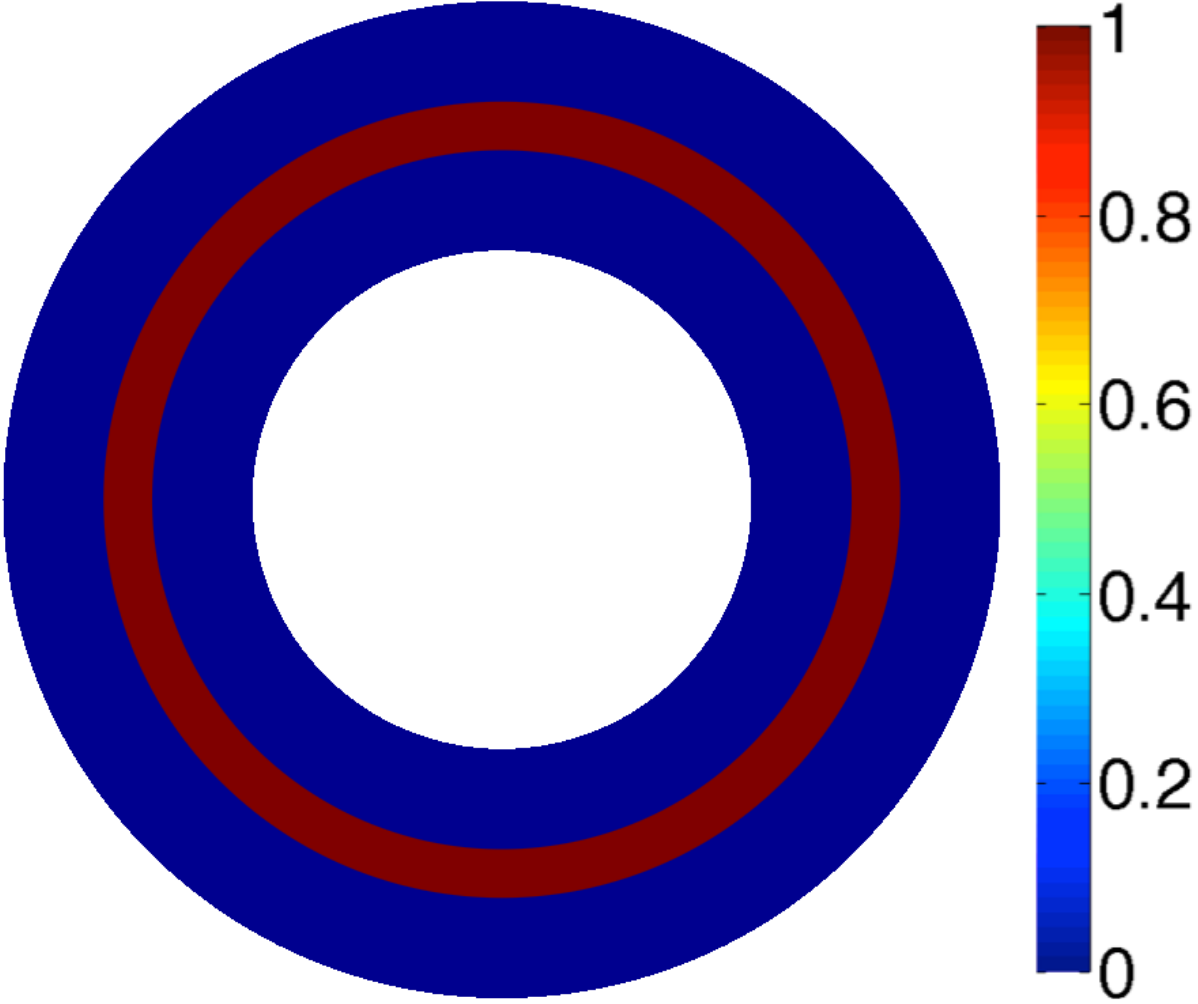}}}
      \label{f:VF10MixS1}}
\end{minipage}
 \begin{minipage}{\textwidth}
\setcounter{subfigure}{0}
\renewcommand*{\thesubfigure}{(b-1)} 
      \hspace{0.5cm}\subfigure[Vesicles at $t = 30$]{\scalebox{0.42}{{\includegraphics{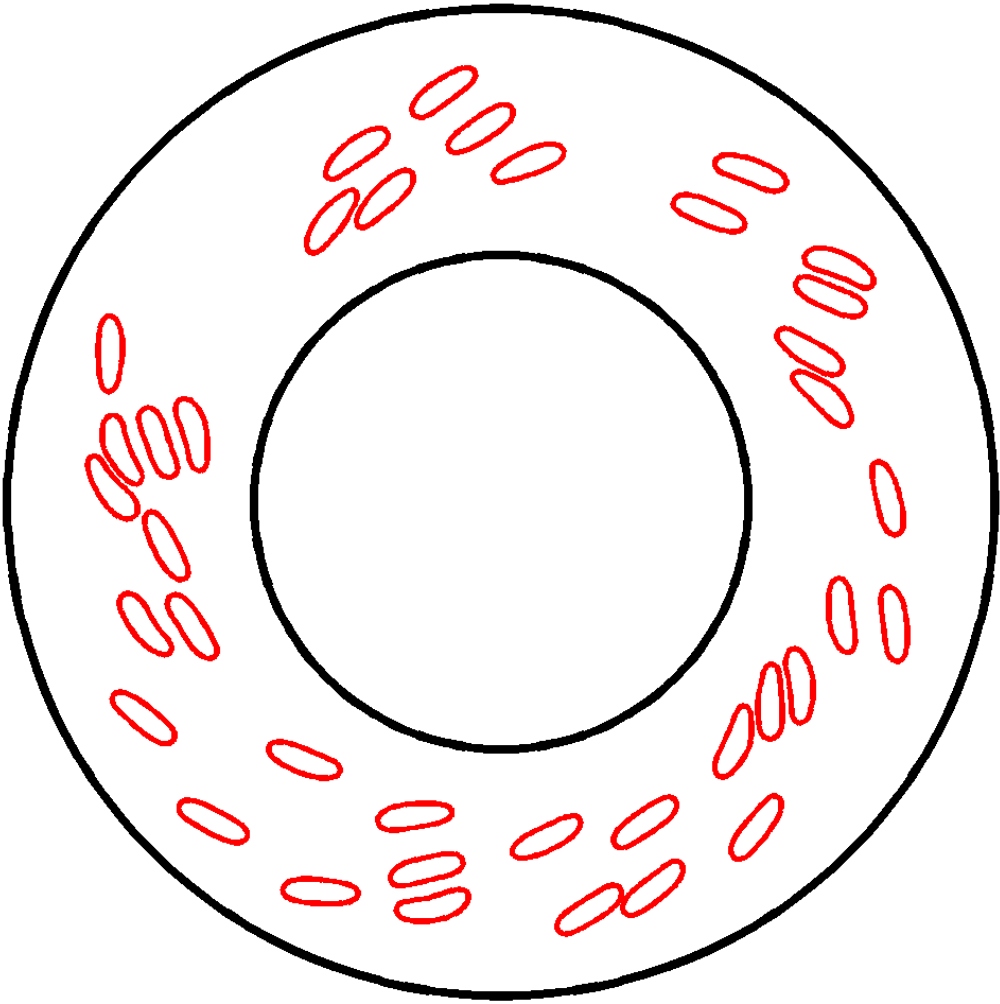}}}	
      \label{f:VF10VesS3}}
\setcounter{subfigure}{0}
\renewcommand*{\thesubfigure}{(b-2)} 
      \hspace{0.2cm}\subfigure[$\| \tilde{\mathbf{v}} \|$  at $t = 30$]{\scalebox{0.43}{{\includegraphics{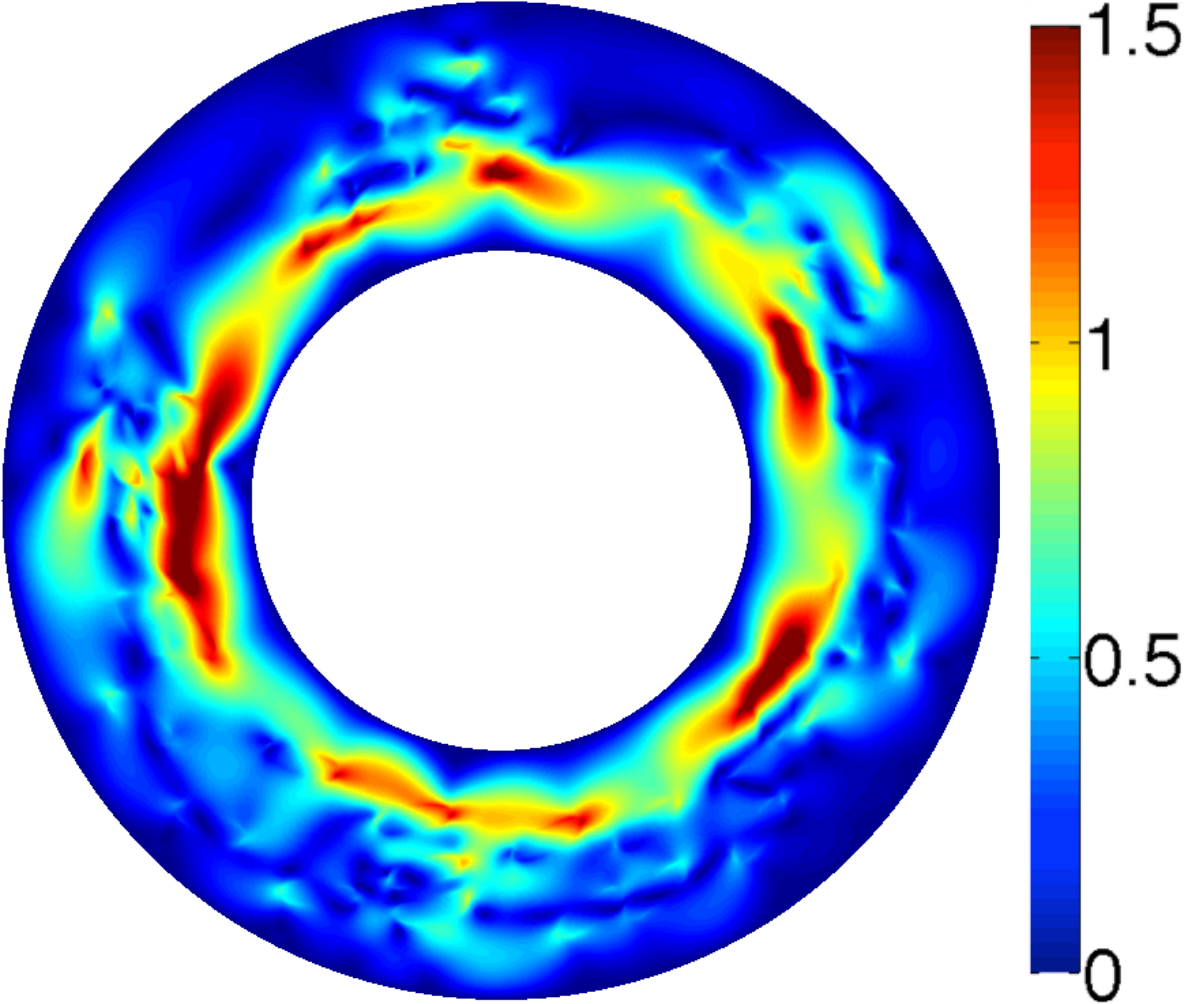}}}
      \label{f:VF10VelS3}}
\setcounter{subfigure}{0}
\renewcommand*{\thesubfigure}{(b-3)} 
      \hspace{0cm}\subfigure[$\phi$ at $t = 30$]{\scalebox{0.43}{{\includegraphics{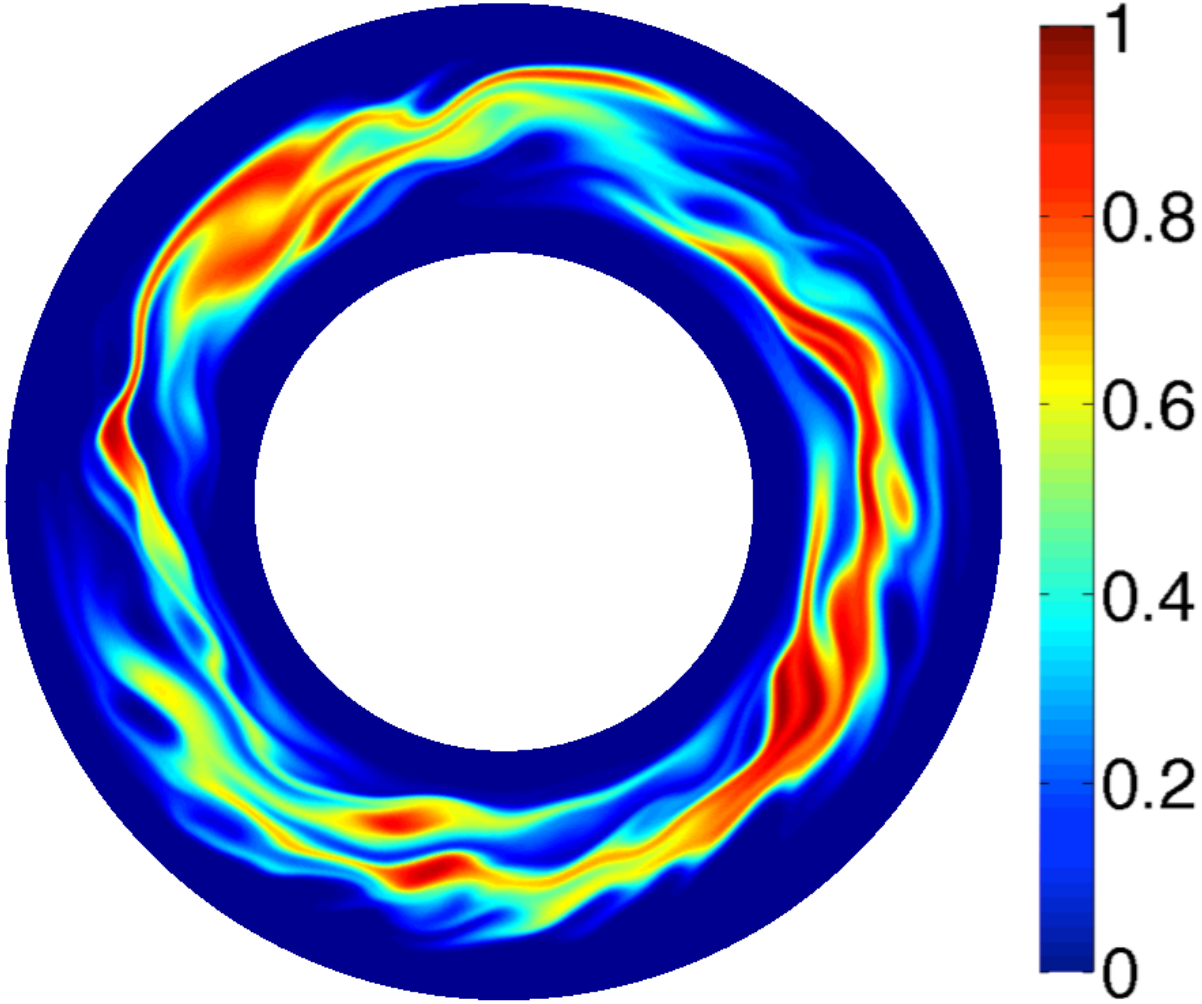}}}
      \label{f:VF10MixS3}}
\end{minipage}
 \begin{minipage}{\textwidth}
\setcounter{subfigure}{0}
\renewcommand*{\thesubfigure}{(c-1)} 
      \hspace{0.5cm}\subfigure[Vesicles at $t = 60$]{\scalebox{0.42}{{\includegraphics{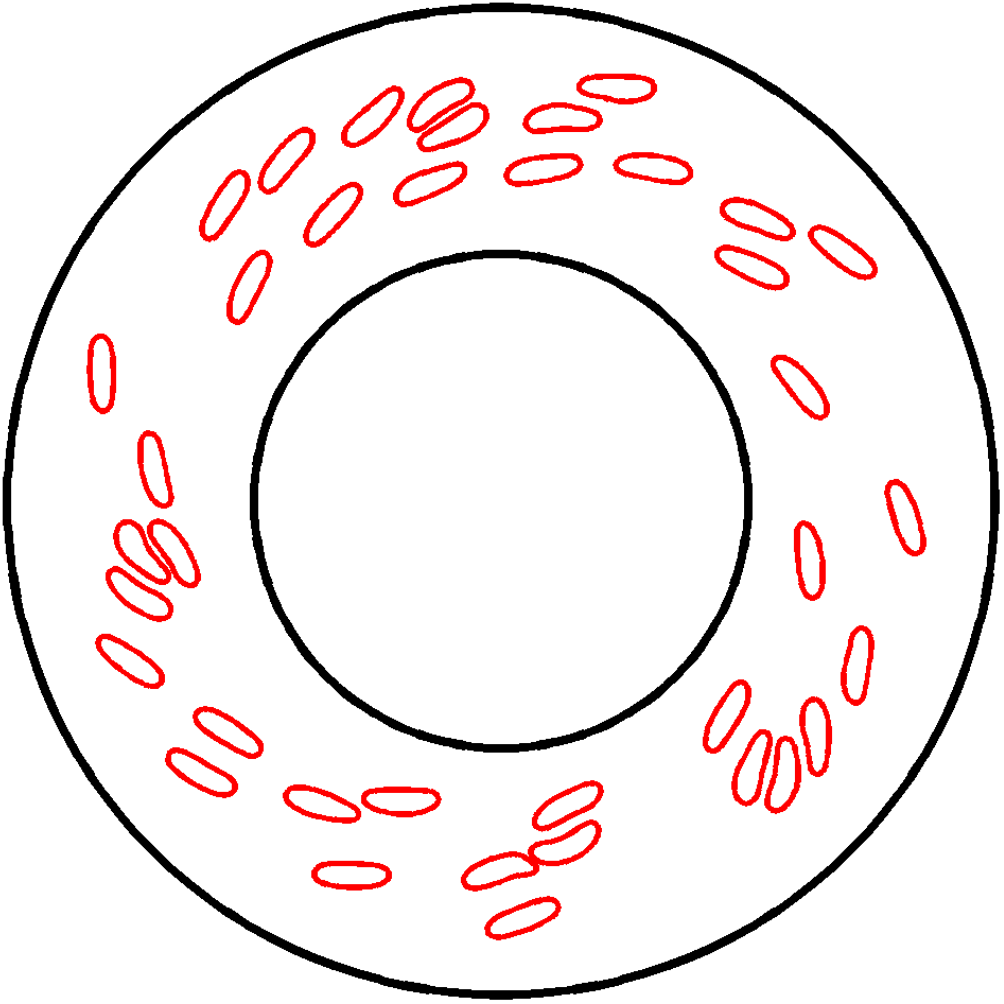}}}	
      \label{f:VF10VesS4}}
\setcounter{subfigure}{0}
\renewcommand*{\thesubfigure}{(c-2)} 
      \hspace{0.2cm}\subfigure[$\| \tilde{\mathbf{v}} \|$  at $t = 60$]{\scalebox{0.43}{{\includegraphics{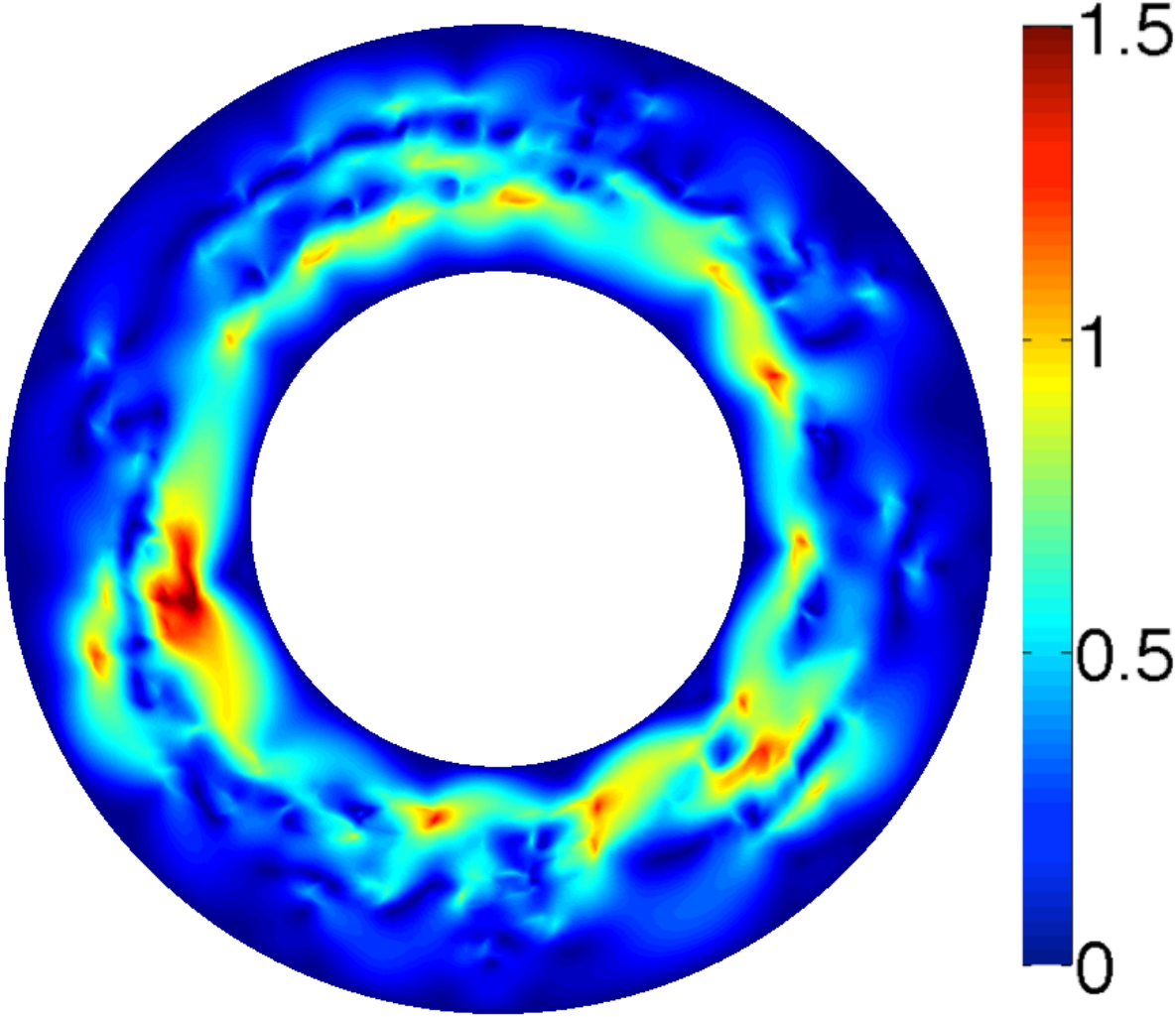}}}
      \label{f:VF10VelS4}}
\setcounter{subfigure}{0}
\renewcommand*{\thesubfigure}{(c-3)} 
      \hspace{0cm}\subfigure[$\phi$ at $t = 60$]{\scalebox{0.43}{{\includegraphics{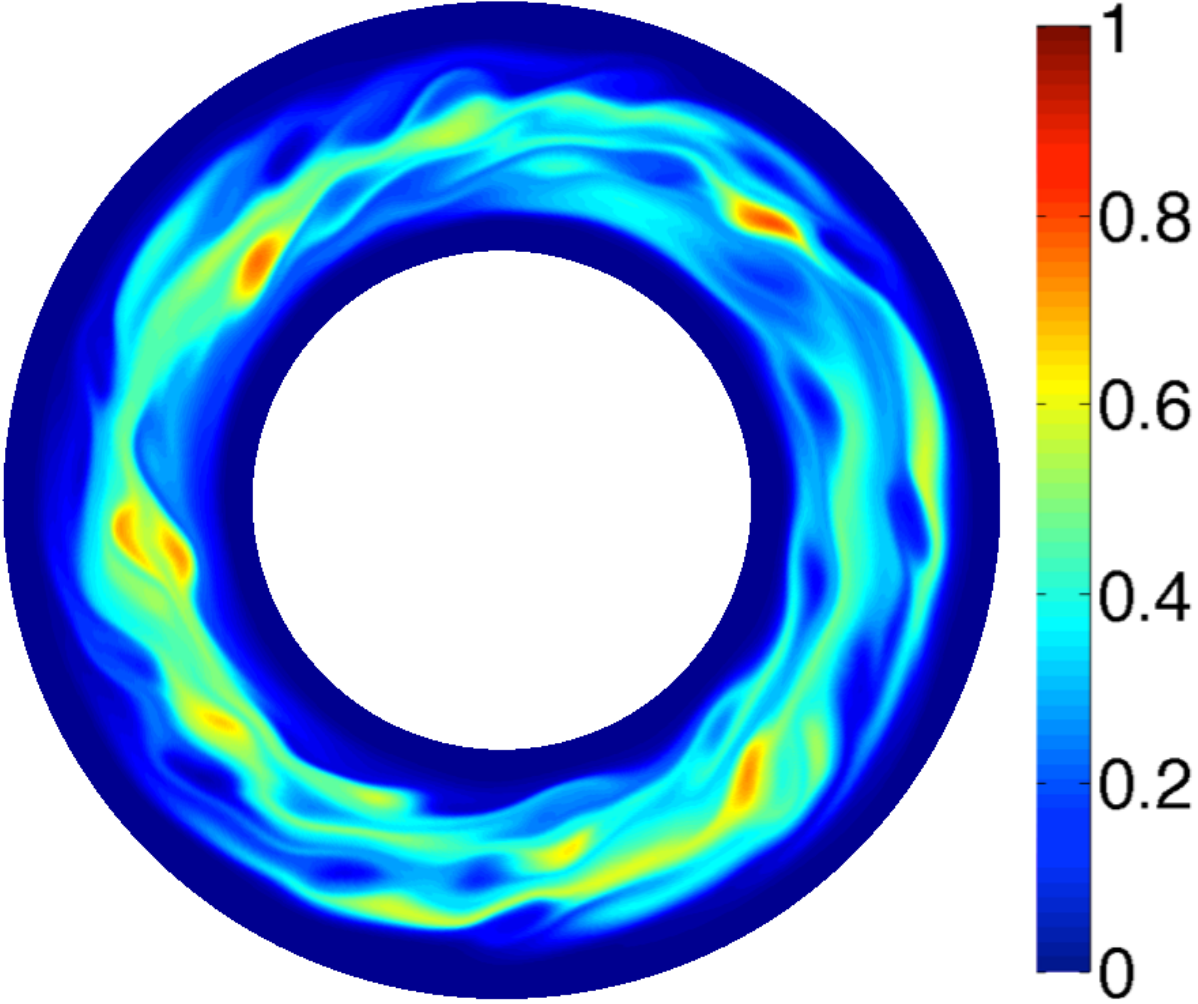}}}
      \label{f:VF10MixS4}}
\end{minipage}
 \begin{minipage}{\textwidth}
\setcounter{subfigure}{0}
\renewcommand*{\thesubfigure}{(d-1)} 
      \hspace{0.5cm}\subfigure[Vesicles at $t = 90$]{\scalebox{0.42}{{\includegraphics{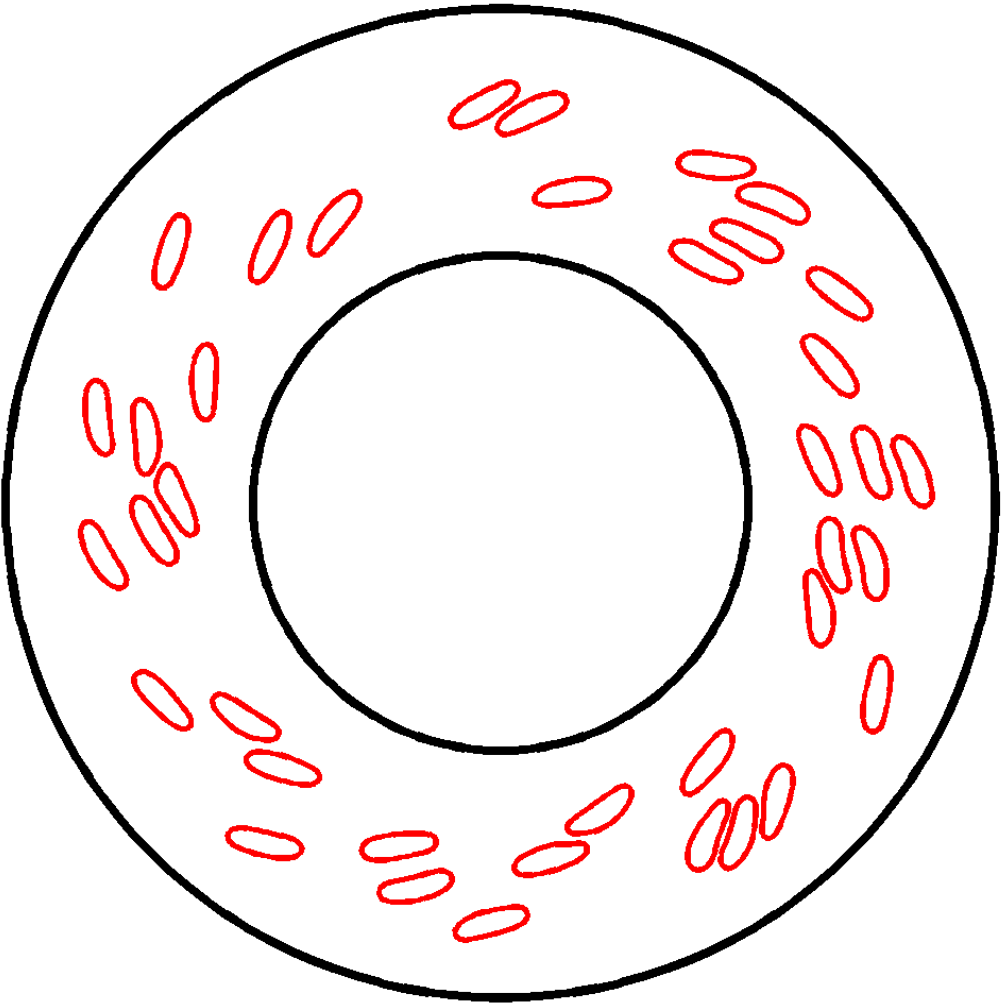}}}	
      \label{f:VF10VesS5}}
\setcounter{subfigure}{0}
\renewcommand*{\thesubfigure}{(d-2)} 
      \hspace{0.2cm}\subfigure[$\| \tilde{\mathbf{v}} \|$  at $t = 90$]{\scalebox{0.43}{{\includegraphics{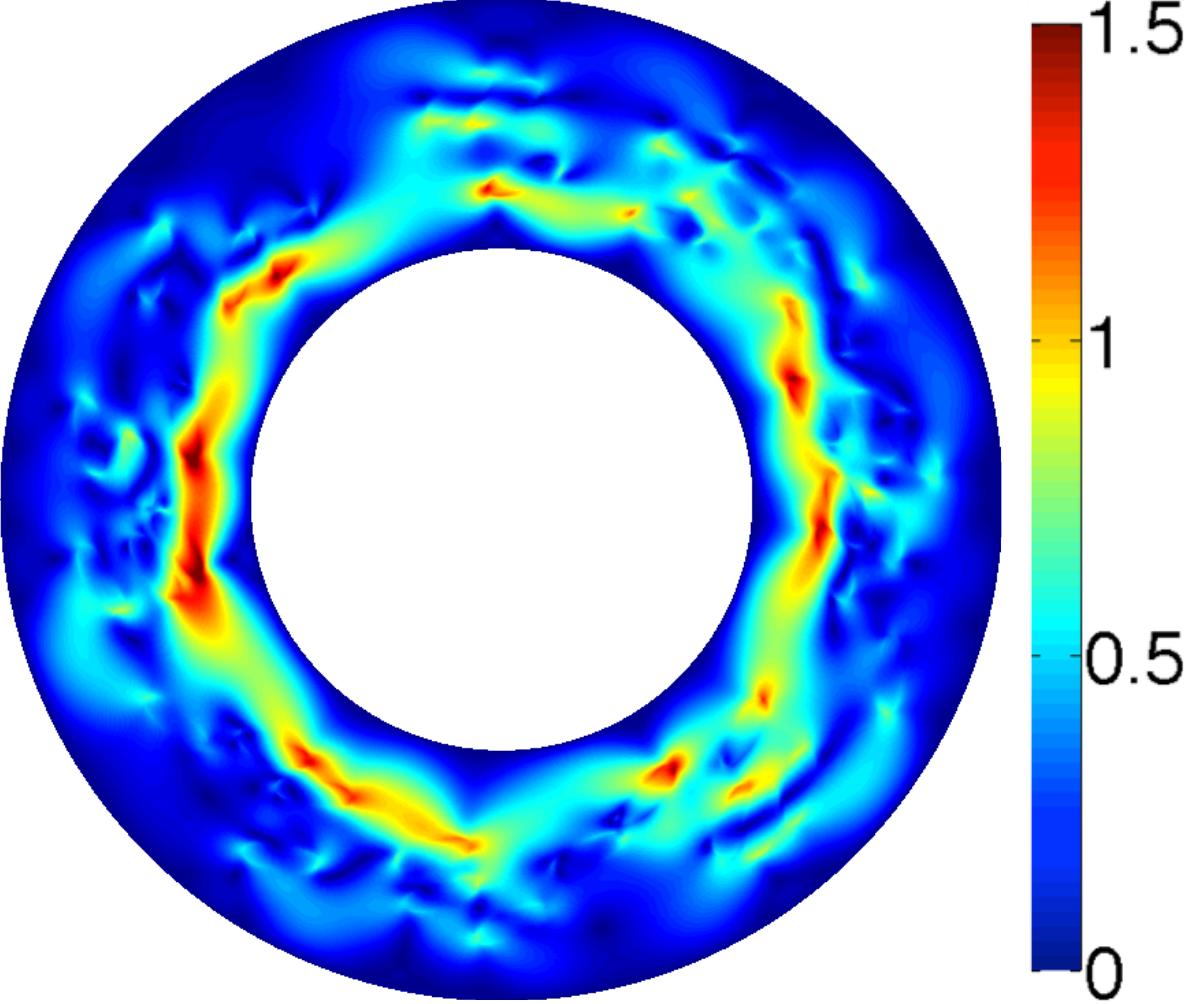}}}
      \label{f:VF10VelS5}}
\setcounter{subfigure}{0}
\renewcommand*{\thesubfigure}{(d-3)} 
      \hspace{0cm}\subfigure[$\phi$ at $t = 90$]{\scalebox{0.43}{{\includegraphics{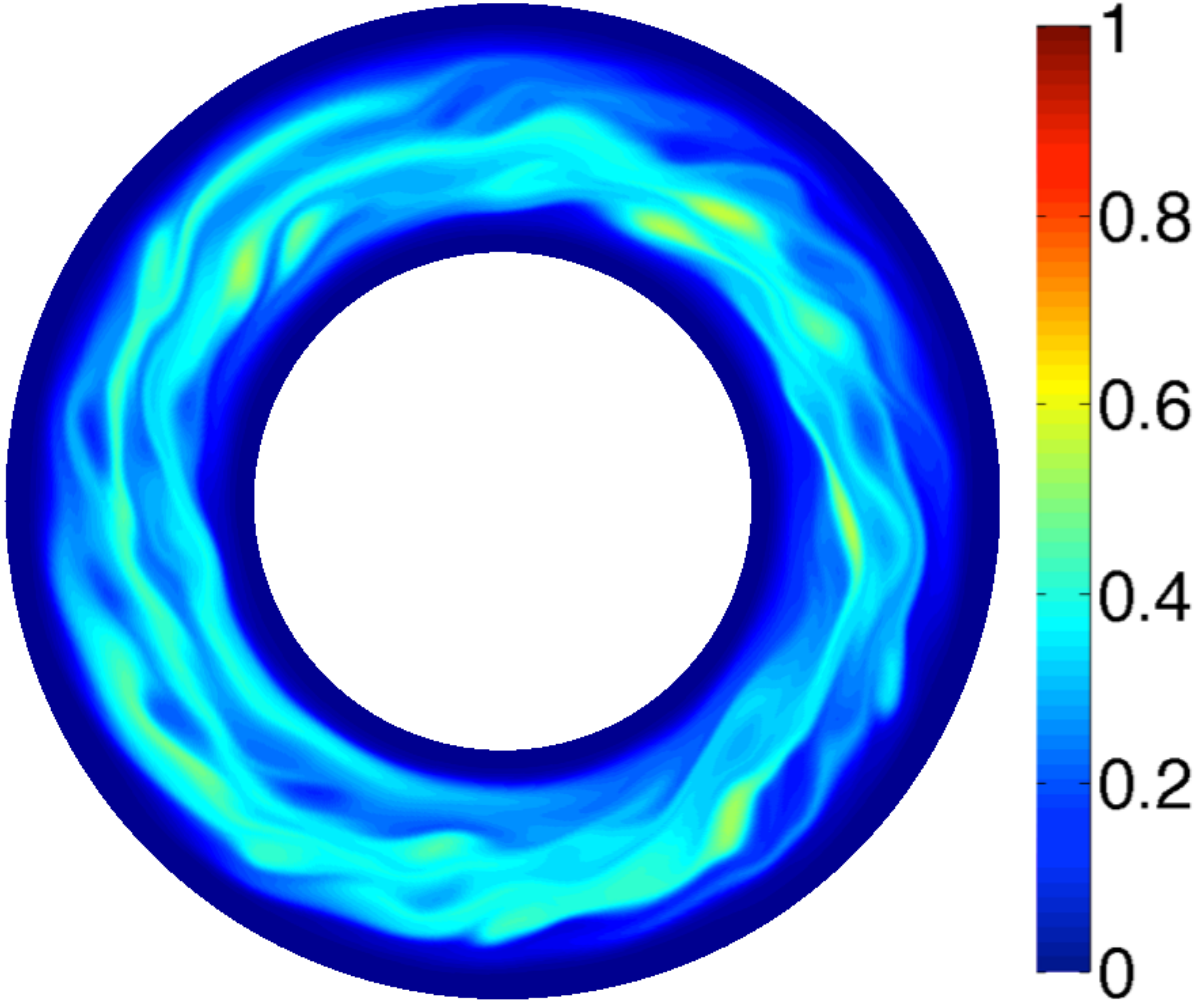}}}
      \label{f:VF10MixS5}}
\end{minipage}
\mcaption{{The} effects of a $10 \%$ area fraction on the velocity field
and mixing (\secref{s:volFrac}).  Here we present the vesicle positions
(left), the magnitude of the velocity field due only to the vesicles
$\|\widetilde{\mathbf{v}}\|$ (middle), and the concentration $\phi$
(right) for the area fraction of $10\%$ and the layer initial
condition.  Each row corresponds to a different time.}{f:VF10Snaps}
\end{figure}

\begin{figure}[H]
\vspace{-0.5cm}
 \begin{minipage}{\textwidth}
\setcounter{subfigure}{0}
\renewcommand*{\thesubfigure}{(a-1)} 
      \hspace{0.5cm}\subfigure[Vesicles at $t = 0$]{\scalebox{0.42}{{\includegraphics{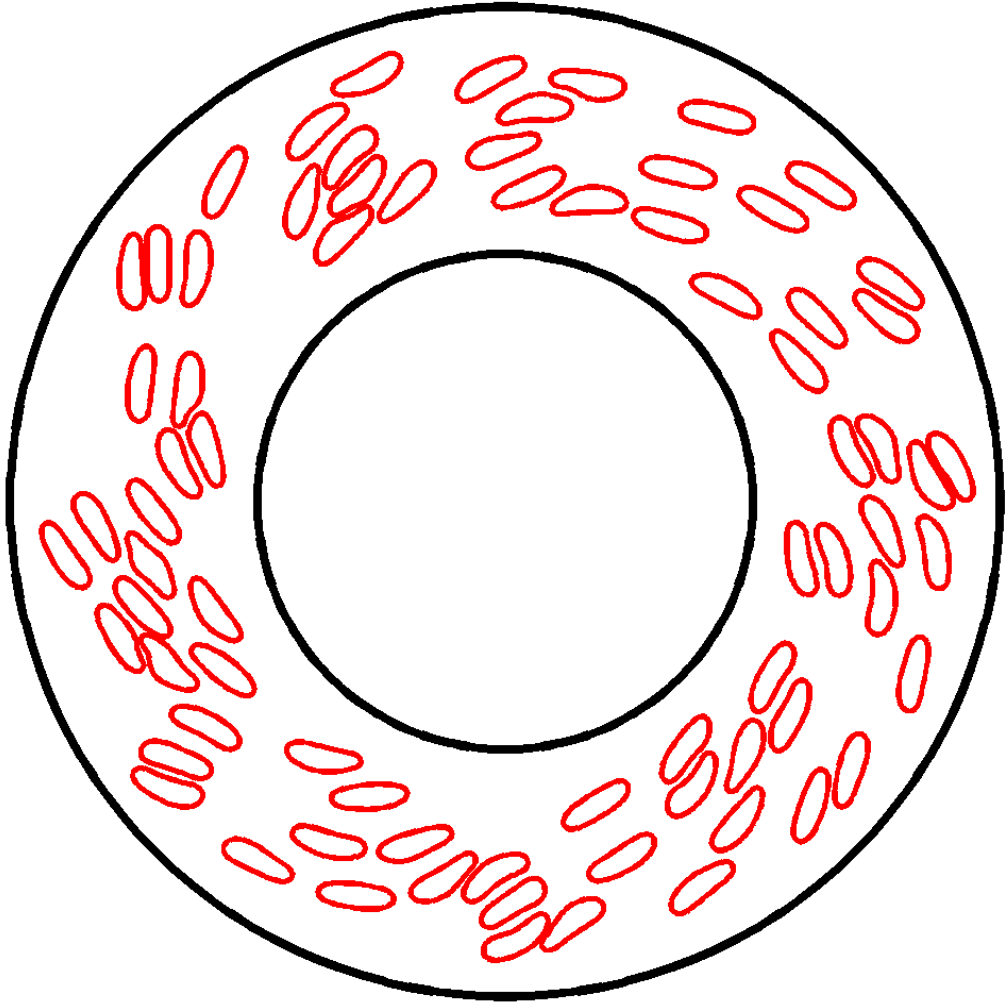}}}	
      \label{f:VF20VesS1}}
\setcounter{subfigure}{0}
\renewcommand*{\thesubfigure}{(a-2)} 
      \hspace{0.2cm}\subfigure[$\| \tilde{\mathbf{v}} \|$ at $t = 0$]{\scalebox{0.43}{{\includegraphics{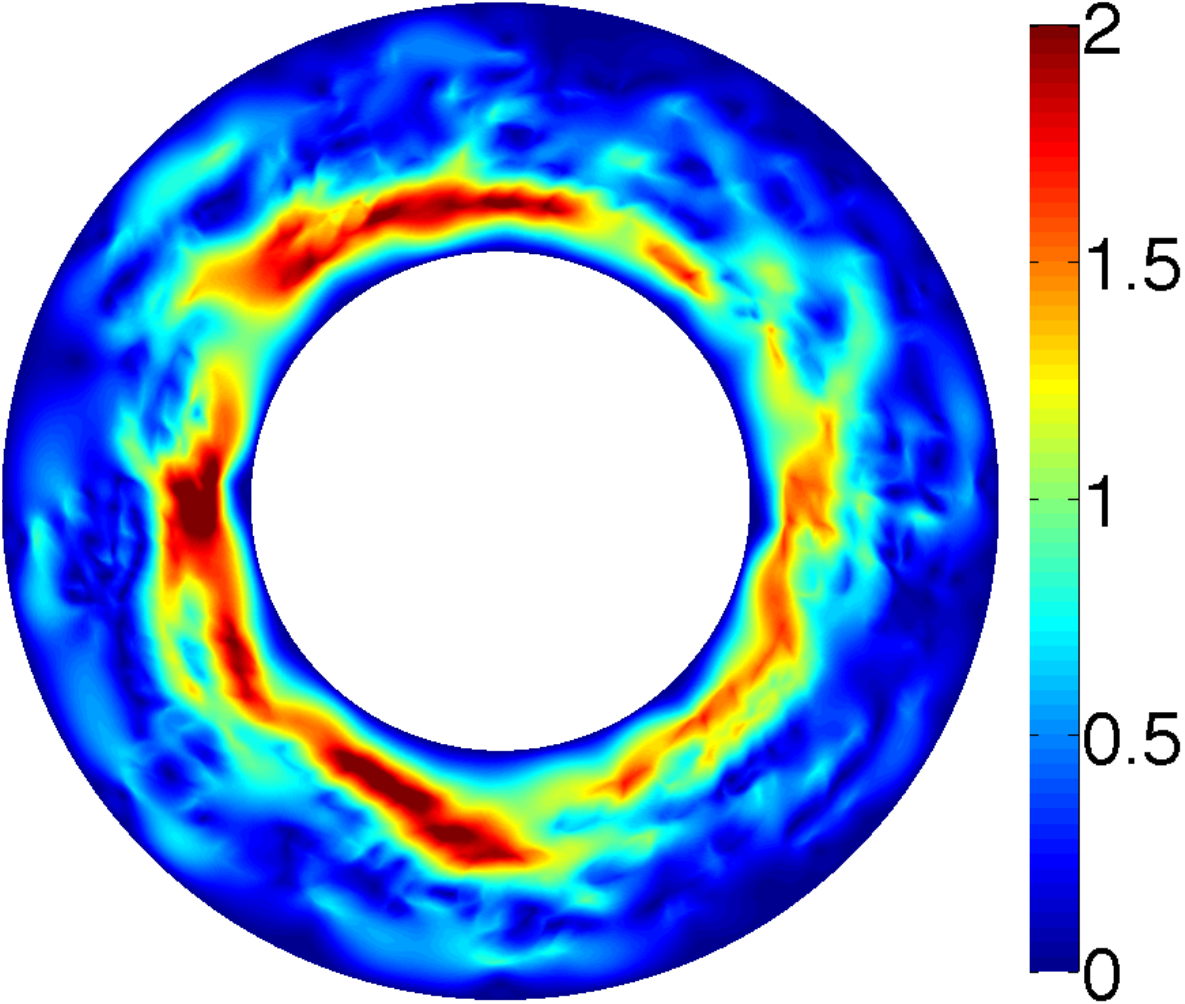}}}
      \label{f:VF20VelS1}}
\setcounter{subfigure}{0}
\renewcommand*{\thesubfigure}{(a-3)} 
      \hspace{0cm}\subfigure[$\phi$ at $t = 0$]{\scalebox{0.43}{{\includegraphics{VF10_MixS1.pdf}}}
      \label{f:VF20MixS1}}
\end{minipage}
 \begin{minipage}{\textwidth}
\setcounter{subfigure}{0}
\renewcommand*{\thesubfigure}{(b-1)} 
      \hspace{0.5cm}\subfigure[Vesicles at $t = 30$]{\scalebox{0.42}{{\includegraphics{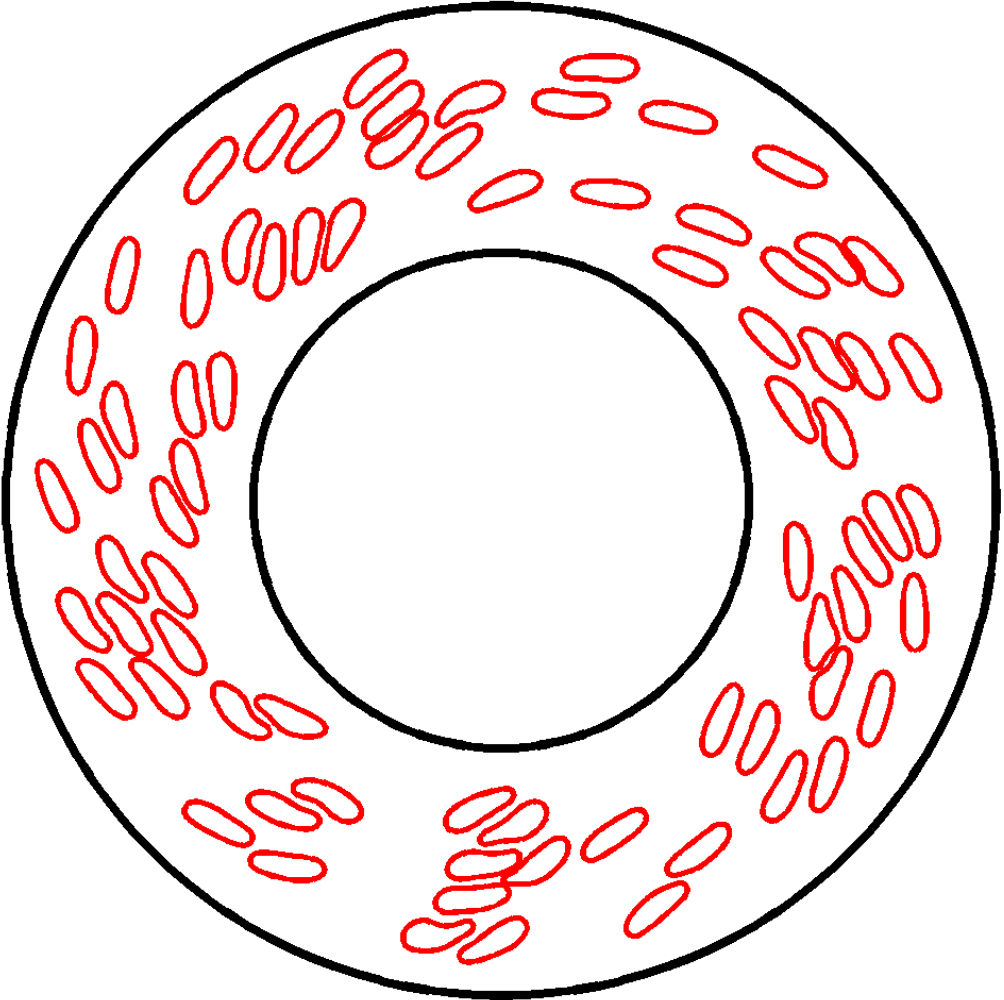}}}	
      \label{f:VF20VesS3}}
\setcounter{subfigure}{0}
\renewcommand*{\thesubfigure}{(b-2)} 
      \hspace{0.2cm}\subfigure[$\| \tilde{\mathbf{v}} \|$  at $t = 30$]{\scalebox{0.43}{{\includegraphics{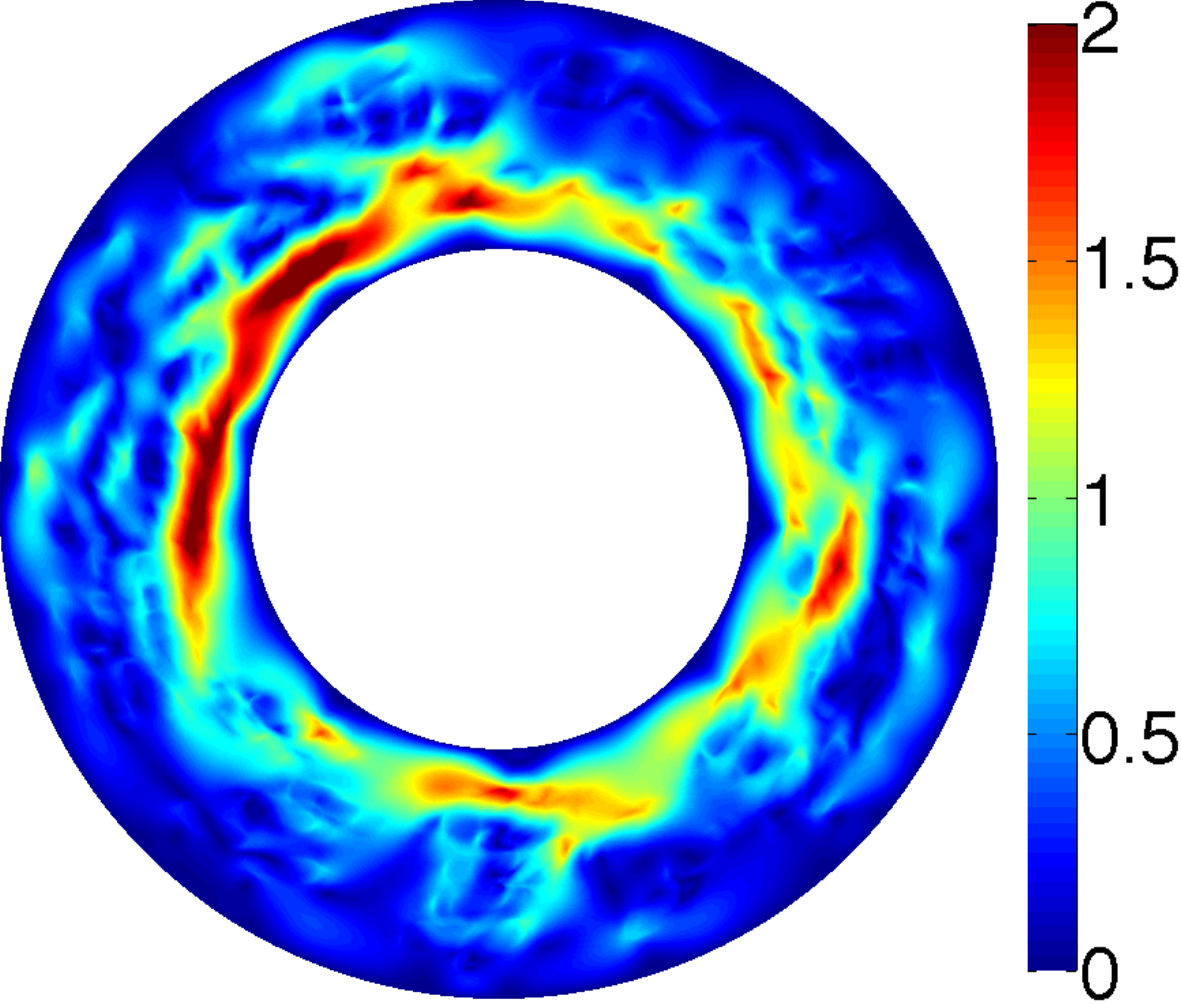}}}
      \label{f:VF20VelS3}}
\setcounter{subfigure}{0}
\renewcommand*{\thesubfigure}{(b-3)} 
      \hspace{0cm}\subfigure[$\phi$ at $t = 30$]{\scalebox{0.43}{{\includegraphics{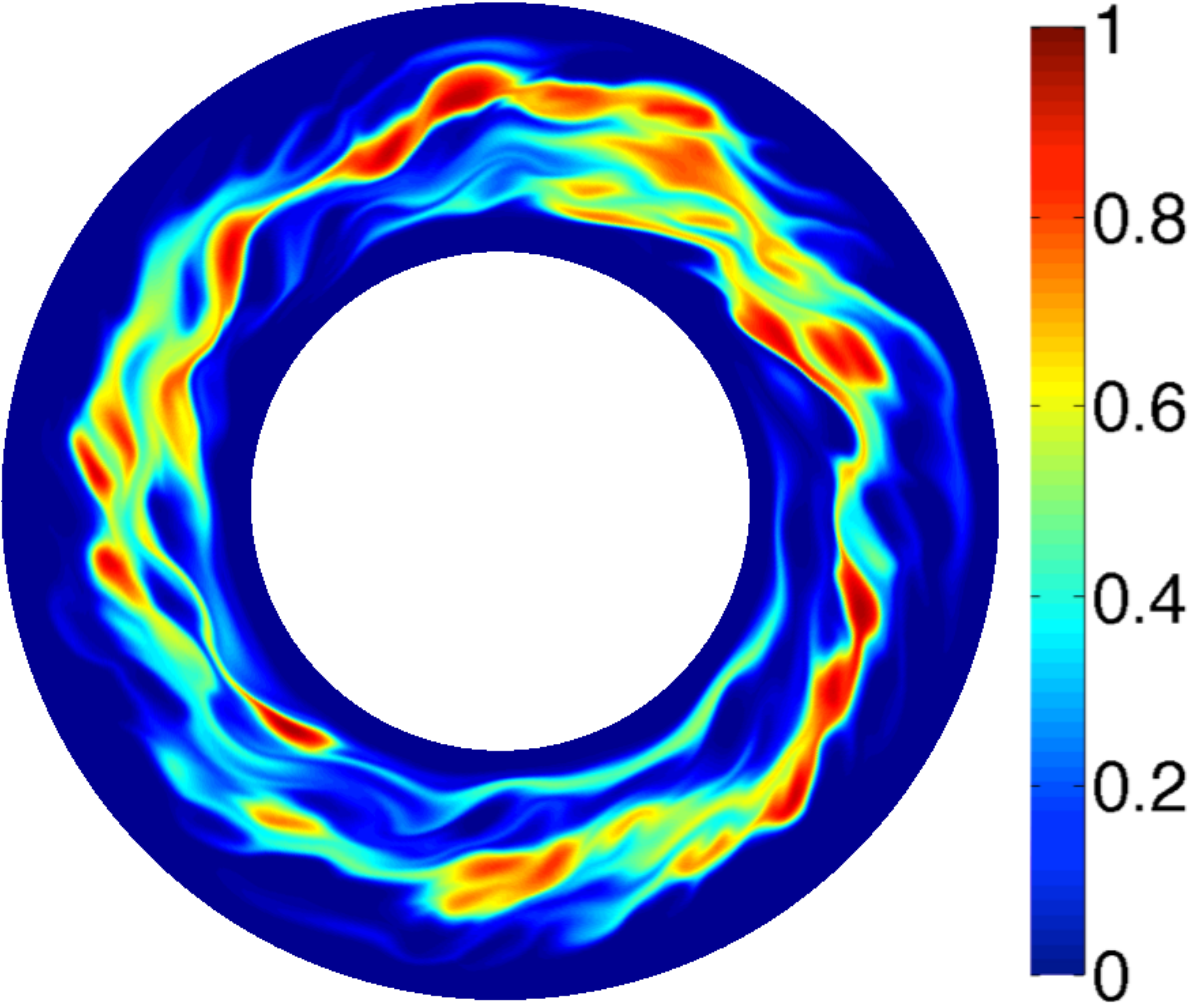}}}
      \label{f:VF20MixS3}}
\end{minipage}
 \begin{minipage}{\textwidth}
\setcounter{subfigure}{0}
\renewcommand*{\thesubfigure}{(c-1)} 
      \hspace{0.5cm}\subfigure[Vesicles at $t = 60$]{\scalebox{0.42}{{\includegraphics{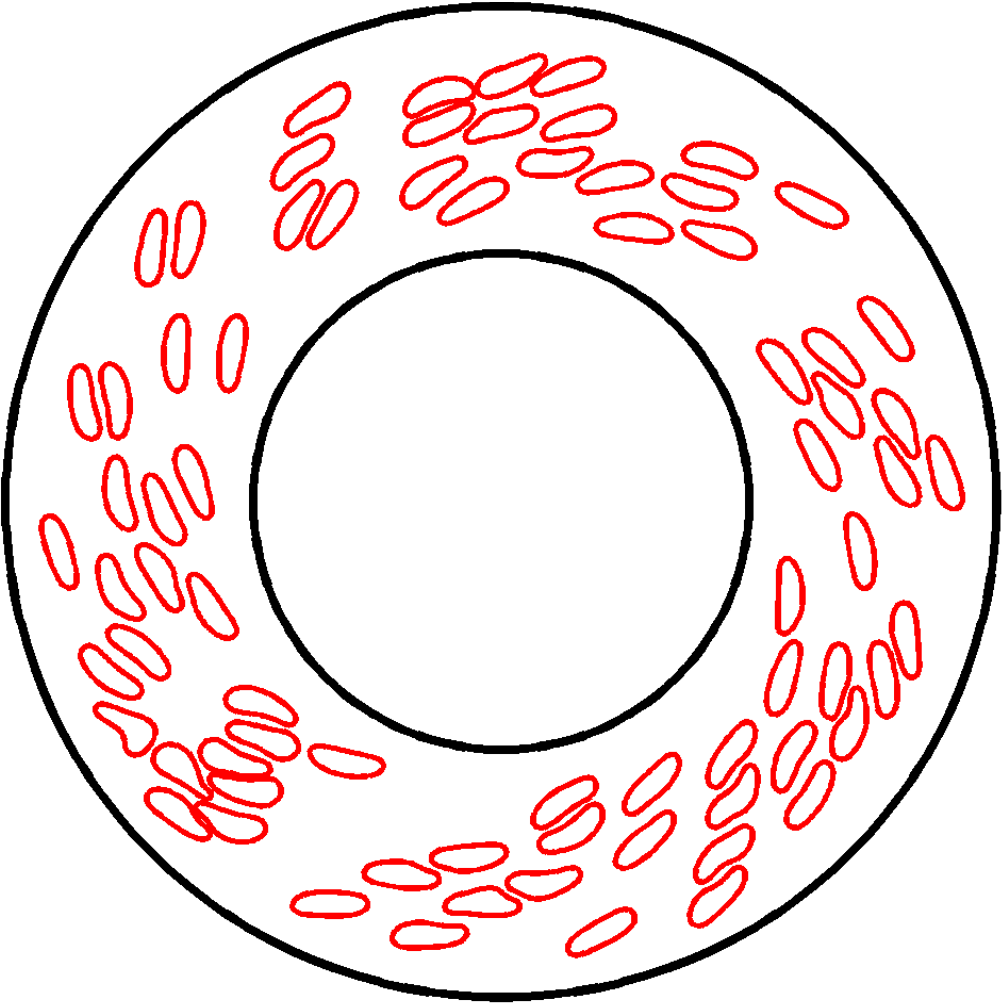}}}	
      \label{f:VF20VesS4}}
\setcounter{subfigure}{0}
\renewcommand*{\thesubfigure}{(c-2)} 
      \hspace{0.2cm}\subfigure[$\| \tilde{\mathbf{v}} \|$  at $t = 60$]{\scalebox{0.43}{{\includegraphics{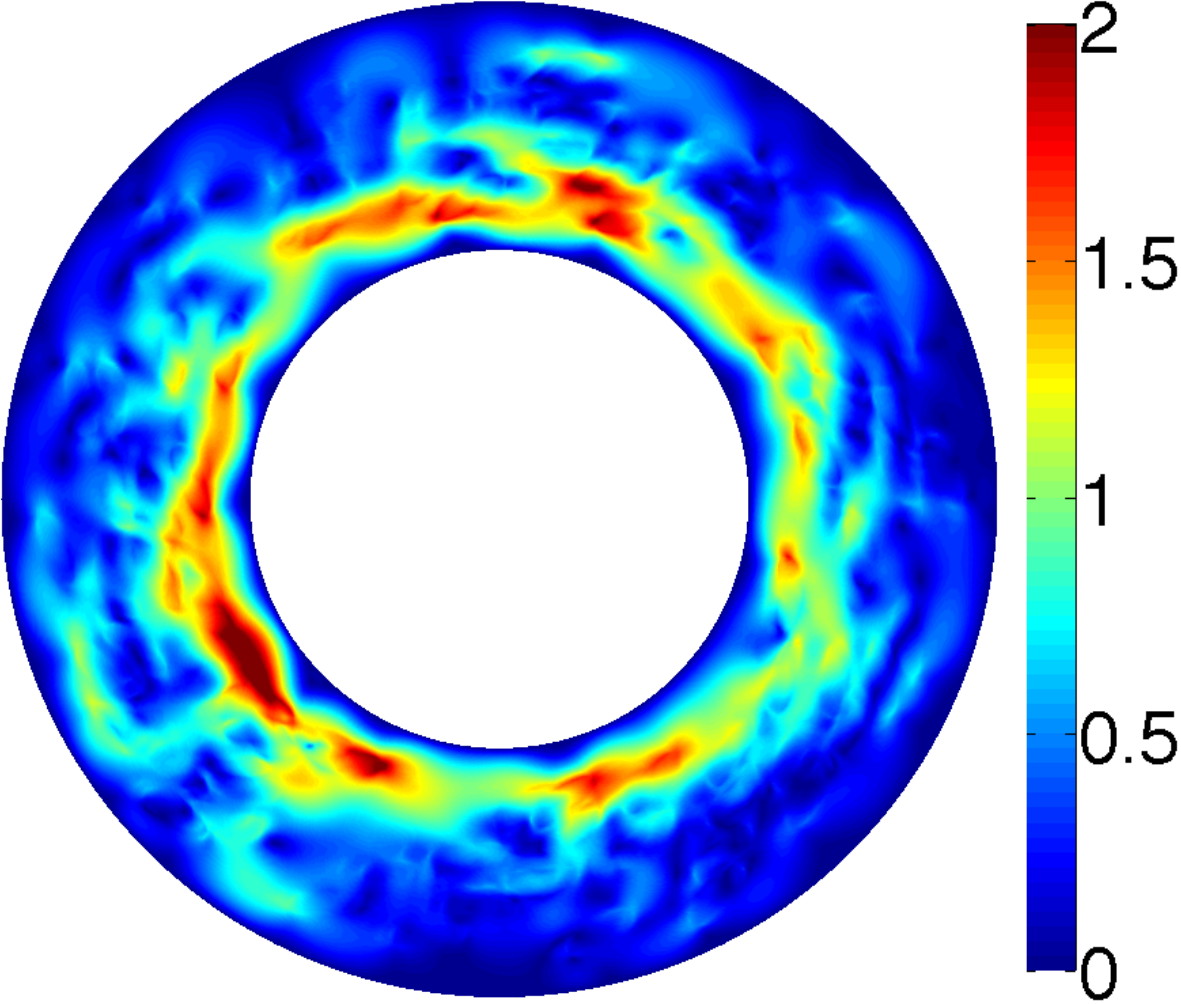}}}
      \label{f:VF20VelS4}}
\setcounter{subfigure}{0}
\renewcommand*{\thesubfigure}{(c-3)} 
      \hspace{0cm}\subfigure[$\phi$ at $t = 60$]{\scalebox{0.43}{{\includegraphics{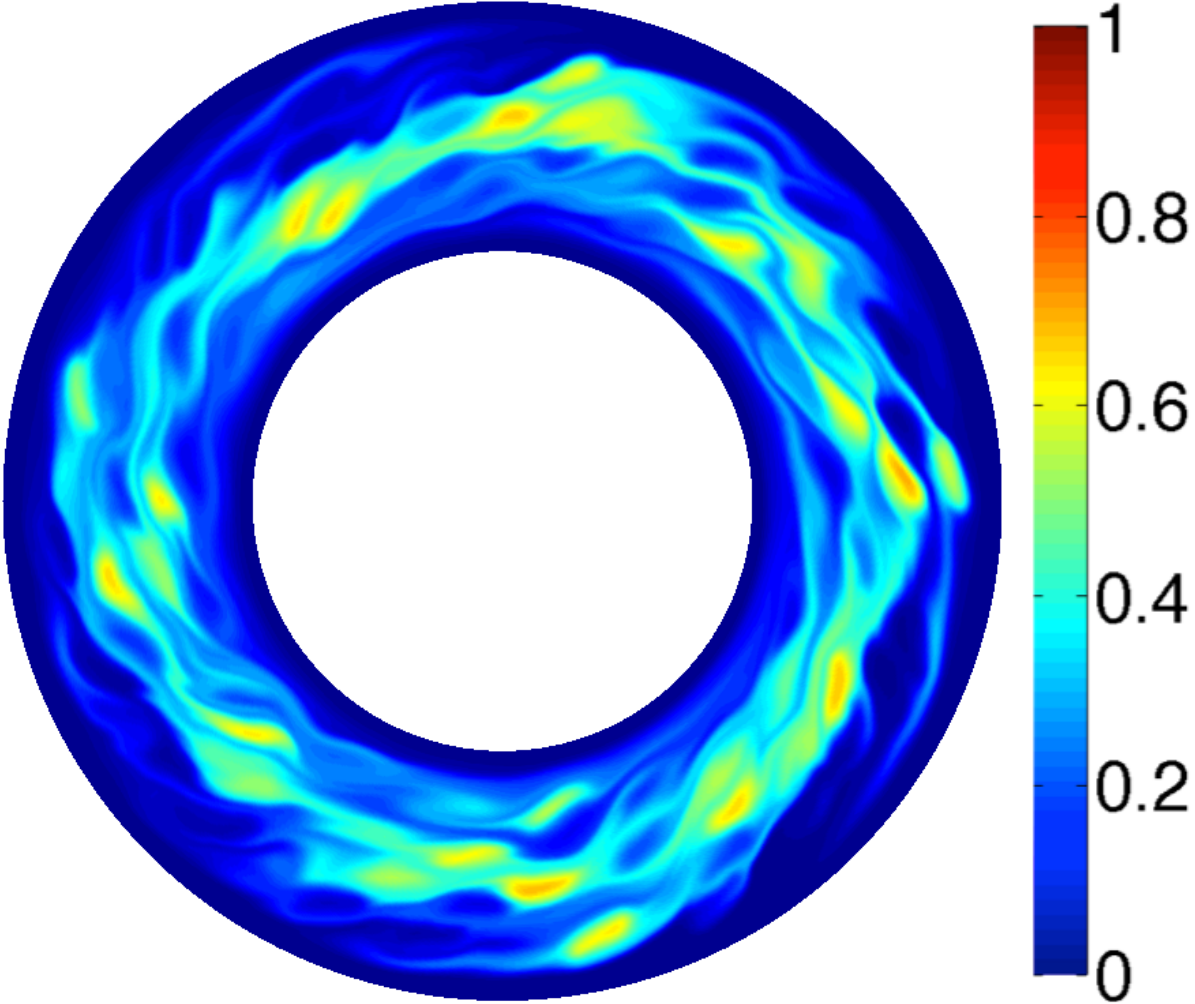}}}
      \label{f:VF20MixS4}}
\end{minipage}
 \begin{minipage}{\textwidth}
\setcounter{subfigure}{0}
\renewcommand*{\thesubfigure}{(d-1)} 
      \hspace{0.5cm}\subfigure[Vesicles at $t = 90$]{\scalebox{0.42}{{\includegraphics{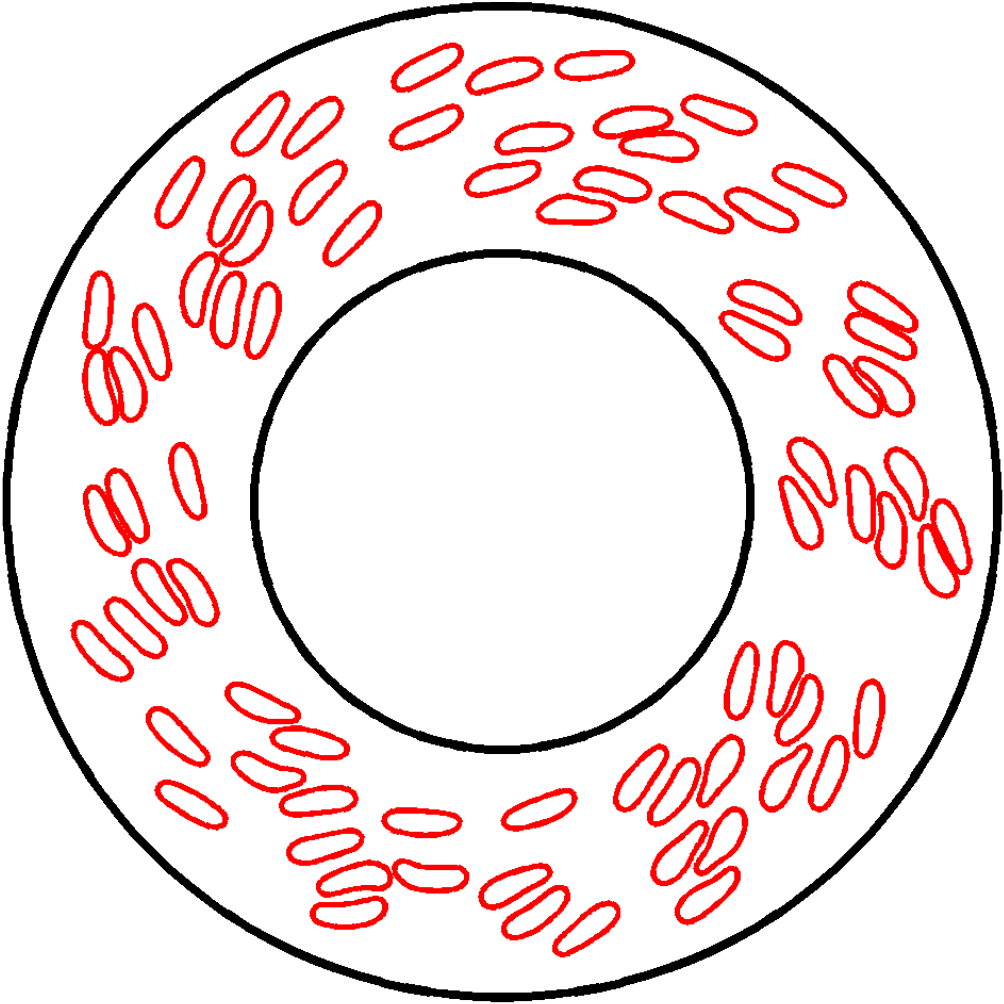}}}	
      \label{f:VF20VesS5}}
\setcounter{subfigure}{0}
\renewcommand*{\thesubfigure}{(d-2)} 
      \hspace{0.2cm}\subfigure[$\| \tilde{\mathbf{v}} \|$  at $t = 90$]{\scalebox{0.43}{{\includegraphics{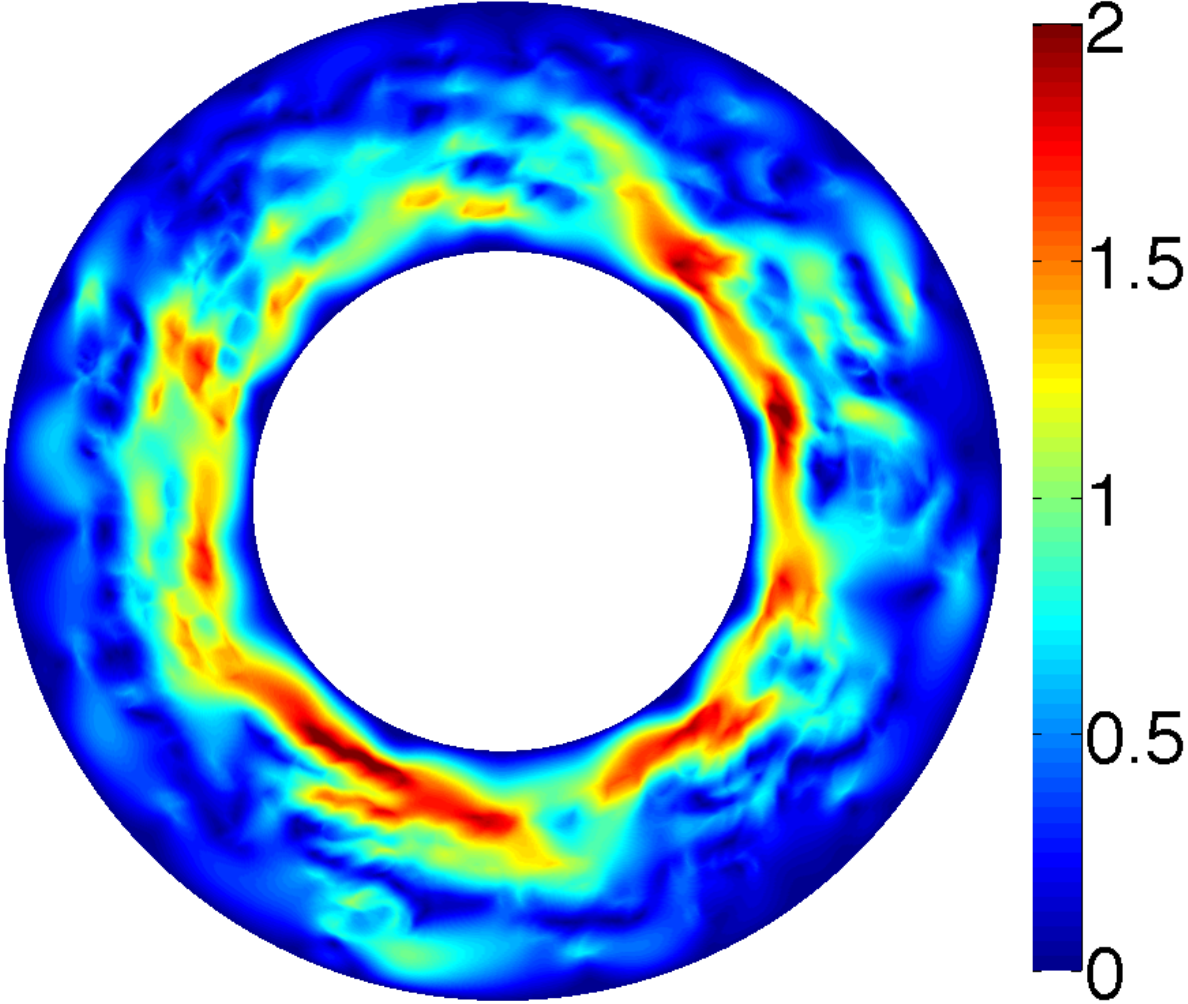}}}
      \label{f:VF20VelS5}}
\setcounter{subfigure}{0}
\renewcommand*{\thesubfigure}{(d-3)} 
      \hspace{0cm}\subfigure[$\phi$ at $t = 90$]{\scalebox{0.43}{{\includegraphics{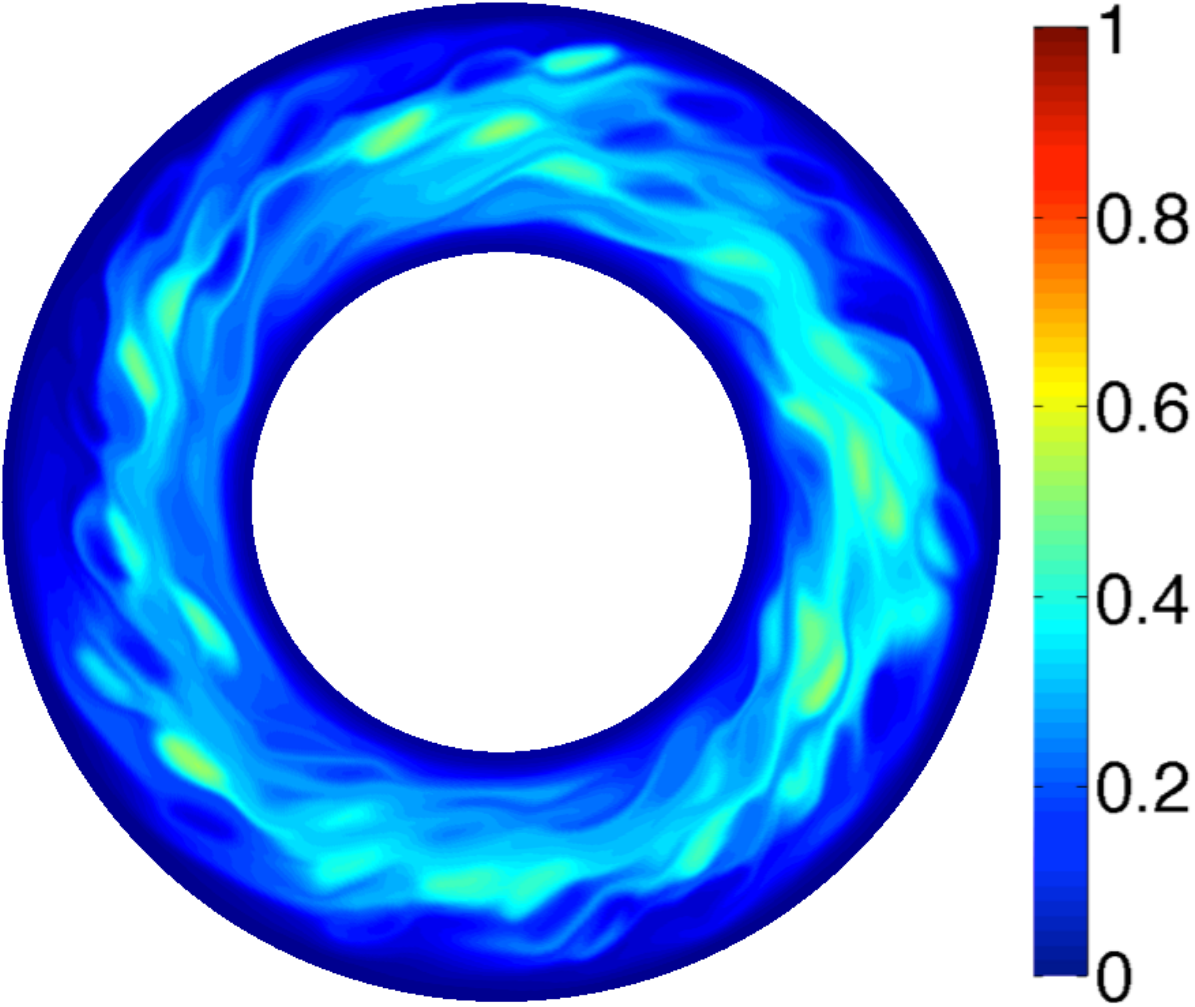}}}
      \label{f:VF20MixS5}}
\end{minipage}
\mcaption{{The} effects of a $20 \%$ area fraction on the velocity field
and mixing (\secref{s:volFrac}).  Here we present the vesicle positions
(left), the magnitude of the velocity field due only to the vesicles
$\|\widetilde{\mathbf{v}}\|$ (middle), and the concentration $\phi$
(right) for the area fraction of $20\%$ and the layer initial
condition.  Each row corresponds to a different time.}{f:VF20Snaps}
\end{figure}

\begin{figure}[H]
\vspace{-0.5cm}
 \begin{minipage}{\textwidth}
\setcounter{subfigure}{0}
\renewcommand*{\thesubfigure}{(a-1)} 
      \hspace{0.5cm}\subfigure[Vesicles at $t = 0$]{\scalebox{0.42}{{\includegraphics{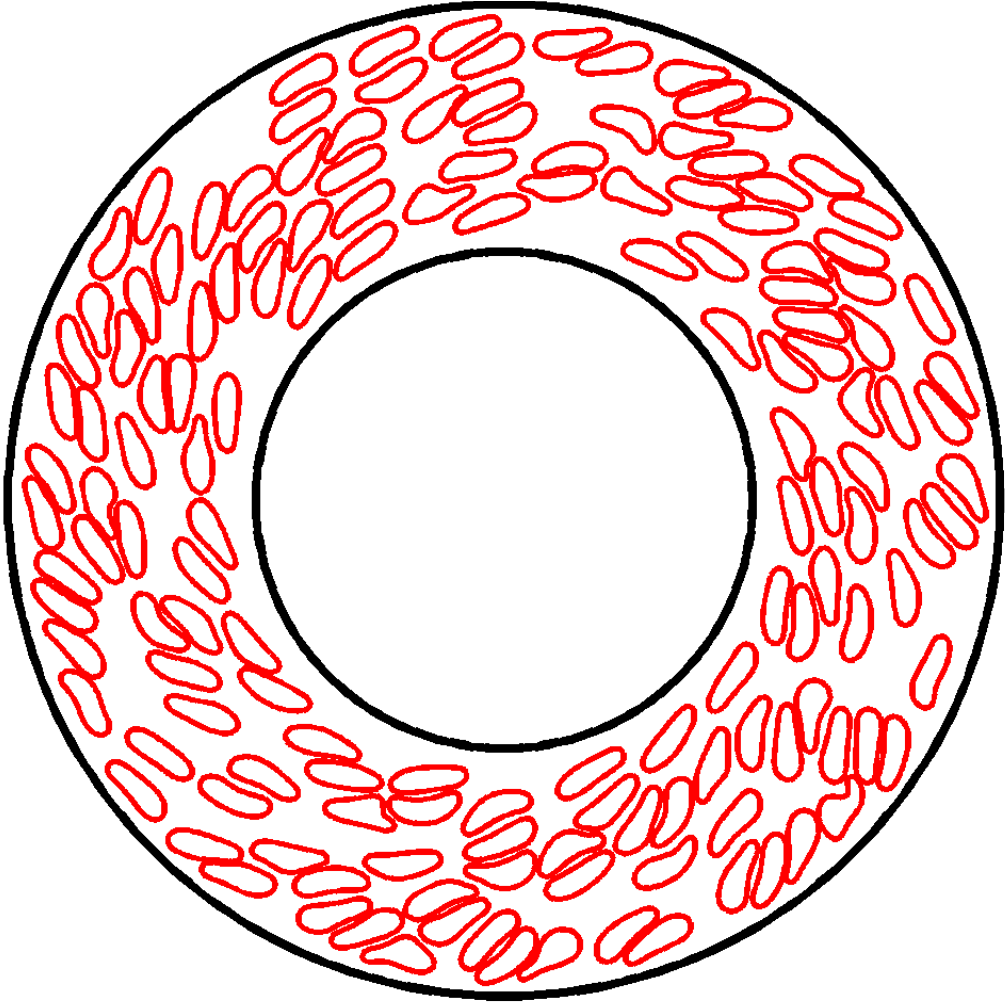}}}	
      \label{f:VF40VesS1}}
\setcounter{subfigure}{0}
\renewcommand*{\thesubfigure}{(a-2)} 
      \hspace{0.2cm}\subfigure[$\| \tilde{\mathbf{v}} \|$  at $t = 0$]{\scalebox{0.43}{{\includegraphics{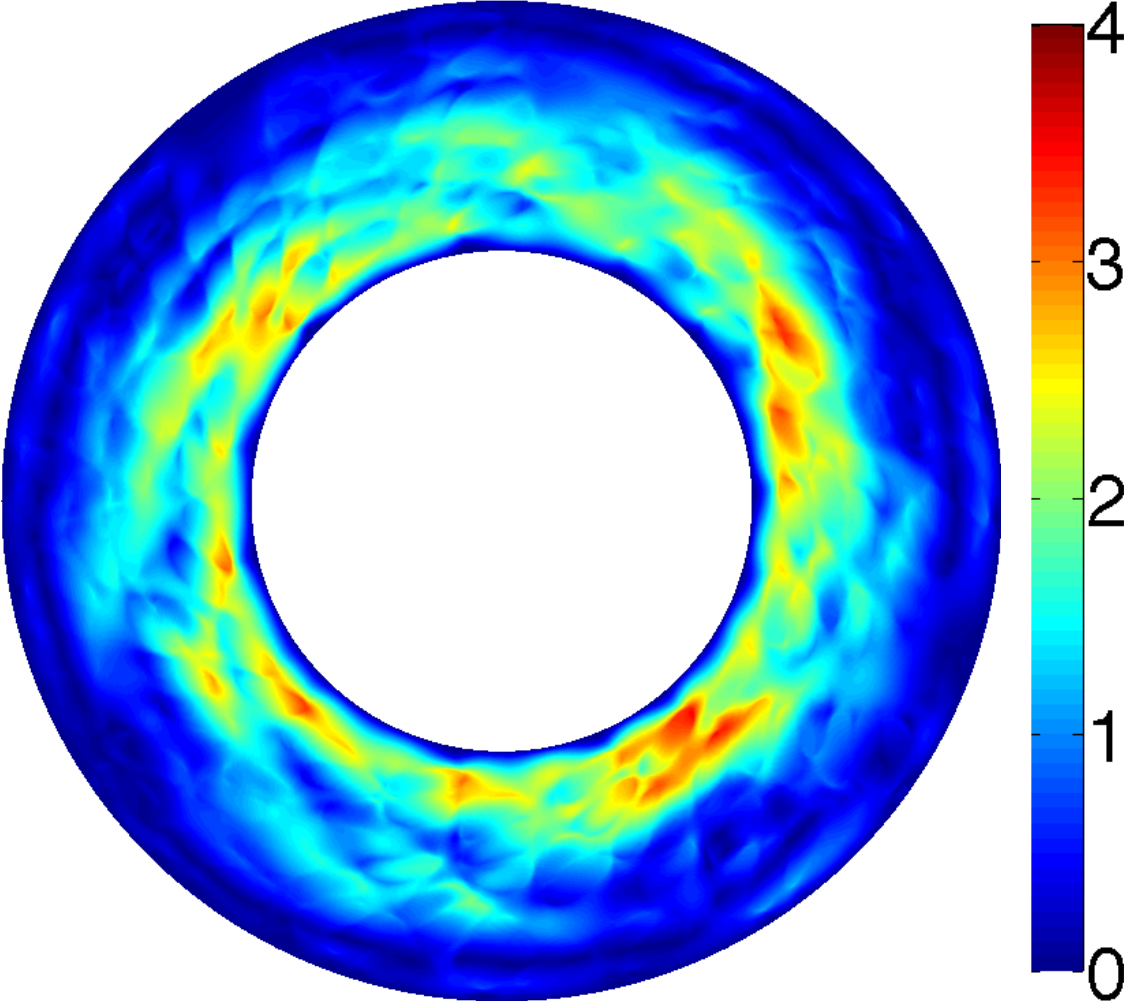}}}
      \label{f:VF40VelS1}}
\setcounter{subfigure}{0}
\renewcommand*{\thesubfigure}{(a-3)} 
      \hspace{0.25cm}\subfigure[$\phi$ at $t = 0$]{\scalebox{0.43}{{\includegraphics{VF10_MixS1.pdf}}}
      \label{f:VF40MixS1}}
\end{minipage}
 \begin{minipage}{\textwidth}
\setcounter{subfigure}{0}
\renewcommand*{\thesubfigure}{(b-1)} 
      \hspace{0.5cm}\subfigure[Vesicles at $t = 30$]{\scalebox{0.42}{{\includegraphics{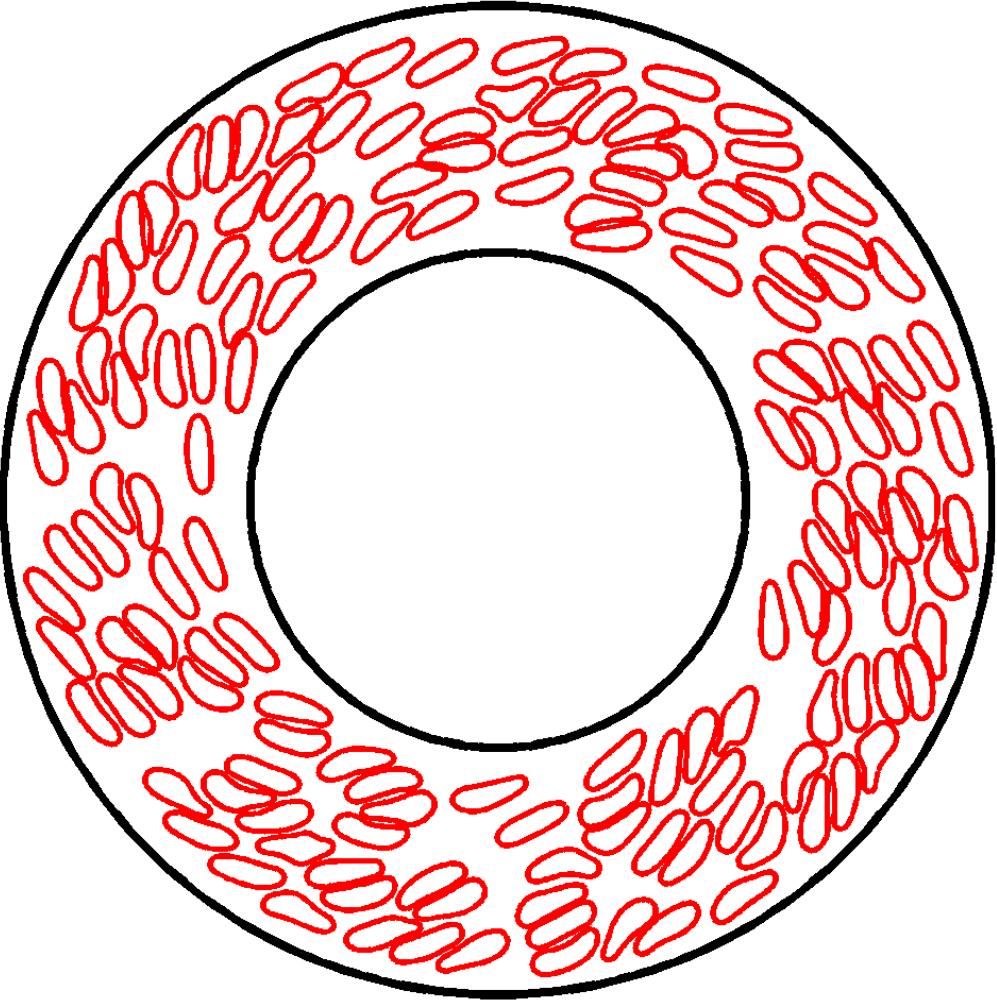}}}	
      \label{f:VF40VesS3}}
\setcounter{subfigure}{0}
\renewcommand*{\thesubfigure}{(b-2)} 
      \hspace{0.2cm}\subfigure[$\| \tilde{\mathbf{v}} \|$  at $t = 30$]{\scalebox{0.43}{{\includegraphics{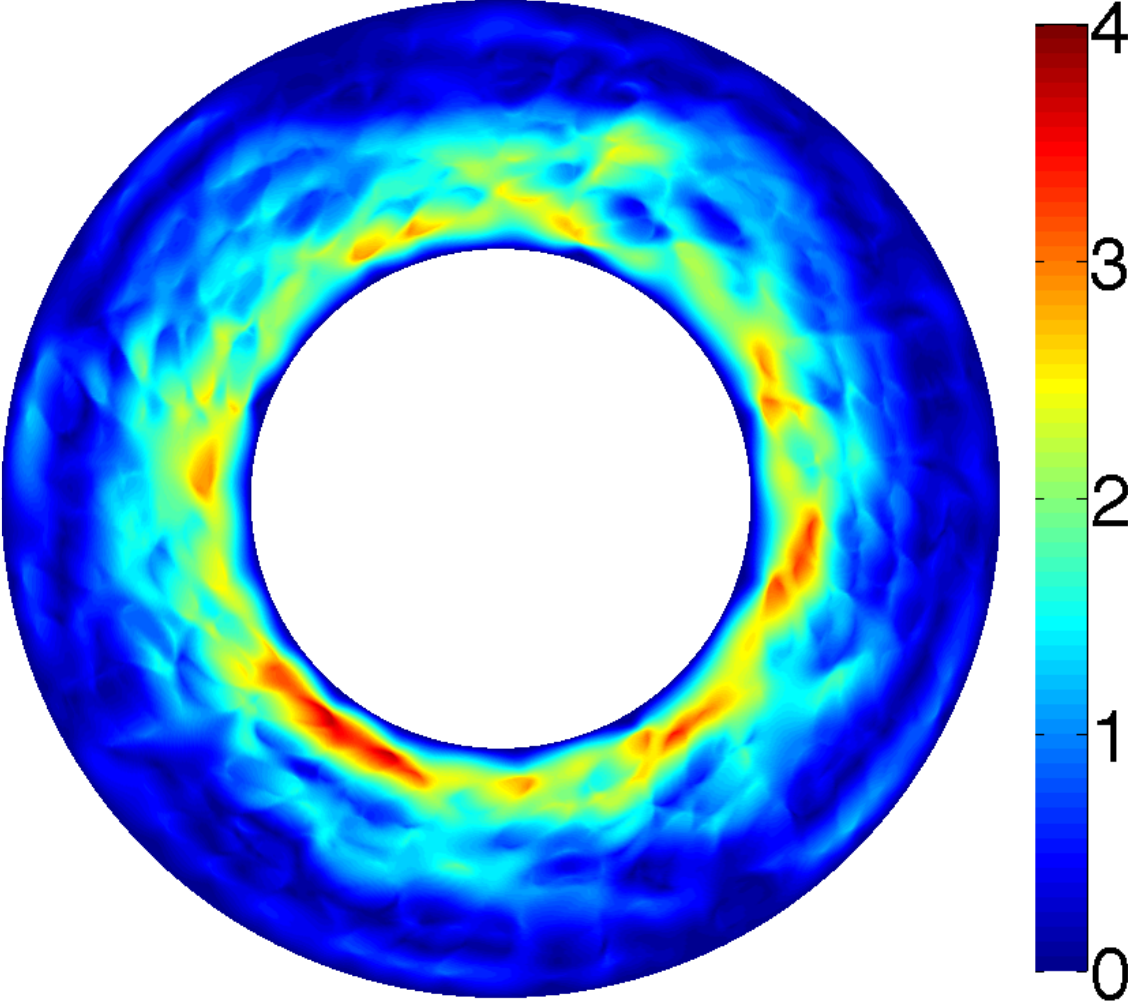}}}
      \label{f:VF40VelS3}}
\setcounter{subfigure}{0}
\renewcommand*{\thesubfigure}{(b-3)} 
      \hspace{0.25cm}\subfigure[$\phi$ at $t = 30$]{\scalebox{0.43}{{\includegraphics{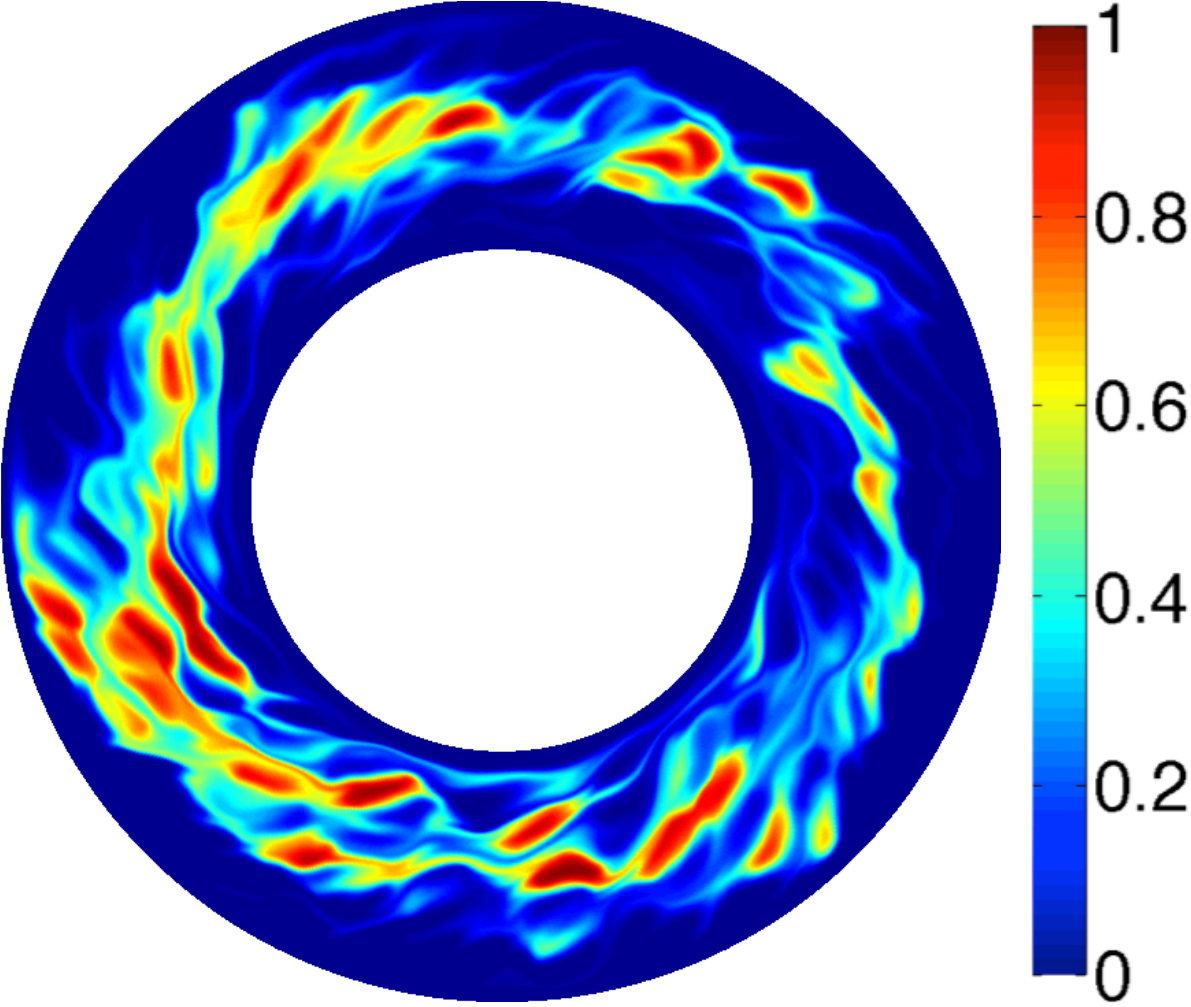}}}
      \label{f:VF40MixS3}}
\end{minipage}
 \begin{minipage}{\textwidth}
\setcounter{subfigure}{0}
\renewcommand*{\thesubfigure}{(c-1)} 
      \hspace{0.5cm}\subfigure[Vesicles at $t = 60$]{\scalebox{0.42}{{\includegraphics{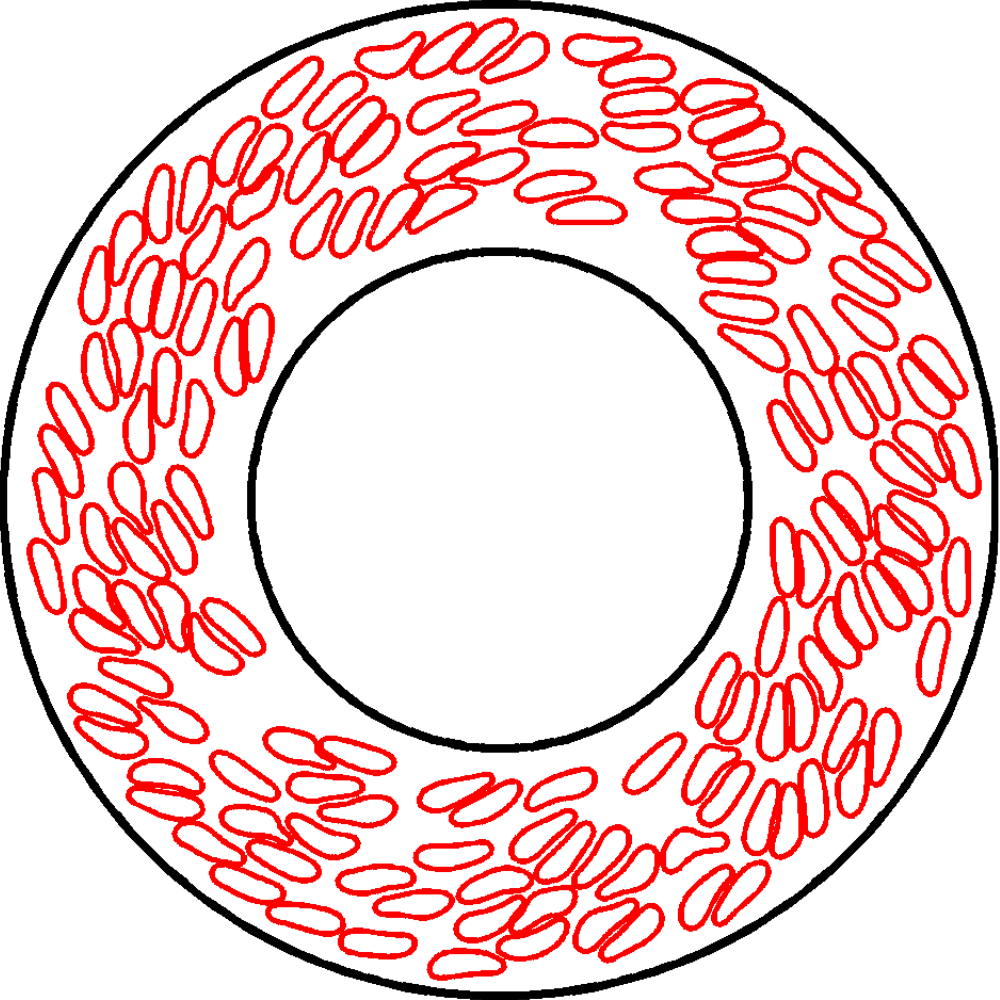}}}	
      \label{f:VF40VesS4}}
\setcounter{subfigure}{0}
\renewcommand*{\thesubfigure}{(c-2)} 
      \hspace{0.2cm}\subfigure[$\| \tilde{\mathbf{v}} \|$  at $t = 60$]{\scalebox{0.43}{{\includegraphics{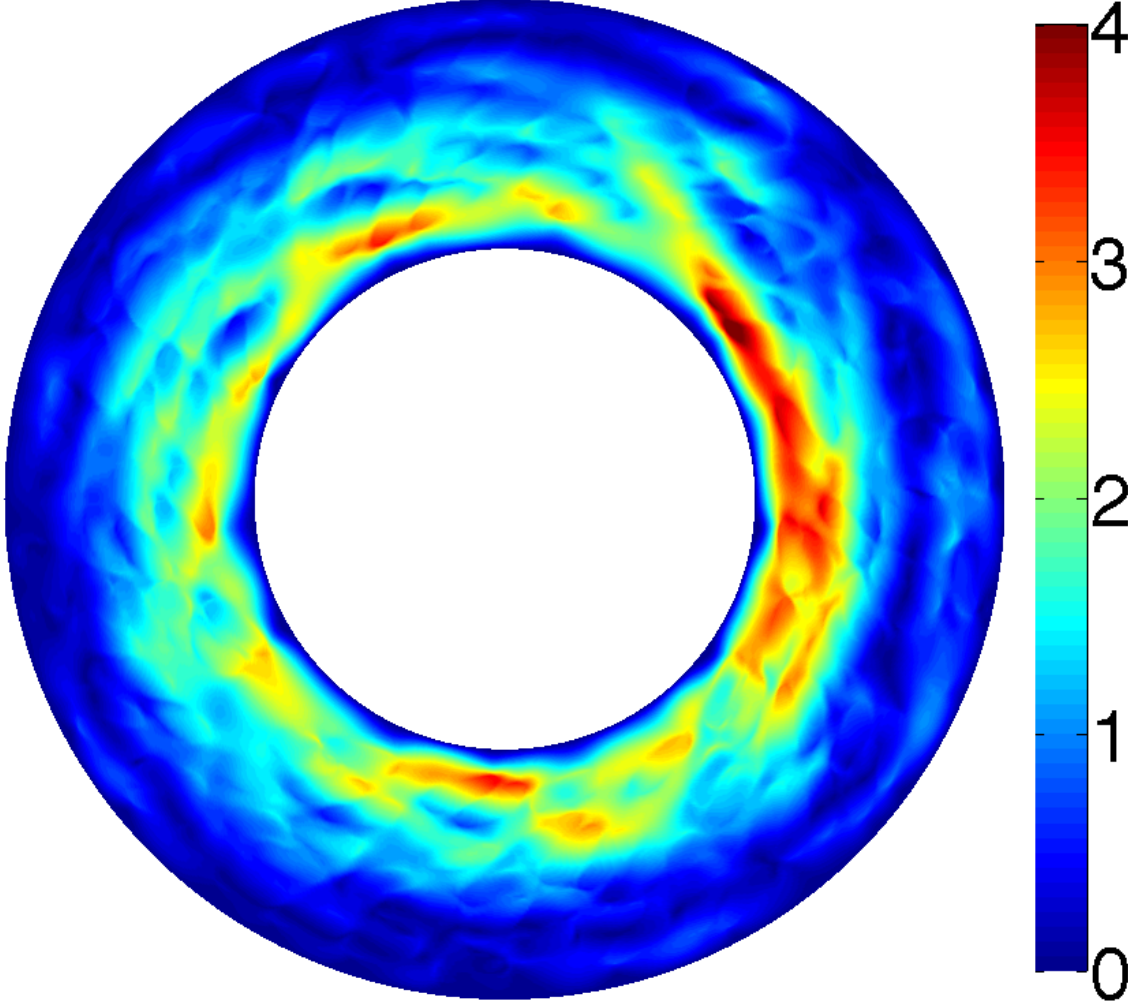}}}
      \label{f:VF40VelS4}}
\setcounter{subfigure}{0}
\renewcommand*{\thesubfigure}{(c-3)} 
      \hspace{0.25cm}\subfigure[$\phi$ at $t = 60$]{\scalebox{0.43}{{\includegraphics{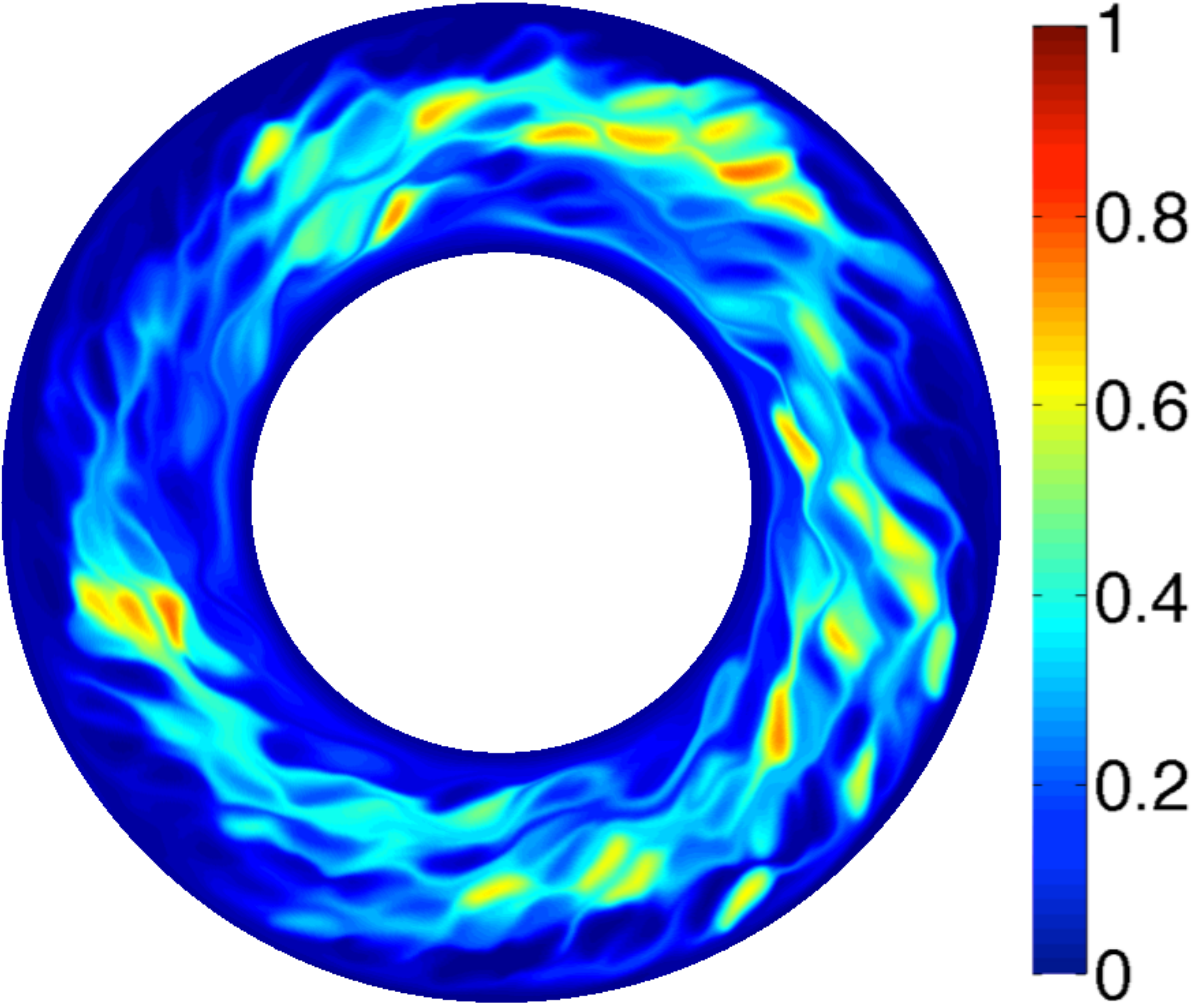}}}
      \label{f:VF40MixS4}}
\end{minipage}
 \begin{minipage}{\textwidth}
\setcounter{subfigure}{0}
\renewcommand*{\thesubfigure}{(d-1)} 
      \hspace{0.5cm}\subfigure[Vesicles at $t = 90$]{\scalebox{0.42}{{\includegraphics{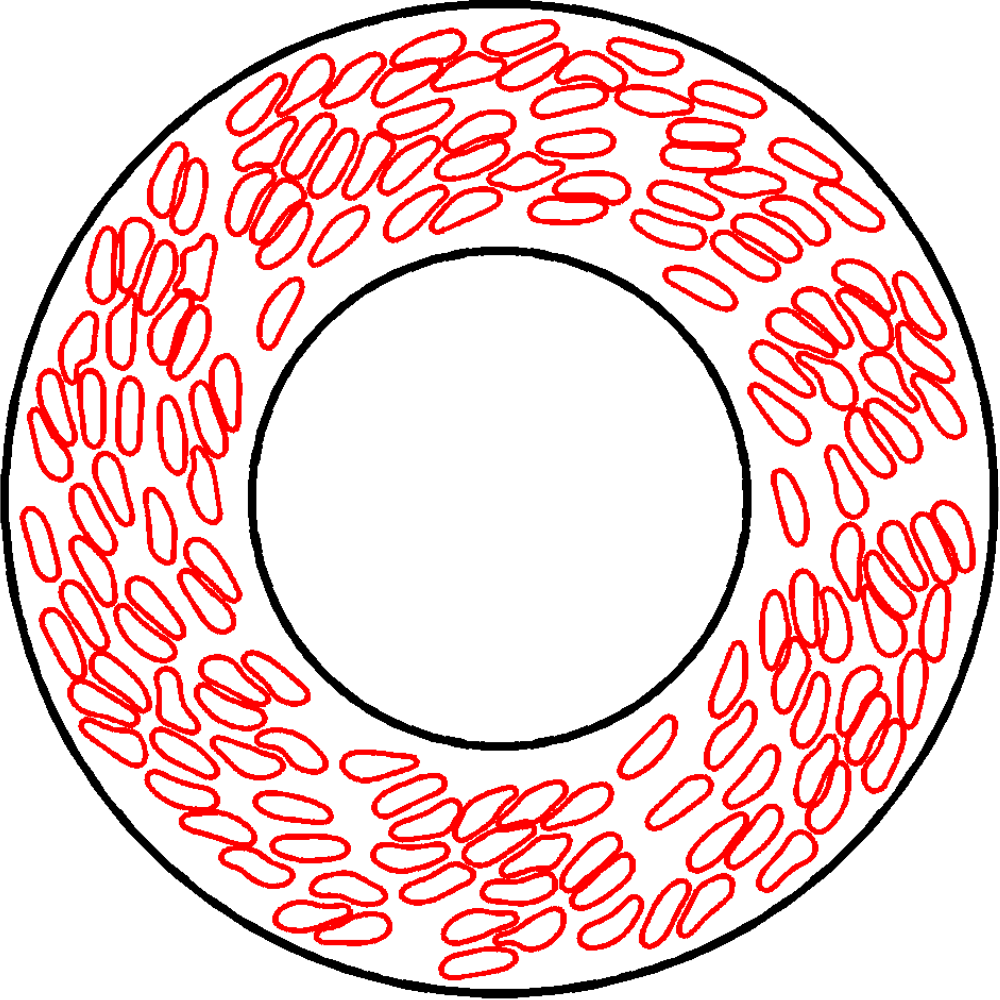}}}	
      \label{f:VF40VesS5}}
\setcounter{subfigure}{0}
\renewcommand*{\thesubfigure}{(d-2)} 
      \hspace{0.2cm}\subfigure[$\| \tilde{\mathbf{v}} \|$  at $t = 90$]{\scalebox{0.43}{{\includegraphics{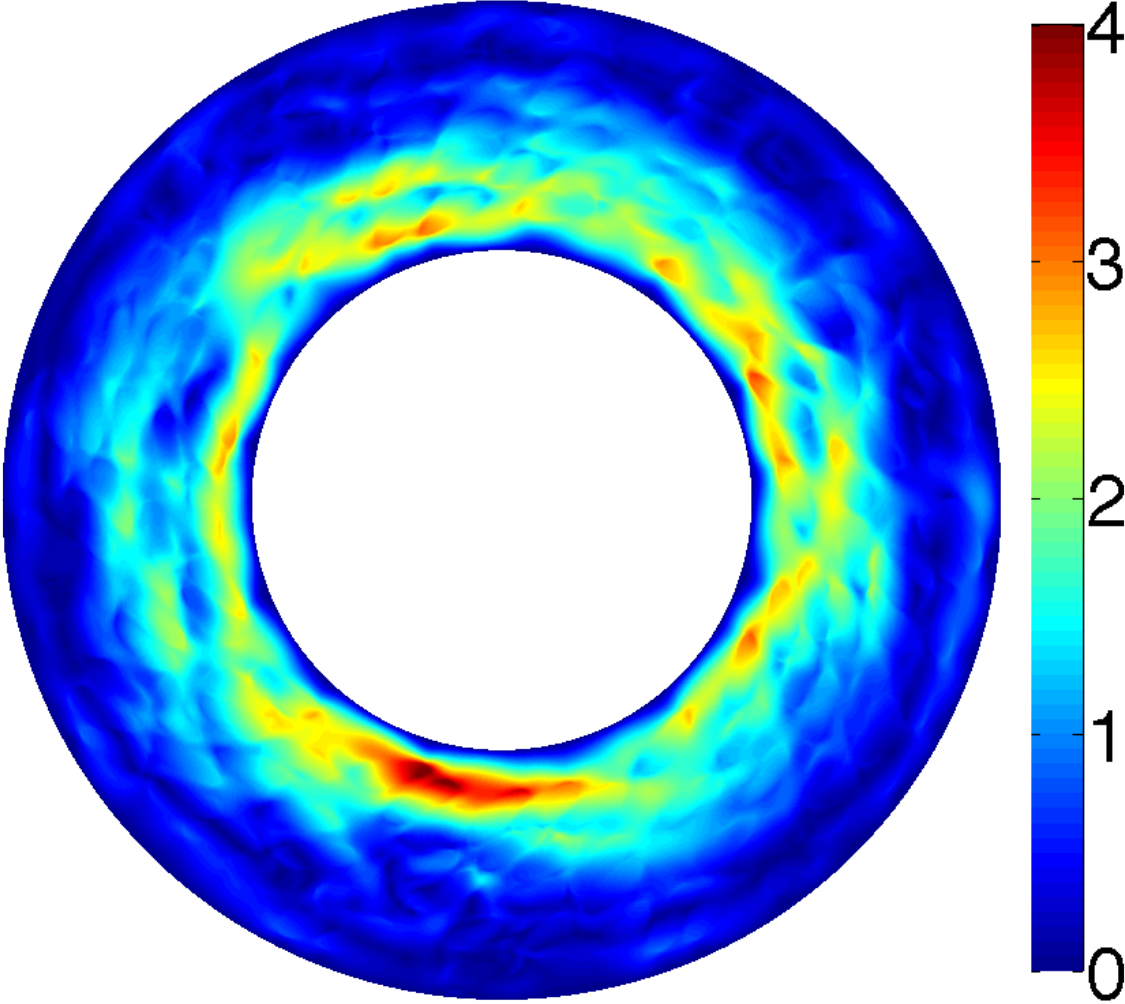}}}
      \label{f:VF40VelS5}}
\setcounter{subfigure}{0}
\renewcommand*{\thesubfigure}{(d-3)} 
      \hspace{0.25cm}\subfigure[$\phi$ at $t = 90$]{\scalebox{0.43}{{\includegraphics{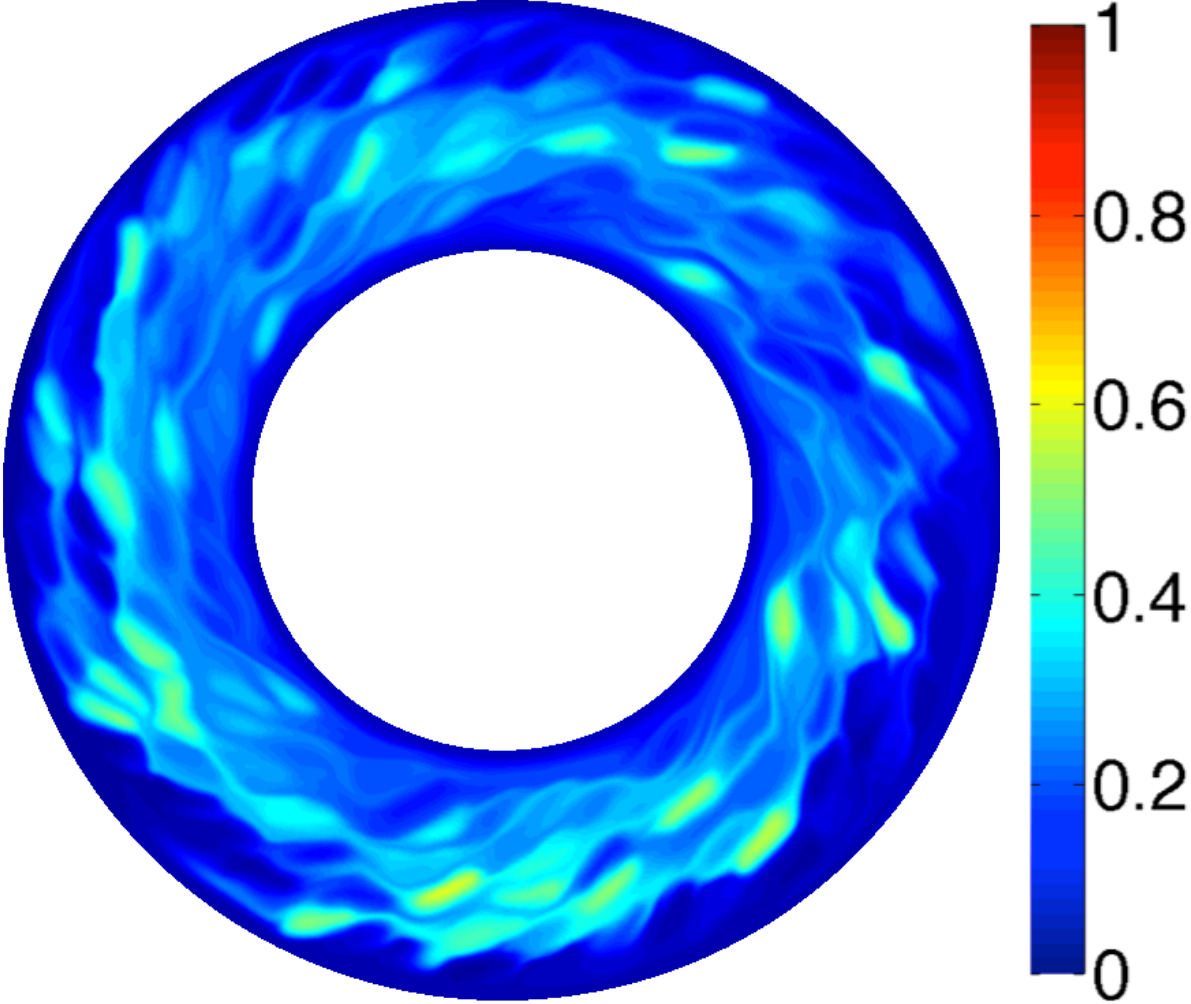}}}
      \label{f:VF40MixS5}}
\end{minipage}
\mcaption{{The} effects of a $40 \%$ area fraction on the velocity field
and mixing (\secref{s:volFrac}).  Here we present the vesicle positions
(left), the magnitude of the velocity field due only to the vesicles
$\|\widetilde{\mathbf{v}}\|$ (middle), and the concentration $\phi$
(right) for the area fraction of $40\%$ and the layer initial
condition.  Each row corresponds to a different time.}{f:VF40Snaps}
\end{figure}

\subsection{Effects of Peclet number and initial condition}\label{s:initialCond}
We investigate the mixing efficiency for various Peclet numbers and
initial conditions. We simulate mixing with the Peclet numbers of
$1e+4$, $5e+3$, $2.5e+3$, $5e+2$, and $5e+1$, and for all four initial
conditions in \figref{f:initConds}. We, then, demonstrate the mixing
efficiency $\eta$ with respect to time in \figref{f:VFMixing1} and
\figref{f:VFMixing2}. The results in \figref{f:VFMixing1} and
\figref{f:VFMixing2} show that the mixing efficiency is close to one
for Pe $ = \mathcal{O}(10)$, but it deviates from one as the Peclet
number increases.  This is expected since the perturbations in the
velocity field become more important in the sense of mixing as the
transport becomes more advection-dominated. 

While, the first row in \figref{f:VFMixing1} shows that the presence of
vesicles enhances mixing (i.e. $\eta > 1$), the second row demonstrates
that vesicles deteriorate mixing. Additionally, as the area fraction
increases (from left to right in \figref{f:VFMixing1} and
\figref{f:VFMixing2}), the maximum efficiency increases for the layer
initial condition and the minimum efficiency increases for the dye
initial condition. The first row in \figref{f:VFMixing2} illustrates
that mixing is better in the absence of vesicles for the vesicle
initial condition, however, the effects of the vesicles on mixing
becomes less important as the area fraction increases. Furthermore, a
Couette flow without vesicles provides the same quality of mixing as
the one with vesicles for the random initial condition for any area
fraction (see the second row in \figref{f:VFMixing2}).

\begin{figure}[H]
 \begin{minipage}{\textwidth}
\setcounter{subfigure}{0}
\renewcommand*{\thesubfigure}{(a-1)} 
      \hspace{0cm}\subfigure[AF $= 10\%$]{\scalebox{0.3}{{\includegraphics{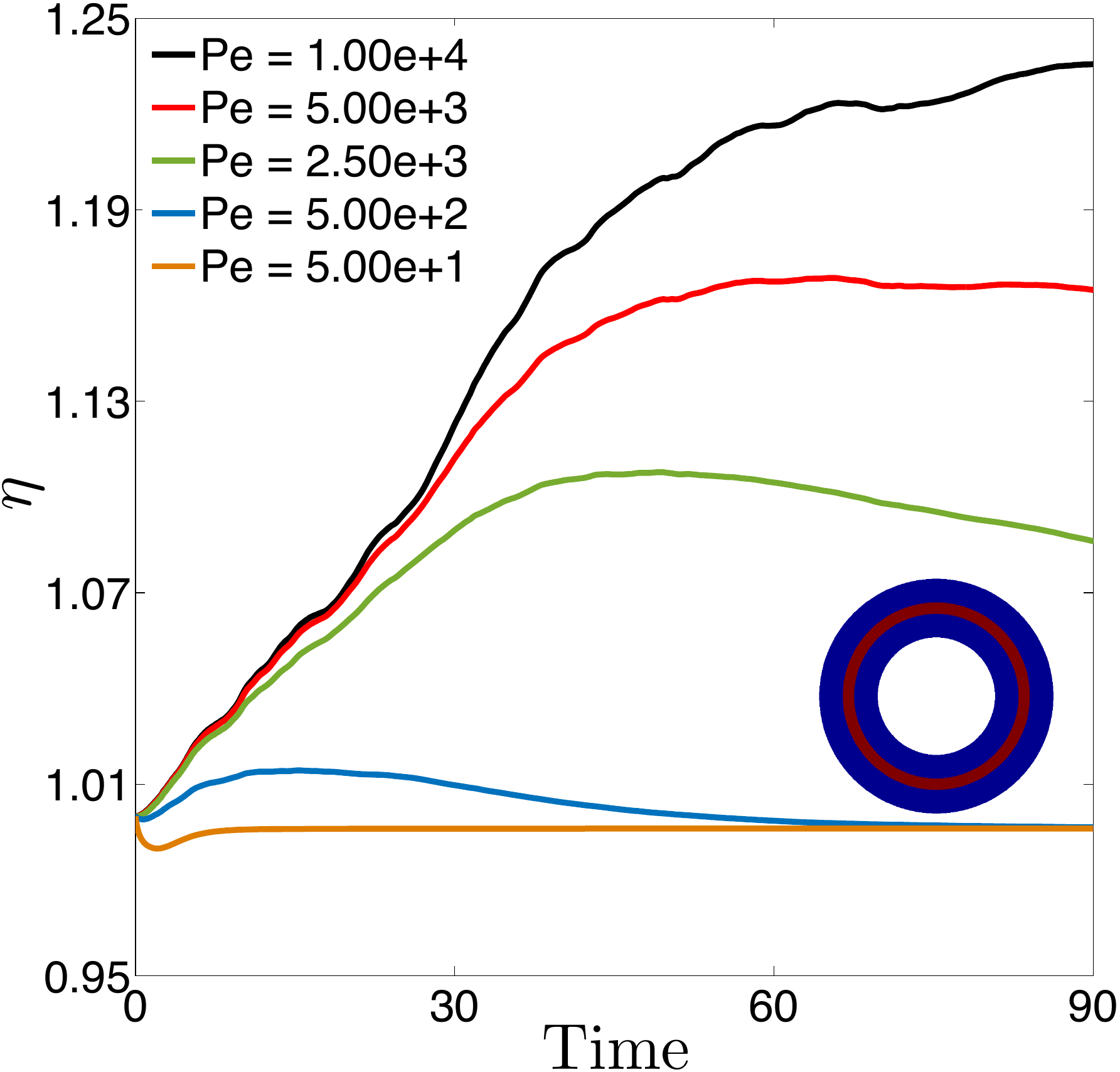}}}	
      \label{f:LayerVF10}}
\setcounter{subfigure}{0}
\renewcommand*{\thesubfigure}{(a-2)} 
      \hspace{-0.4cm}\subfigure[AF $= 20\%$]{\scalebox{0.3}{{\includegraphics{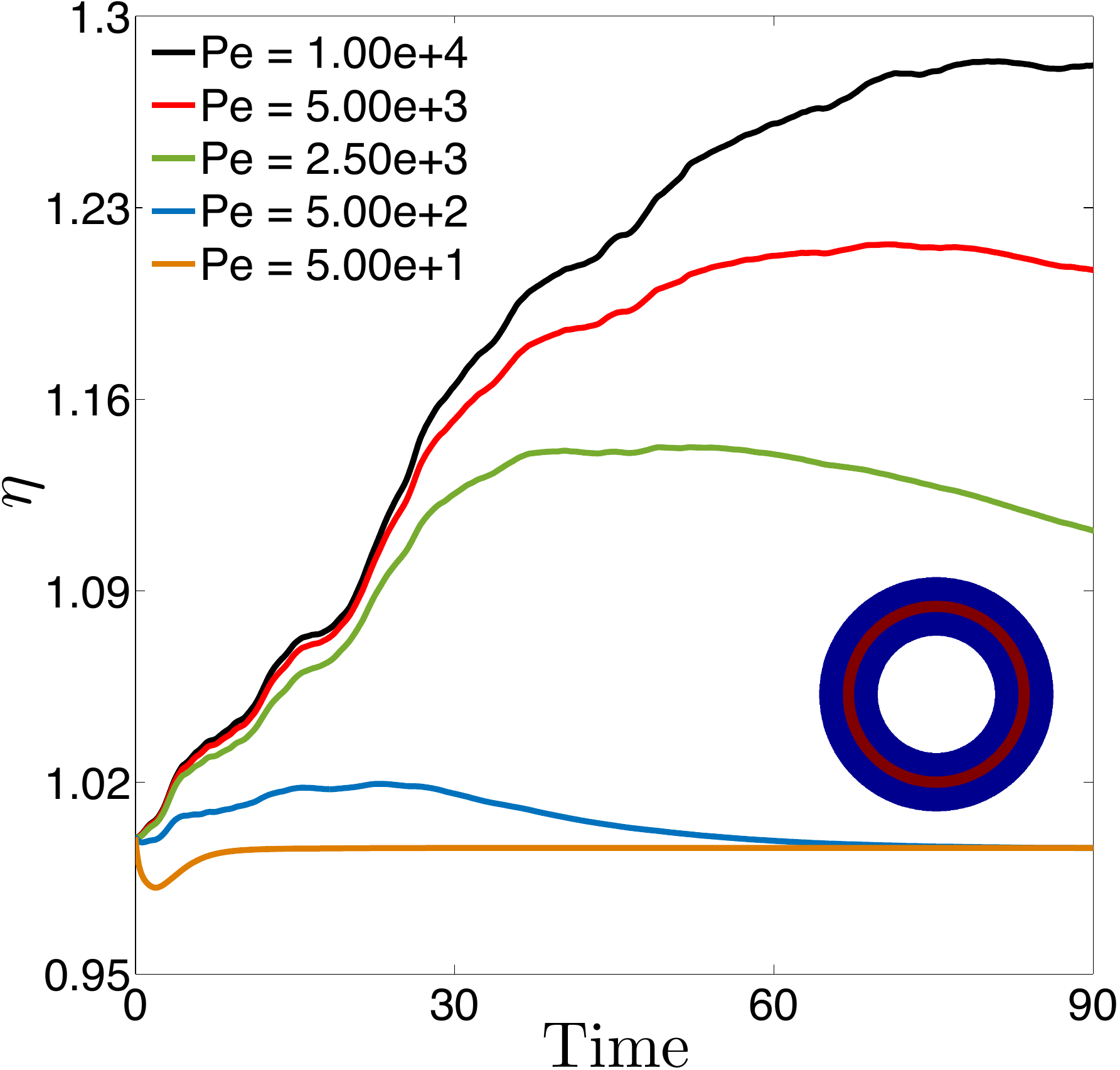}}}
      \label{f:LayerVF20}}
\setcounter{subfigure}{0}
\renewcommand*{\thesubfigure}{(a-3)} 
      \hspace{-0.4cm}\subfigure[AF $= 40\%$]{\scalebox{0.3}{{\includegraphics{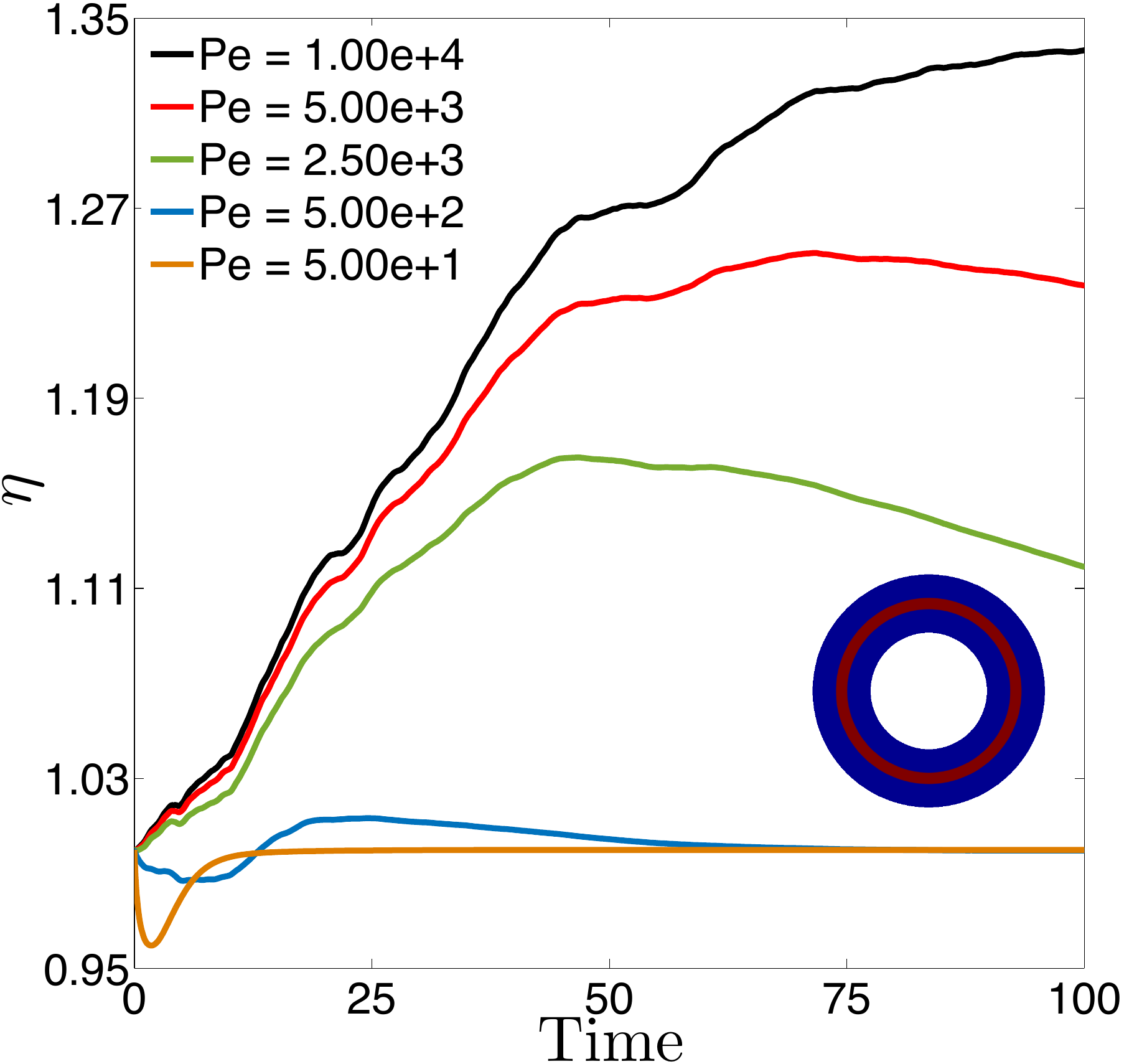}}}
      \label{f:LayerVF40}}
  \end{minipage}
  \begin{minipage}{\textwidth}
\setcounter{subfigure}{0}
\renewcommand*{\thesubfigure}{(b-1)} 
      \hspace{0cm}\subfigure[AF $= 10\%$]{\scalebox{0.295}{{\includegraphics{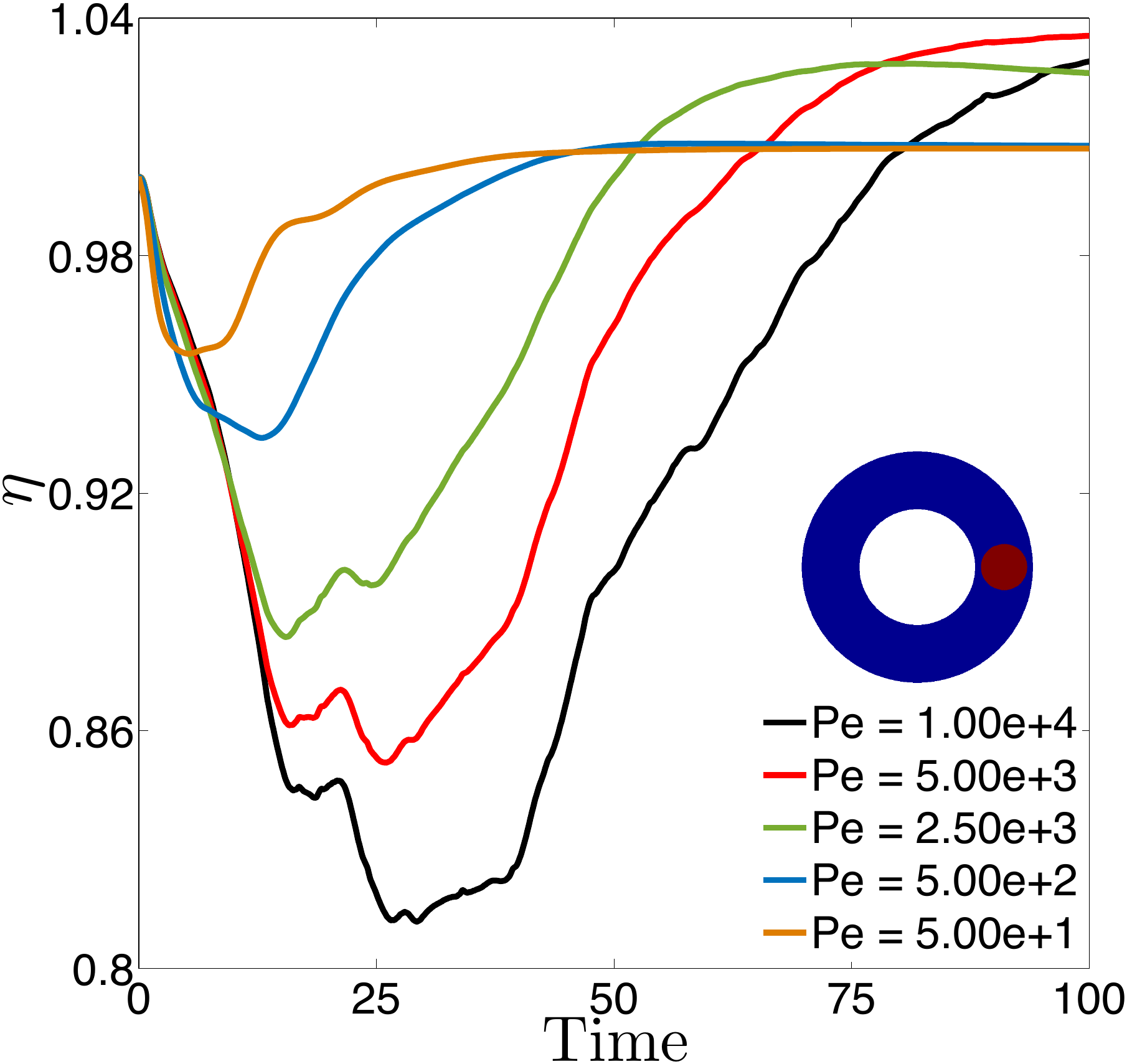}}}
      \label{f:DyeVF10}}
\setcounter{subfigure}{0}
\renewcommand*{\thesubfigure}{(b-2)} 
      \hspace{-0.4cm}\subfigure[AF $= 20\%$]{\scalebox{0.295}{{\includegraphics{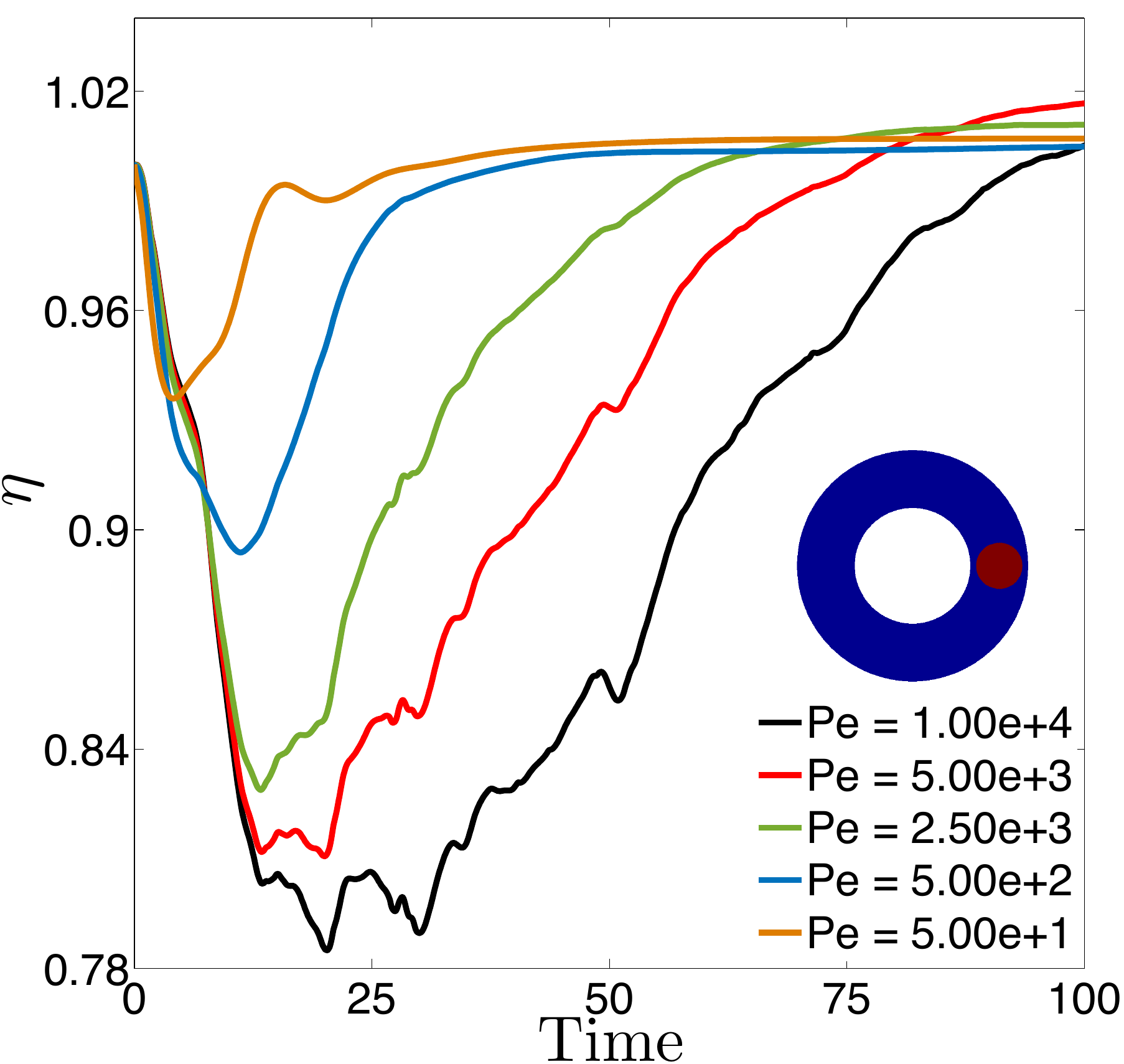}}}
      \label{f:DyeVF20}}
\setcounter{subfigure}{0}
\renewcommand*{\thesubfigure}{(b-3)} 
      \hspace{-0.4cm}\subfigure[AF $= 40\%$]{\scalebox{0.295}{{\includegraphics{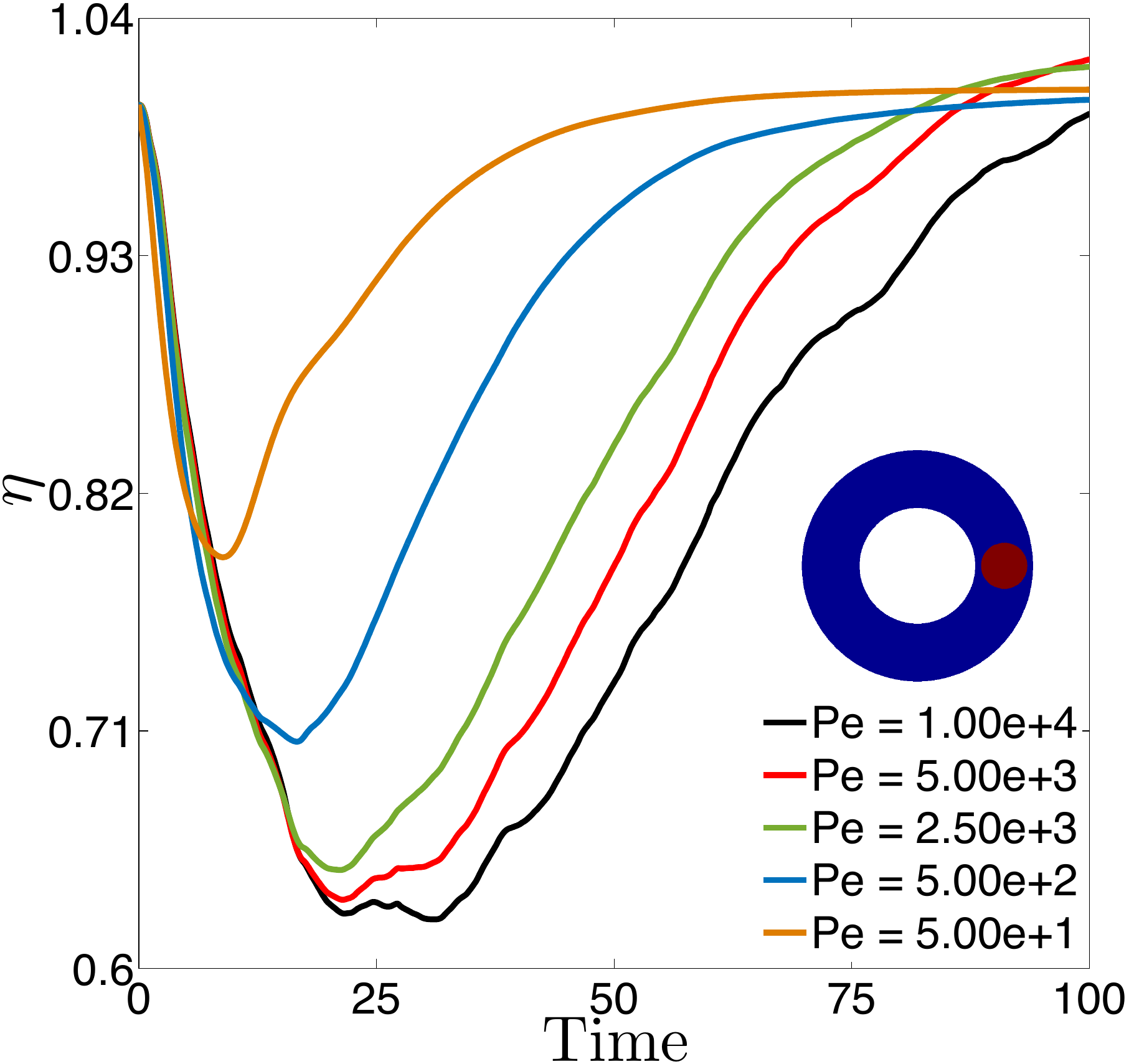}}}
      \label{f:DyeVF40}} 
    \end{minipage}                  
\mcaption{{The} effects of the area fraction on the degree of mixing for
various Peclet numbers and initial conditions
(\secref{s:initialCond}).  The first row demonstrates that the mixing
efficiency $\eta$ increases with increasing area fraction of the
vesicles for the layer initial condition. The second row shows that
$\eta$ decreases when the initial condition is switched to the dye.}
{f:VFMixing1}
\end{figure}

\begin{figure}[H]
\begin{minipage}{\textwidth} 
\setcounter{subfigure}{0}
\renewcommand*{\thesubfigure}{(c-1)} 
      \hspace{0cm}\subfigure[AF $= 10\%$]{\scalebox{0.3}{{\includegraphics{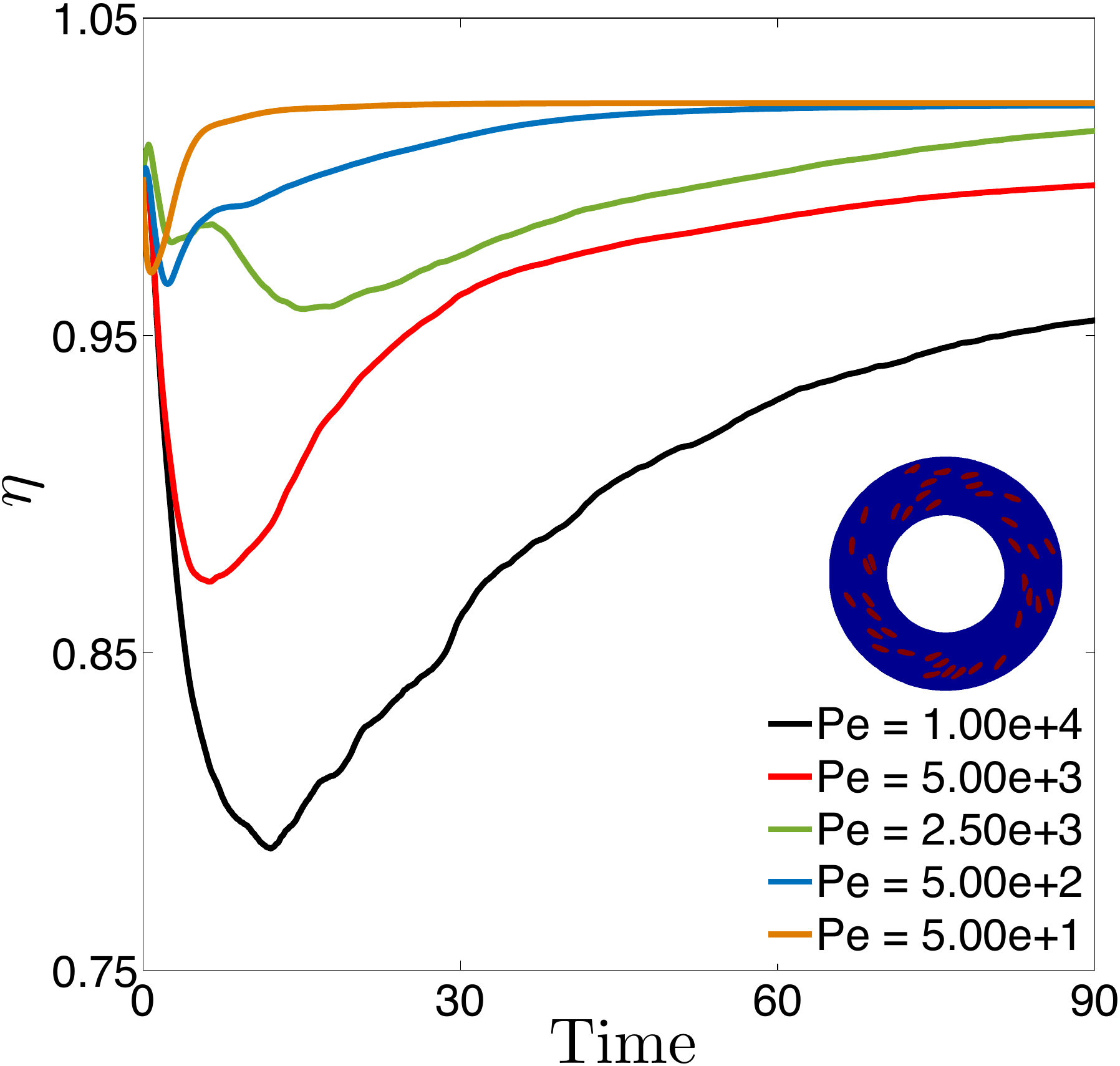}}}
      \label{f:VesVF10}}
\setcounter{subfigure}{0}
\renewcommand*{\thesubfigure}{(c-2)} 
      \hspace{-0.4cm}\subfigure[AF $= 20\%$]{\scalebox{0.3}{{\includegraphics{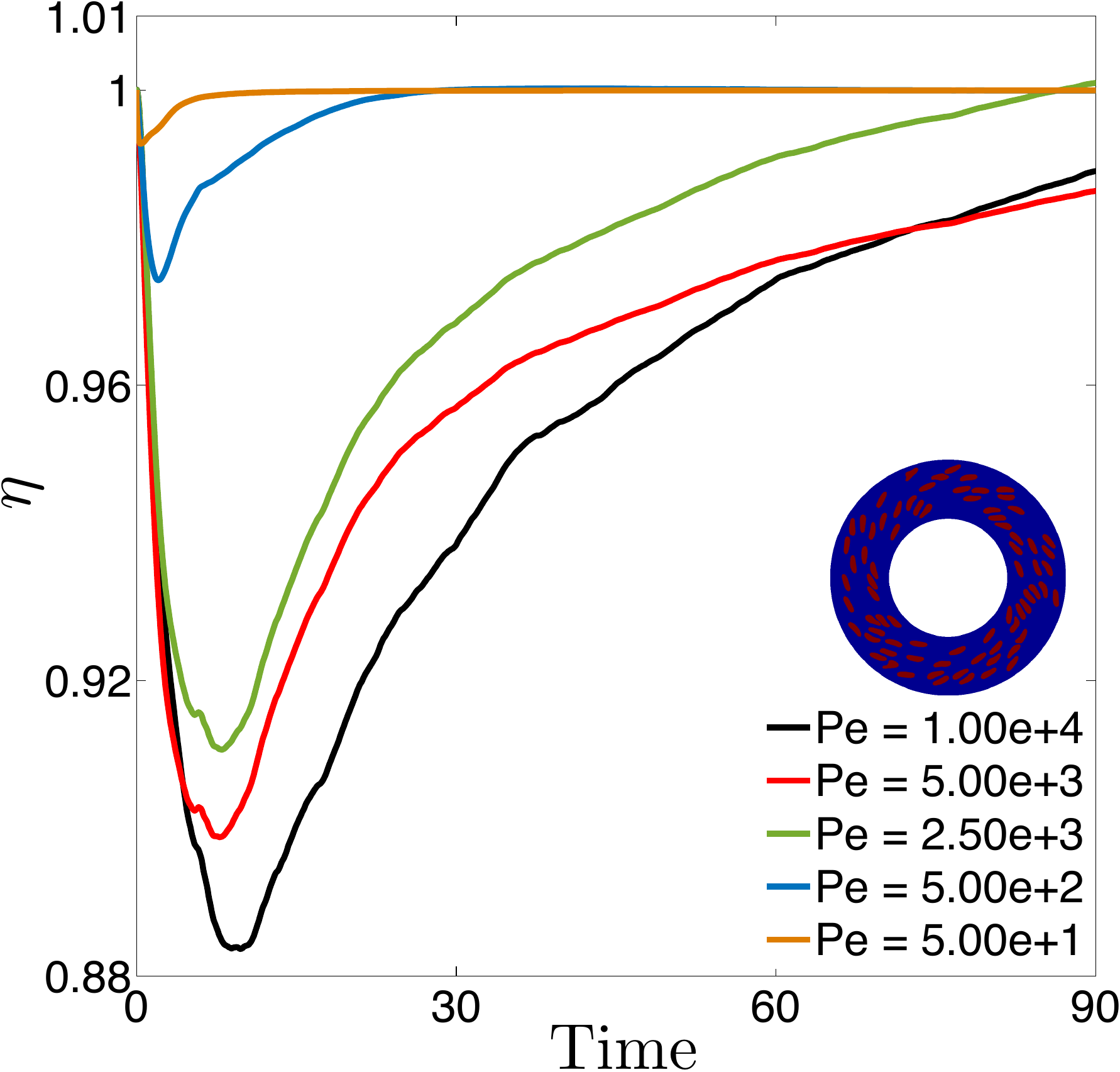}}}
      \label{f:VesVF20}}
\setcounter{subfigure}{0}
\renewcommand*{\thesubfigure}{(c-3)} 
      \hspace{-0.4cm}\subfigure[AF $= 40\%$]{\scalebox{0.3}{{\includegraphics{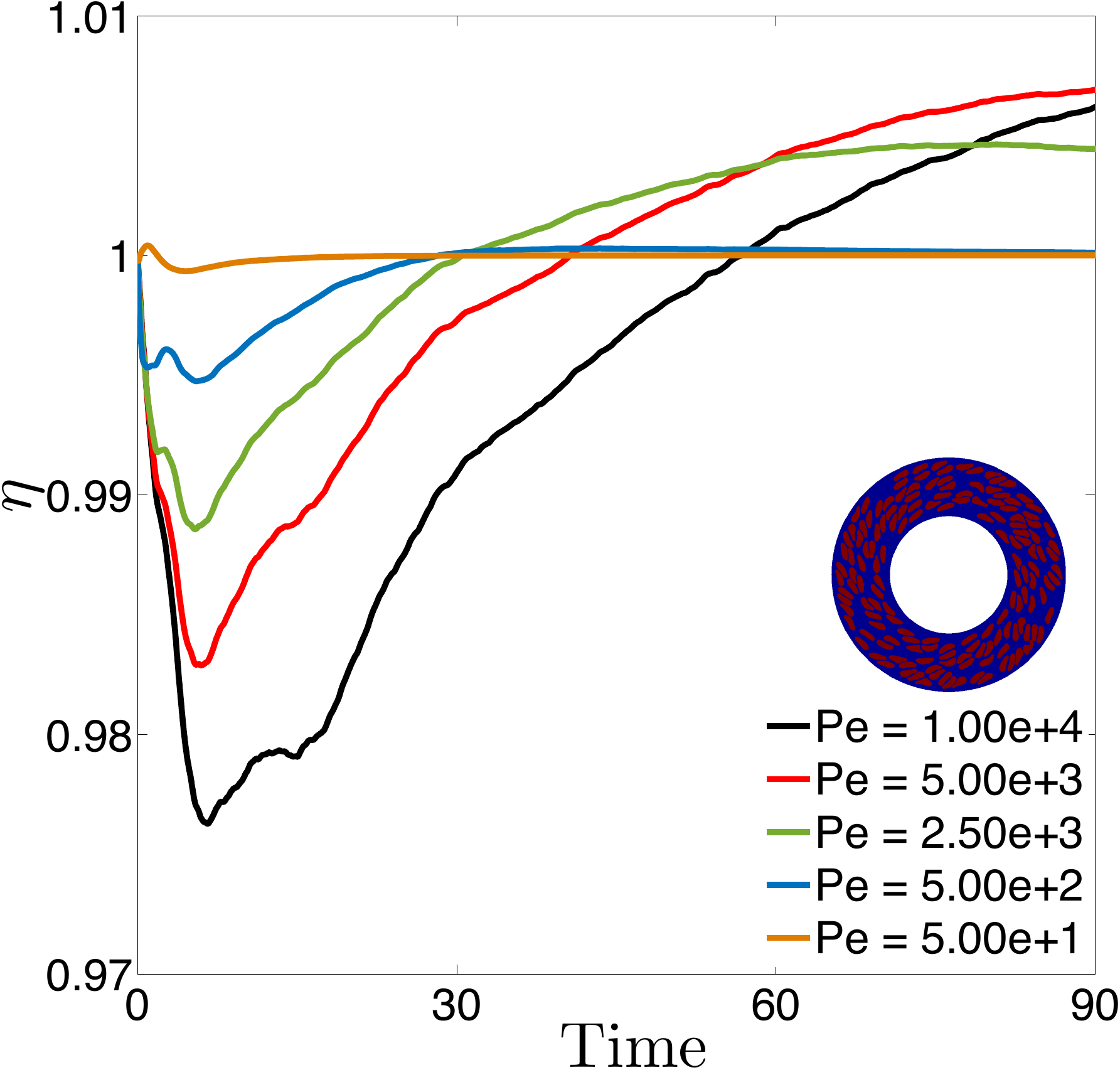}}}
      \label{f:VesVF40}} 
    \end{minipage}
\begin{minipage}{\textwidth} 
\setcounter{subfigure}{0}
\renewcommand*{\thesubfigure}{(d-1)} 
      \hspace{0cm}\subfigure[AF $= 10\%$]{\scalebox{0.3}{{\includegraphics{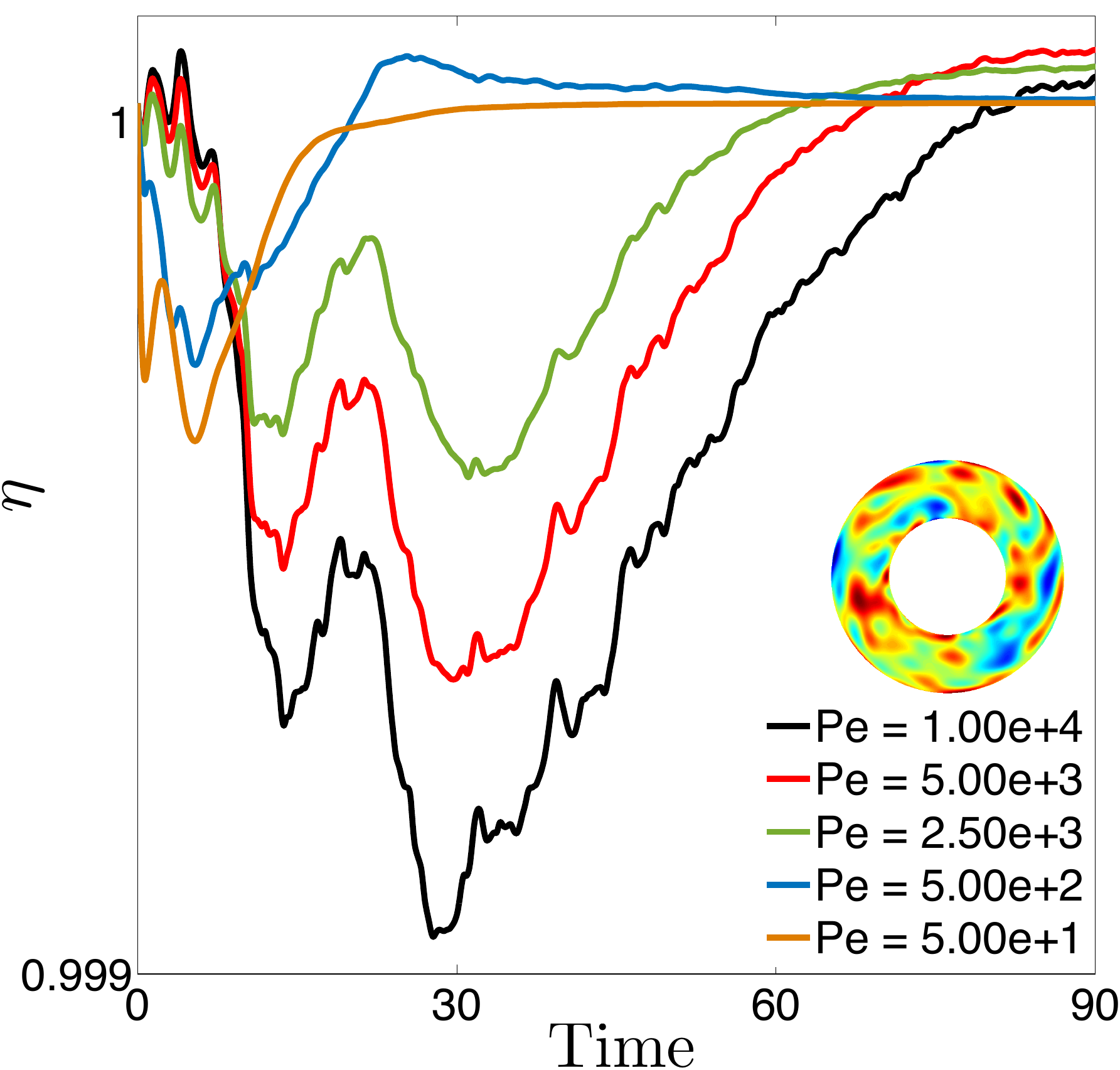}}}
      \label{f:RandVF10}}
\setcounter{subfigure}{0}
\renewcommand*{\thesubfigure}{(d-2)} 
      \hspace{-0.4cm}\subfigure[AF $= 20\%$]{\scalebox{0.3}{{\includegraphics{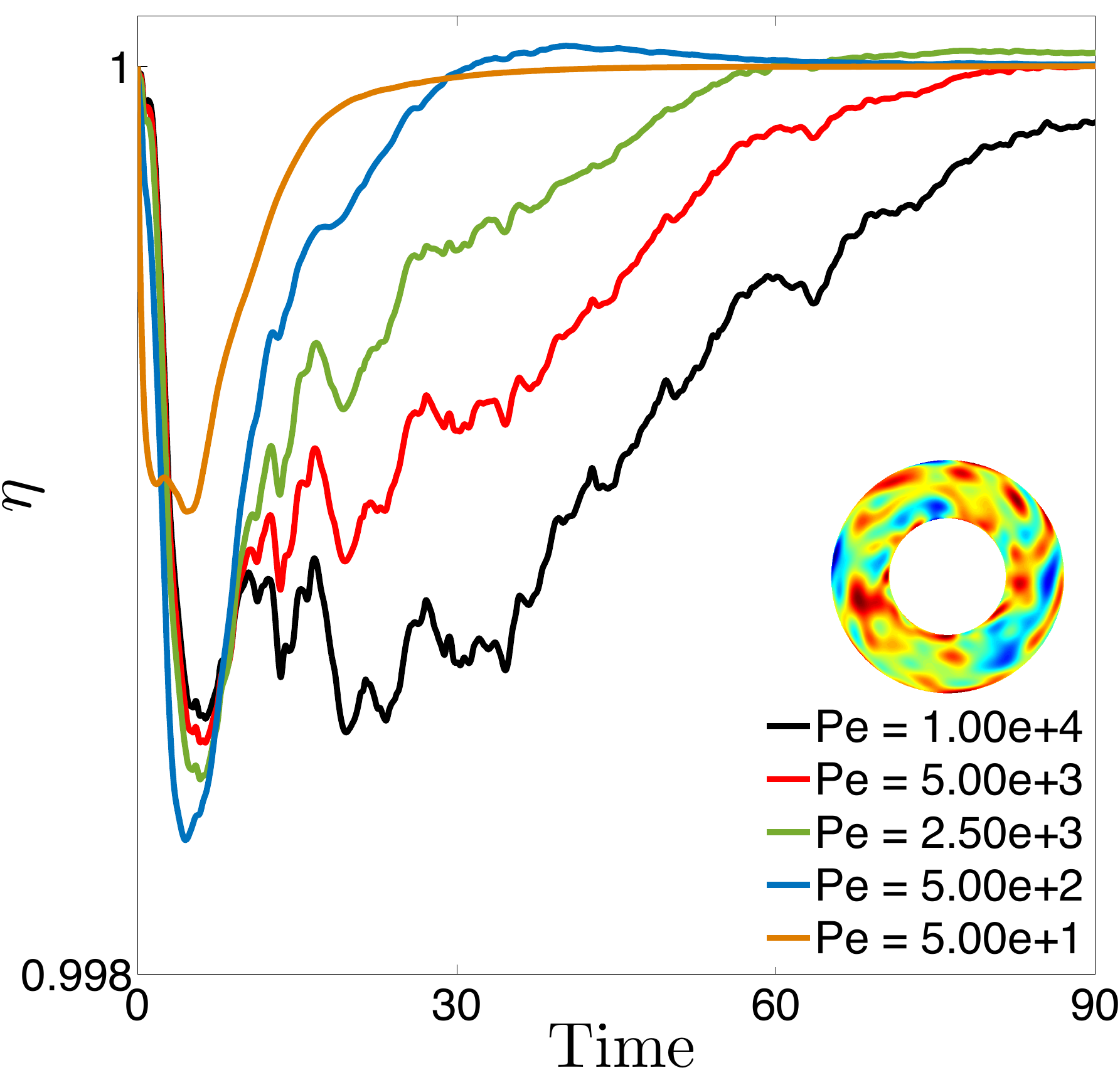}}}
      \label{f:RandVF20}}
\setcounter{subfigure}{0}
\renewcommand*{\thesubfigure}{(d-3)} 
      \hspace{-0.4cm}\subfigure[AF $= 40\%$]{\scalebox{0.3}{{\includegraphics{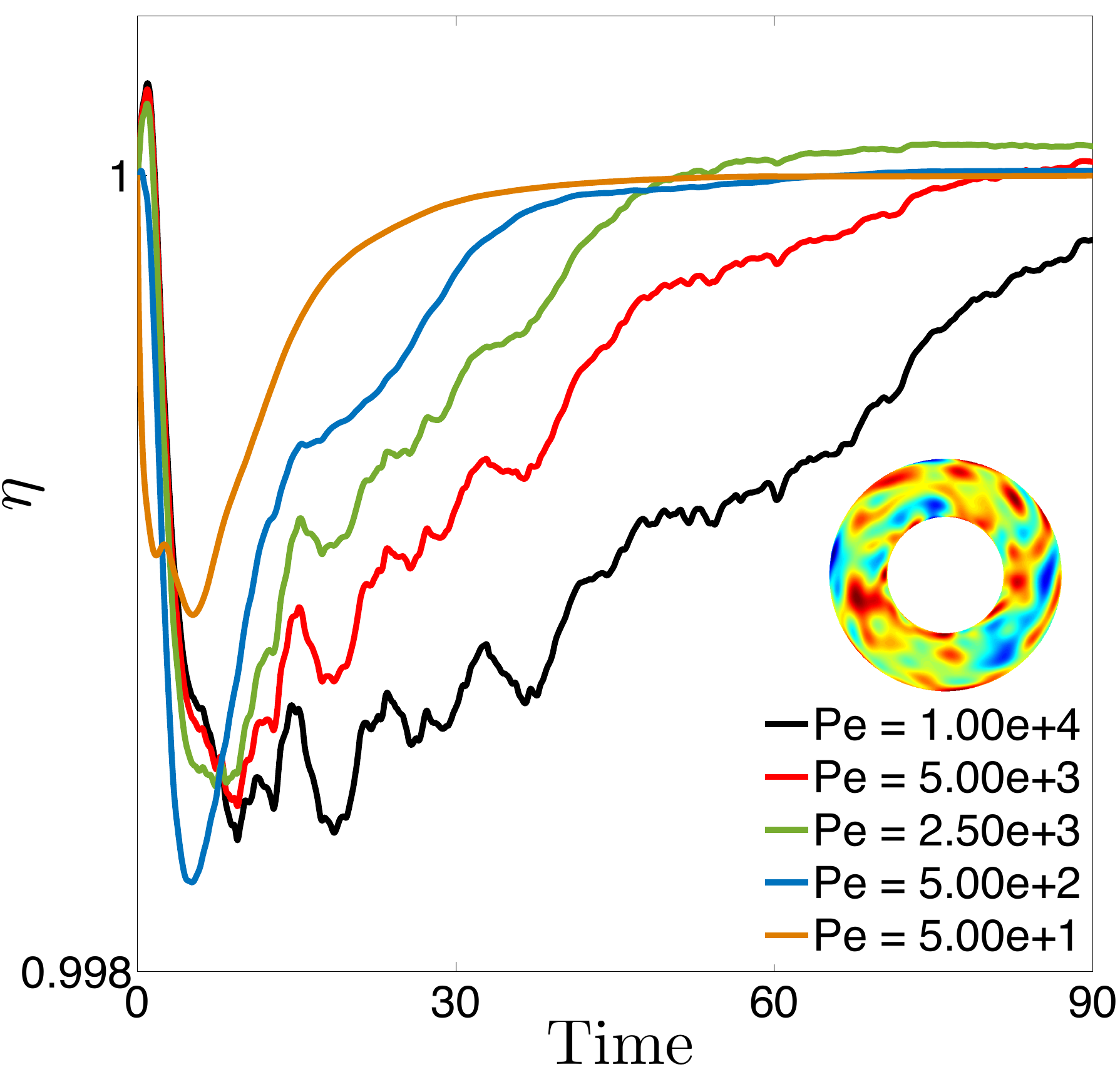}}}
      \label{f:RandVF40}} 
 \end{minipage}
\mcaption{{The} effects of the area fraction on the degree of mixing for
various Peclet numbers and the initial conditions
(\secref{s:initialCond}). The first row shows the mixing efficiency
$\eta$ with respect to time for the vesicle initial conditions for each
area fractions and the second row is for the random initial condition.
The results illustrate that there is no clear effect of the presence
mixing if the random initial condition is used.}{f:VFMixing2}
\end{figure}

Even though we observe a correlation between the maximum mixing
efficiency in time, the Peclet number, and the area fraction for some
initial conditions, this study suggests that it is not possible to
generalize this correlation for any initial condition. In
\secref{s:sumOfExperiments} we have a simple analysis on why this
happens, but as other researchers \cite{Khakhar-Ottino86, Ottino90,
Mathew-Petzold-e05, Fourer-Schmid-e14} have observed, mixing can be
quite difficult to characterize. In addition, the $H^{-1}$ norm, the
norm that we use, depends on the initial condition.

\subsection{Effects of viscosity contrast}\label{s:viscCont}

Vesicles manifest different dynamics under simple shear flow: either a
tank-treading rotation or a tumbling motion. An increase in the
viscosity contrast leads to  a transition from a tank-treading to a
tumbling motion~\cite{Keller-Skalak82}. In order to identify the
effects of the viscosity contrast on the degree of mixing, we study
mixing in vesicle suspensions with the area fraction of $5 \%$ and the
viscosity contrasts of 1, 5 and 8.  We run the simulations only for the
layer initial condition and demonstrate the mixing efficiency $\eta$
with respect to time in \figref{f:VCMixing}.

\begin{figure}[H]
 \begin{minipage}{\textwidth}
\setcounter{subfigure}{0}
\renewcommand*{\thesubfigure}{(a-1)} 
      \hspace{0cm}\subfigure[VC $= 1$]{\scalebox{0.3}{{\includegraphics{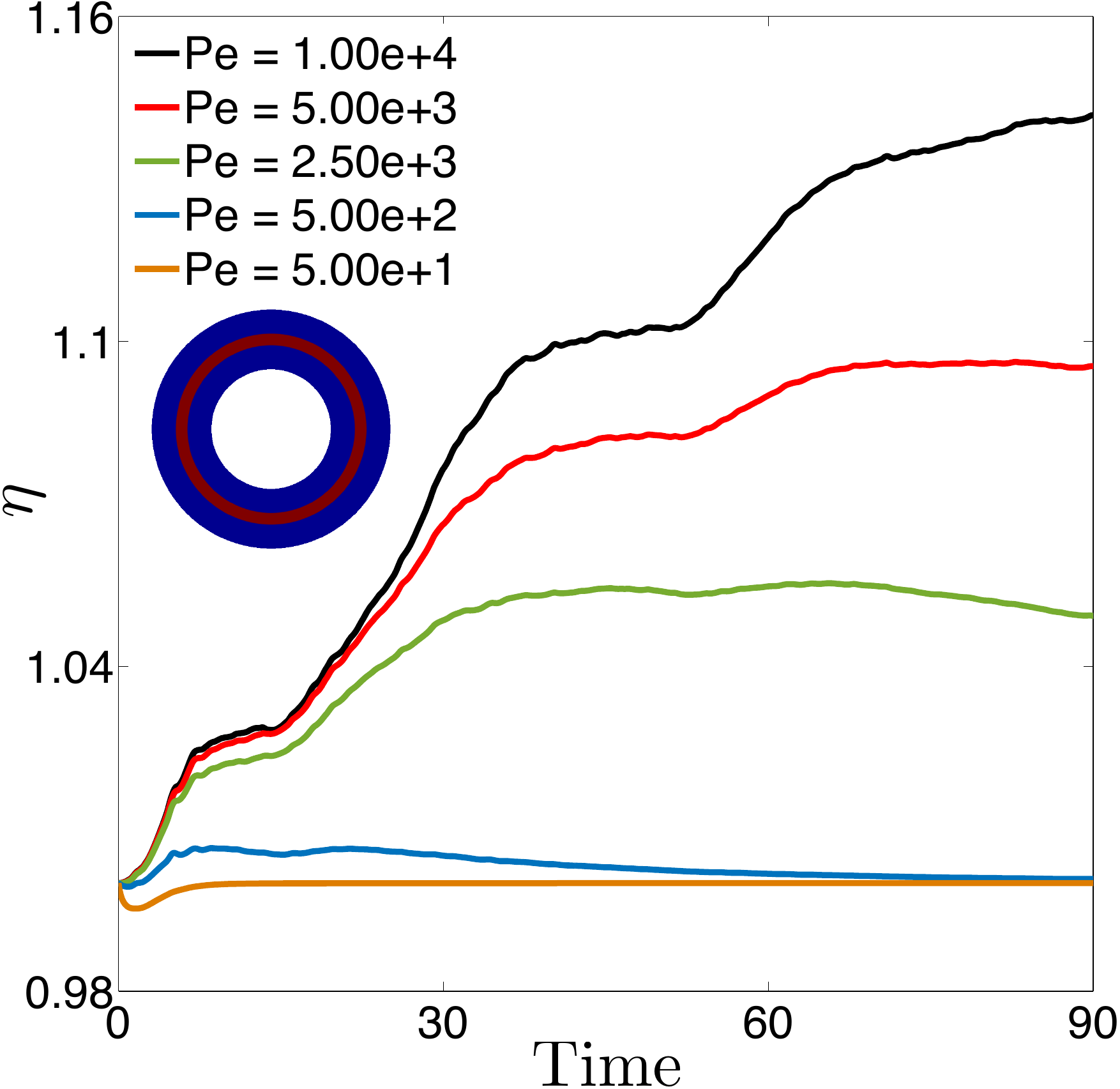}}}	
      \label{f:LayerVC1}}
\setcounter{subfigure}{0}
\renewcommand*{\thesubfigure}{(a-2)} 
      \hspace{-0.3cm}\subfigure[VC $= 5$]{\scalebox{0.3}{{\includegraphics{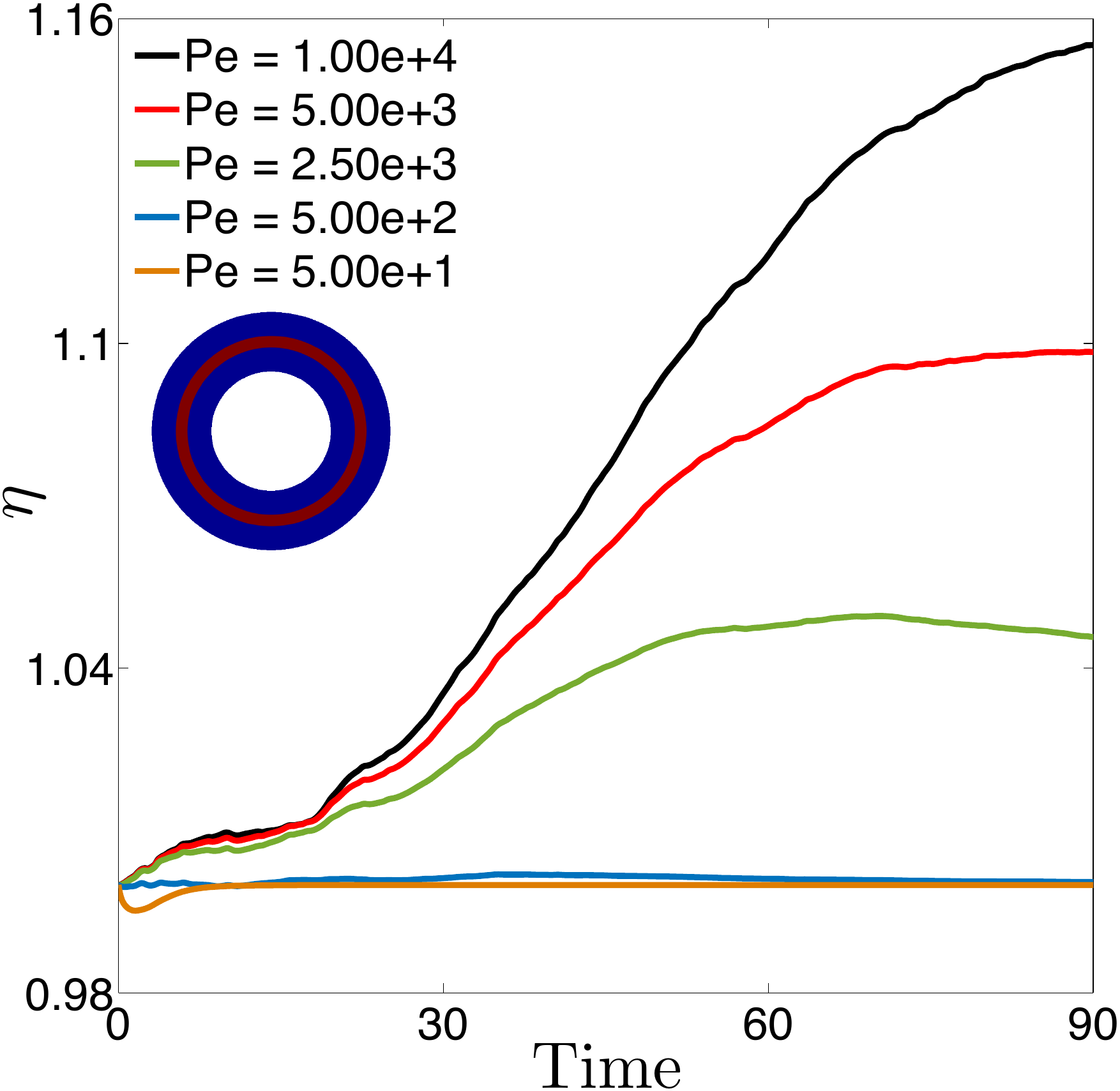}}}
      \label{f:LayerVC5}}
\setcounter{subfigure}{0}
\renewcommand*{\thesubfigure}{(a-3)} 
      \hspace{-0.3cm}\subfigure[VC $= 8$]{\scalebox{0.3}{{\includegraphics{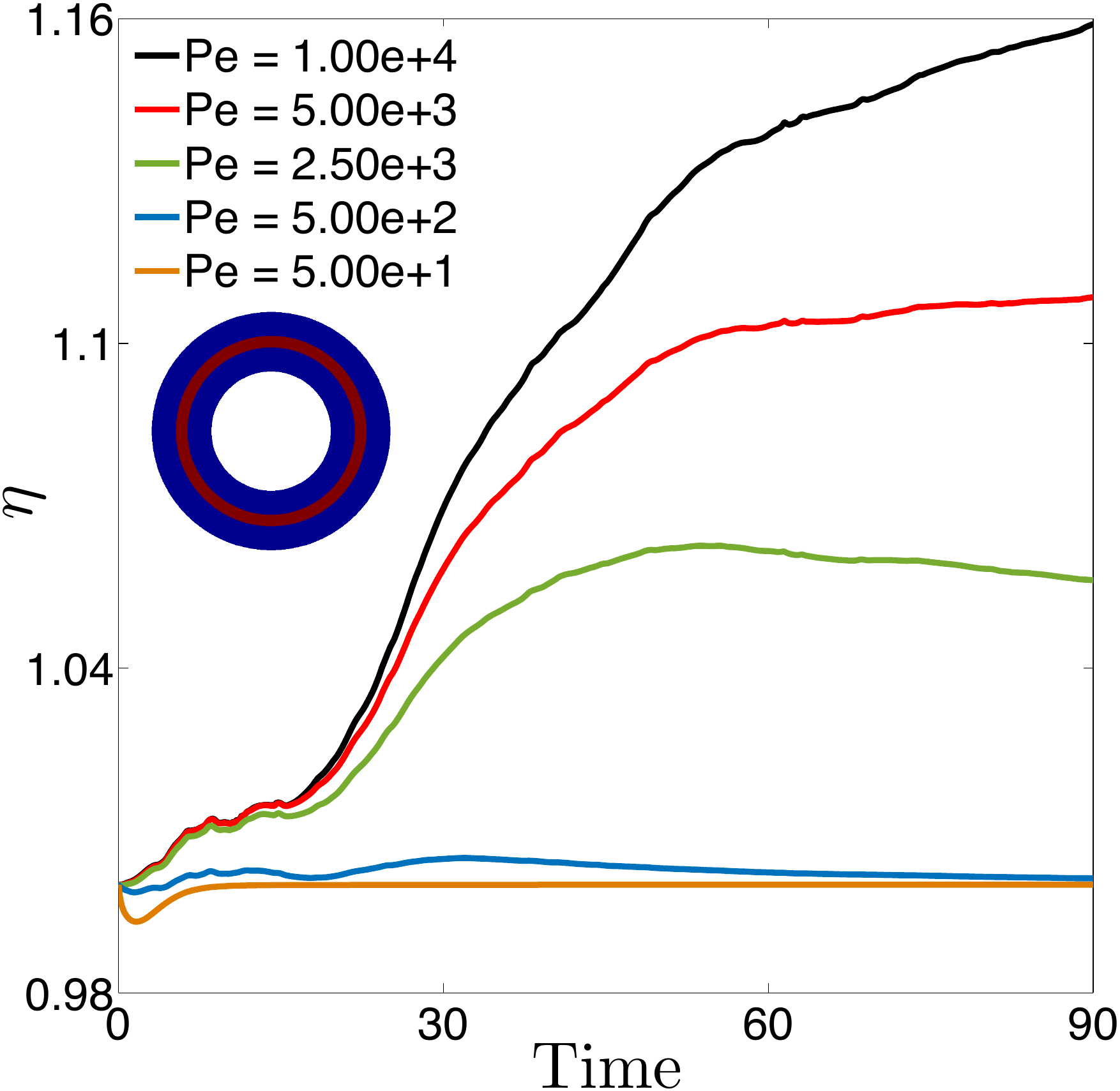}}}
      \label{f:LayerVC8}}        
\mcaption{{The} effects of the viscosity contrast on mixing.  We use the
layer initial condition and the area fraction of $5\%$.  The results
show that the viscosity contrast has less of an effect on the mixing
efficiency than the area fraction.  We do observe an increase in the
maximum mixing efficiency $\eta$ as the viscosity contrasts increases
from 1 to 8 (see \figref{f:LayerVC1} and \figref{f:LayerVC8}), and
this might stem from the fact that vesicles start to tumble for VC $>5$
(see~\cite{Keller-Skalak82}).}{f:VCMixing}
\end{minipage}
\end{figure}

\figref{f:LayerVC1}, \figref{f:LayerVC5} and \figref{f:LayerVC8} show
that an increase in the viscosity contrast results in additional mixing
efficiency.  However, for this initial condition, the viscosity
contrast has less of an effect than the area fraction on the mixing
efficiency.

\subsection{Summary}\label{s:sumOfExperiments}
\begin{figure}[H]
 \begin{minipage}{\textwidth}
\setcounter{subfigure}{0}
\renewcommand*{\thesubfigure}{(a)} 
      \hspace{0cm}\subfigure[Mixing efficiency vs. time]{\scalebox{0.3}{{\includegraphics{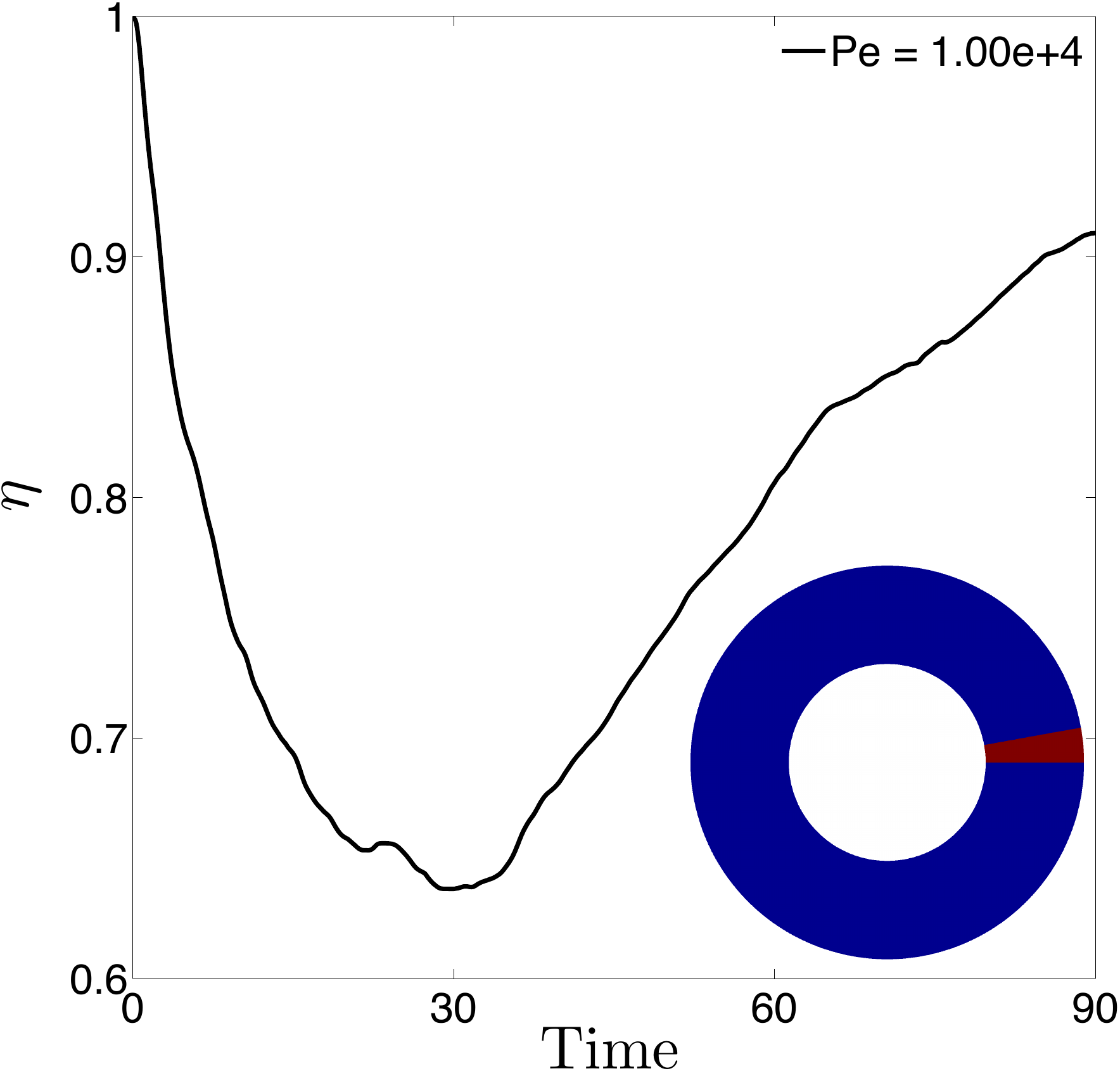}}}	
      \label{f:mixEffPatch}}
\setcounter{subfigure}{0}
\renewcommand*{\thesubfigure}{(b)} 
      \hspace{-0.4cm}\subfigure[Mixing efficiency vs. time]{\scalebox{0.3}{{\includegraphics{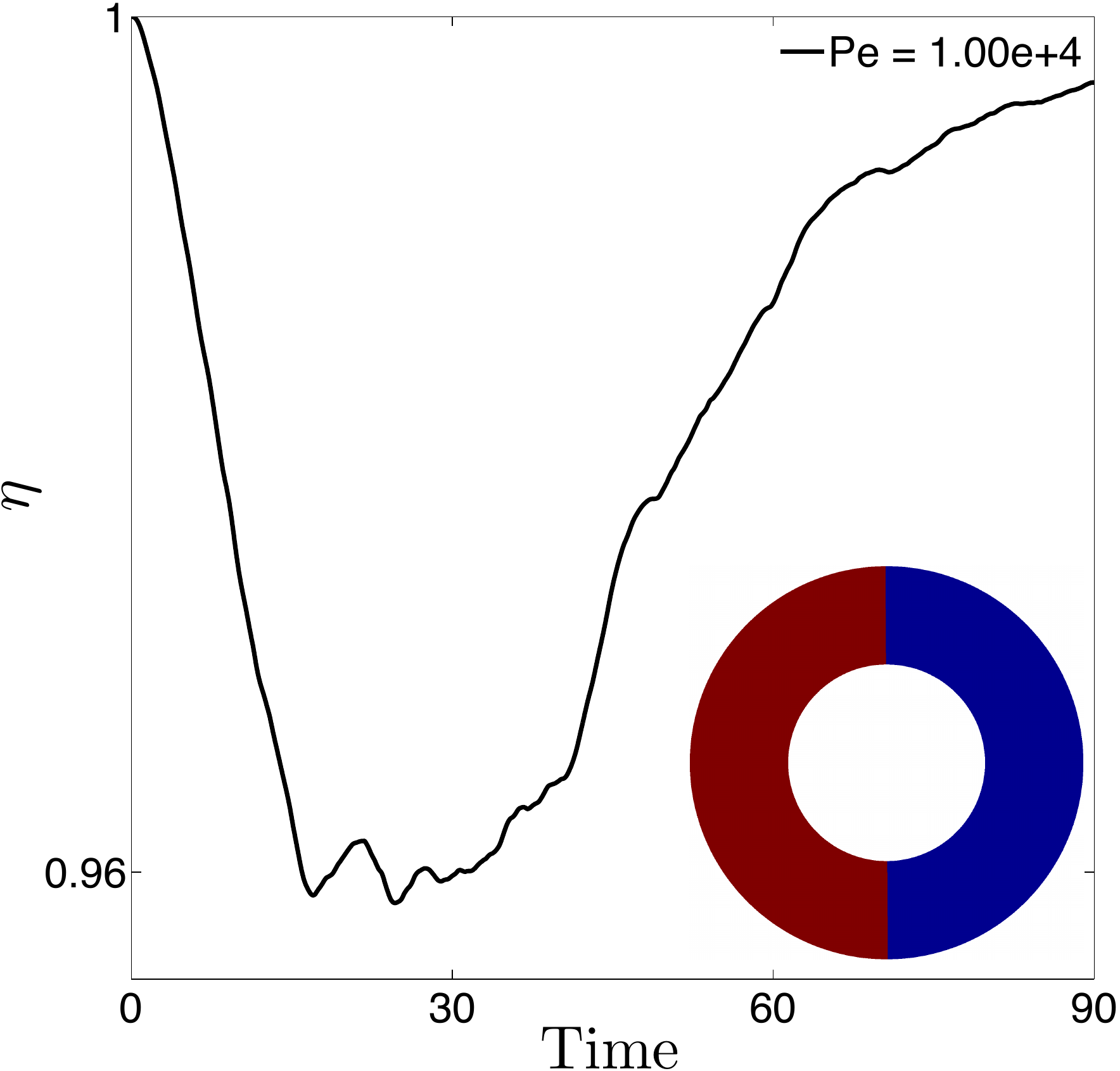}}}
      \label{f:mixEffMoon}}
\setcounter{subfigure}{0}
\renewcommand*{\thesubfigure}{(c)} 
      \hspace{-0.4cm}\subfigure[Mixing efficiency vs. time]{\scalebox{0.3}{{\includegraphics{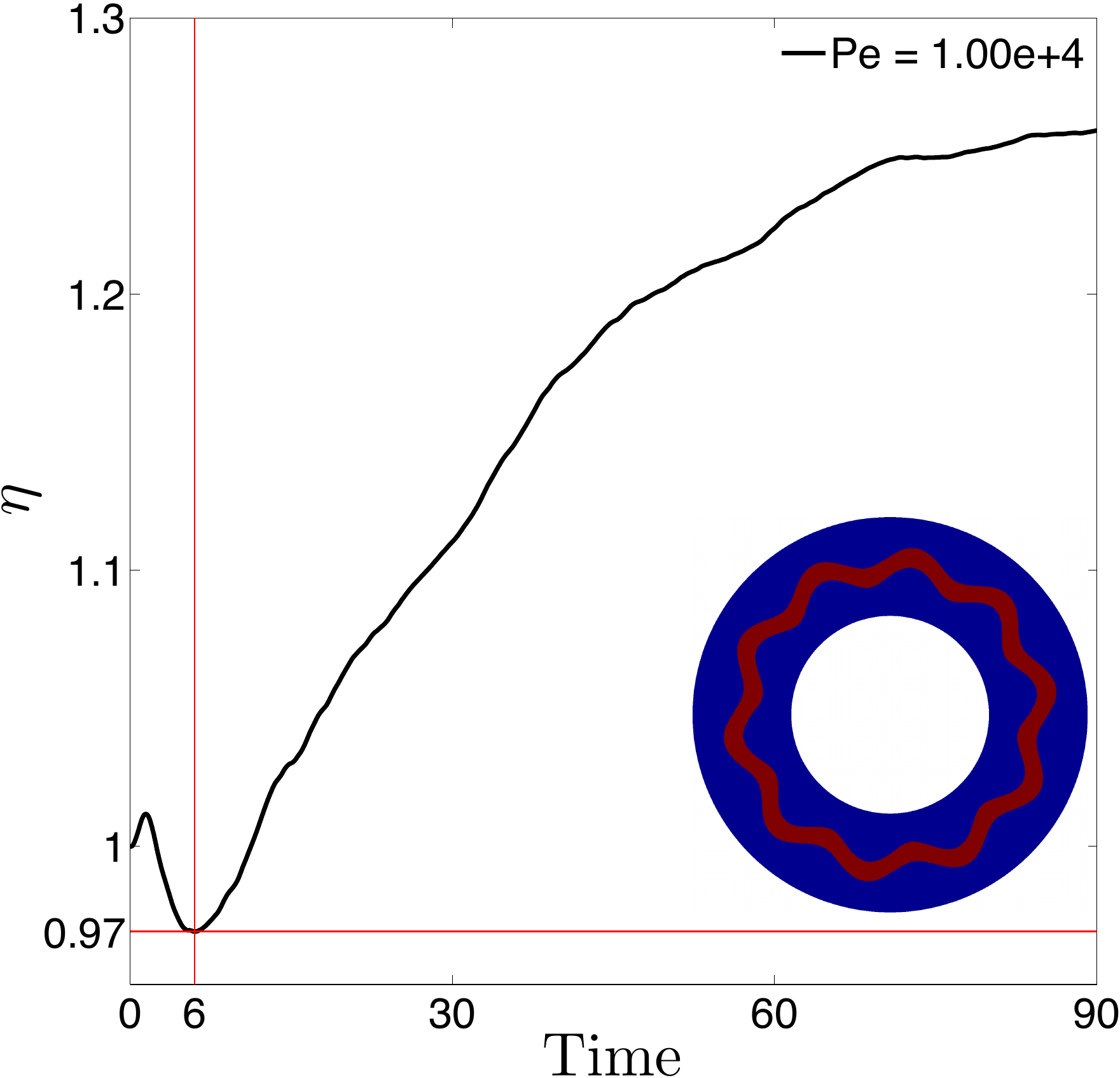}}}
      \label{f:mixEffSerp}}
  \end{minipage}    
 \mcaption{{We} show the mixing efficiency $\eta$ as a function of time delivered 
at $\pec = 1e+4$ by two initial concentrations \figref{f:mixEffPatch}, \figref{f:mixEffMoon} 
that we show in \tabref{t:m4measureForIcs}. Here, the vesicle flow is the same for all 
initial concentrations and has a area fraction AF = $40\%$ and viscosity contrast VC = $1$.
 Additionally, we perturb the initial concentration \figref{f:LayerVF40} and present the
 mixing efficiency in \figref{f:mixEffSerp}. We also show frames from the mixing simulation
 of \figref{f:mixEffSerp} in \figref{f:FramesICserp}. The proposed measure \eqref{e:m4measure}
 for this initial concentration is $M = 0.97$.}{f:newICsEff}
\end{figure}

In an attempt to predict the mixing efficiency based on the initial
concentration ($\phi_{\text{IC}}$), we introduce a measure $M$ of the ability of mixing for
the default Couette flow
\begin{equation}\label{e:m4measure}
M = \frac{\int_{\Omega} \left | \nabla \phi_{\text{IC}} \cdot \mathbf{v} \right | 
d\Omega}{\|\nabla \phi_{\text{IC}}\|_{L^2} \|\mathbf{v}\|_{L^2}}
\end{equation}
where $\nabla \phi_{\text{IC}} = (\frac{\partial \phi_{\text{IC}}}{\partial r},
\frac{1}{r}\frac{\partial \phi_{\text{IC}}}{\partial \theta})$ and $\mathbf{v} =
(v_r, v_{\theta})$ is the velocity field of the Couette flow (without
vesicles).  Equation~\eqref{e:m4measure} is a normalized $L^1$ norm of
the advective term.  We tabulate various initial conditions, the
corresponding $M$ values, and the minimum and the maximum efficiencies
${\eta}_{\min}$, ${\eta}_{\max}$ in \tabref{t:m4measureForIcs}. For all
the initial concentrations that we consider except {\em LAYER}, $M$ is
initially non-zero meaning that mixing will occur due to advection. For
these initial concentrations, the vesicle flow suppresses mixing by
creating trapped regions. For the {\em LAYER} initial concentration, the
advective term is initially zero and hence mixing occurs only due to
diffusion. The vesicle flow provides better stirring and hence better
mixing of this initial concentration than the default Couette flow. In
order to verify this observation, we consider a slightly perturbed
initial concentration in \figref{f:mixEffSerp}, which has a
concentration gradient such that the advective term is initially
non-zero. The measure \eqref{e:m4measure} for this initial concentration
is $M = 0.97$. We simulate mixing of this initial concentration with the
vesicle flow at $\text{AF} = 40\%$ and $\text{VC} = 1$. The mixing efficiency $\eta$
is shown in \figref{f:mixEffSerp}. For $t \in [0,6]$, the vesicle flow
suppresses mixing, but for $t \in [6,T_{h}]$ the vesicle flow promotes
mixing.  To explain this behavior we show frames from the mixing
simulations with the vesicle flow and the Couette flow in
\figref{f:FramesICserp}. Before $t = 6$, mixing occurs due to advection
in the Couette flow; however, after $t = 6$, the concentration field
approaches the {\em LAYER} intial condition whose gradient only depends
on $r$.  Therefore, the advective term approaches zero, and mixing is
dominated by diffusion in the Couette flow.  Consequently, the Couette
flow provides better mixing than the vesicle flow does as long as
advective mixing occurs with the Couette flow since the vesicle flow has
trapped regions.

\begin{table}[H]
\mcaption{{We} report the proposed measure \eqref{e:m4measure} for various
initial concentrations $\phi_{\text{IC}}$ together with the minimum and maximum
mixing efficiencies ${\eta}_{\min}$, ${\eta}_{\max}$ they deliver with
the vesicle flow of area fraction $\text{AF} = 40\%$ and viscosity contrast
$\text{VC} = 1$ at $\pec = 1e+4$. Here, red is for $\phi_{\text{IC}} = 1$ and blue is for
$\phi_{\text{IC}} = 0$.}{t:m4measureForIcs}
  \centering
  \begin{tabular}{  c  c  c  c | c  c  c  c}
    \hline
    $\phi_{\text{IC}}$ & $M$ & ${\eta}_{\min}$ & ${\eta}_{\max}$ &
$\phi_{\text{IC}}$ & $M$ & ${\eta}_{\min}$ & ${\eta}_{\max}$ \\ \hline
    \begin{minipage}{.2\textwidth}
      \centering
      \includegraphics[scale = 0.2]{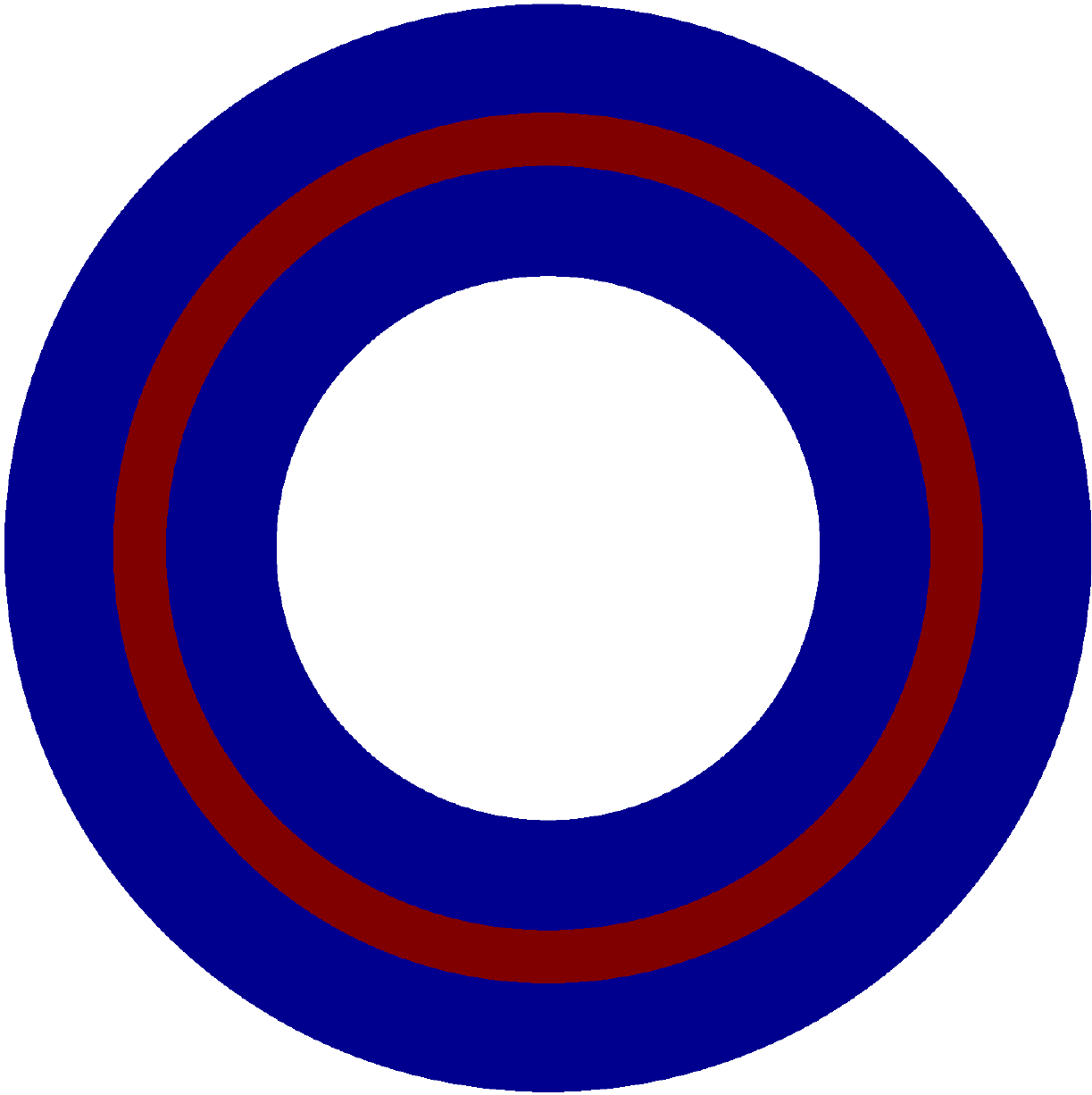}
    \end{minipage}
    &
      0
    & 
      1
     &
     1.34 
    &
    \begin{minipage}{.2\textwidth}
      \centering
      \includegraphics[scale = 0.2]{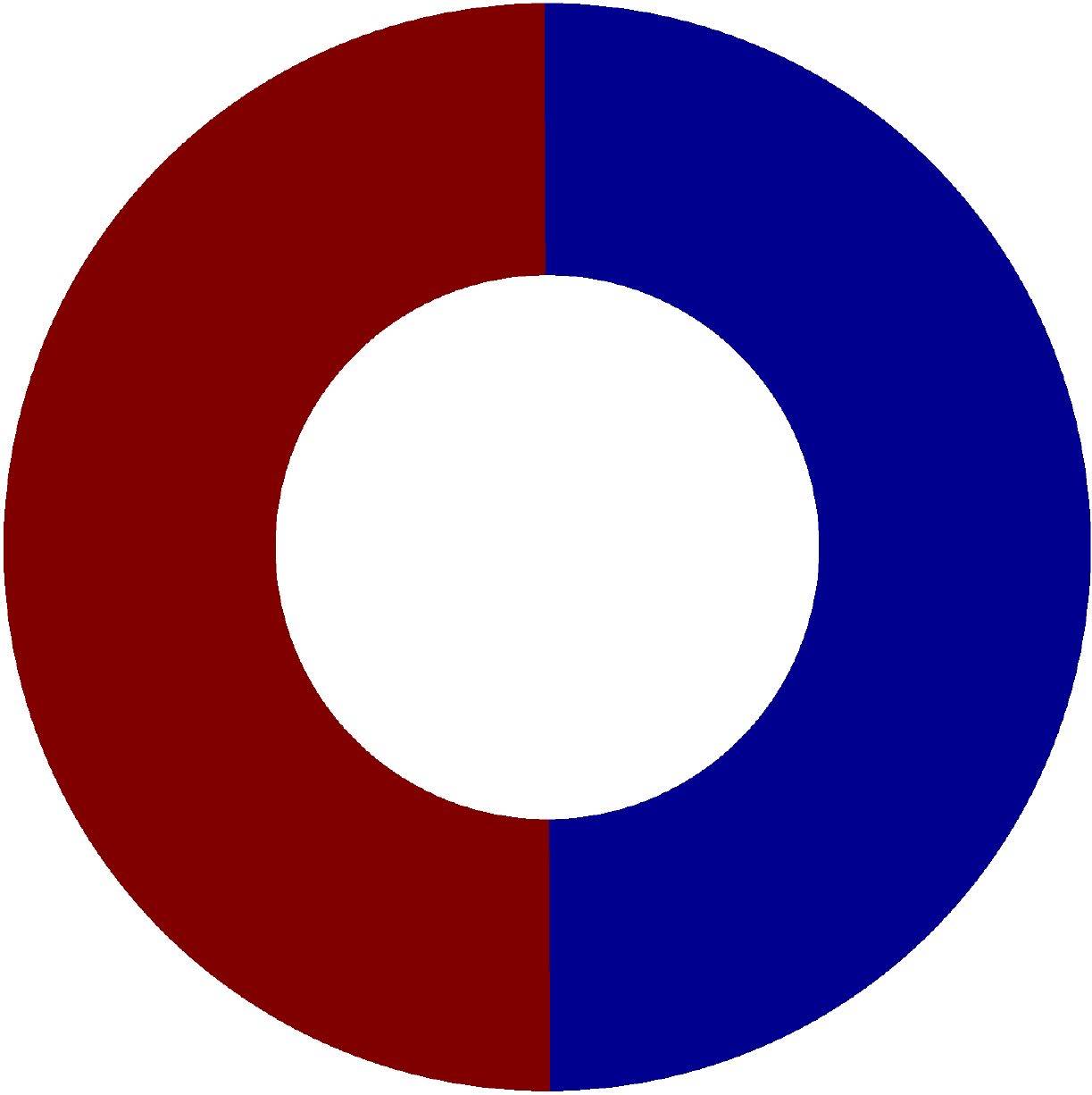}
    \end{minipage}
    &
      0.98
    & 
      0.96
     &
     1 
    \\ 
    \begin{minipage}{.2\textwidth}
      \centering
      \includegraphics[scale = 0.2]{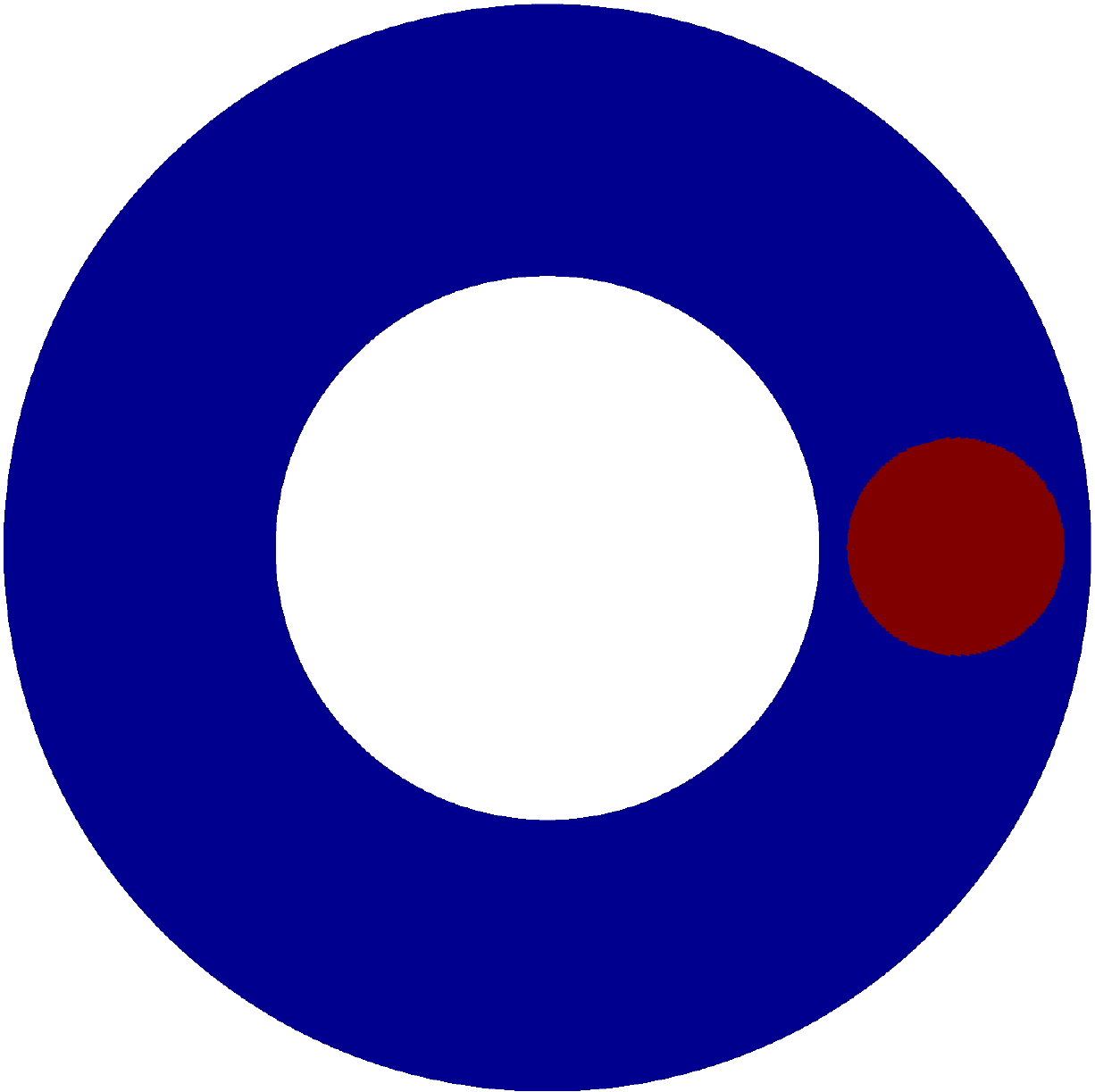}
    \end{minipage}
    &
      0.52
    & 
      0.63
     &
     1 
    &
     \begin{minipage}{.2\textwidth}
      \centering
      \includegraphics[scale = 0.2]{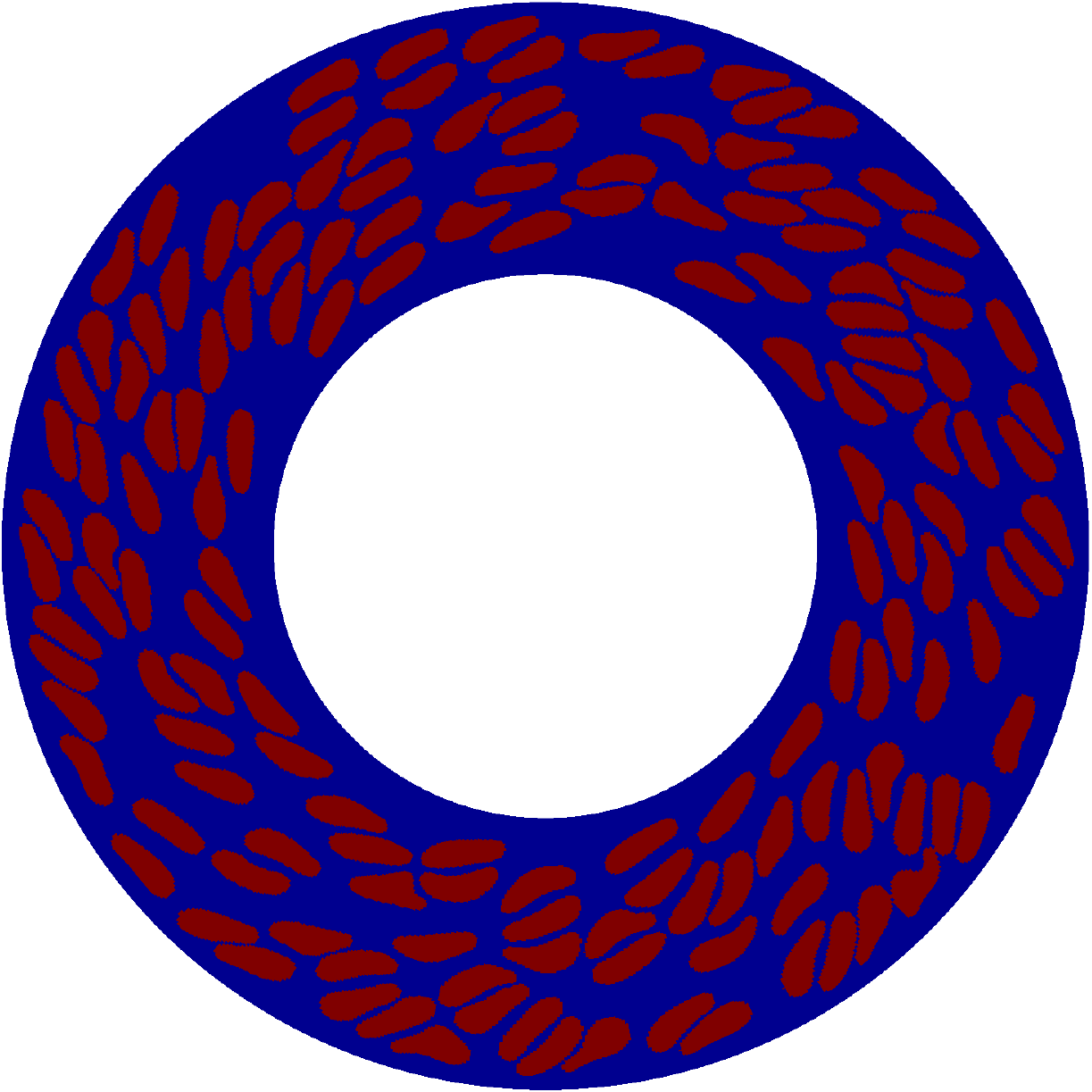}
    \end{minipage}
    &
      1.86
    & 
      0.98
     &
     1
   \\ 
    \begin{minipage}{.2\textwidth}
      \centering
      \includegraphics[scale = 0.2]{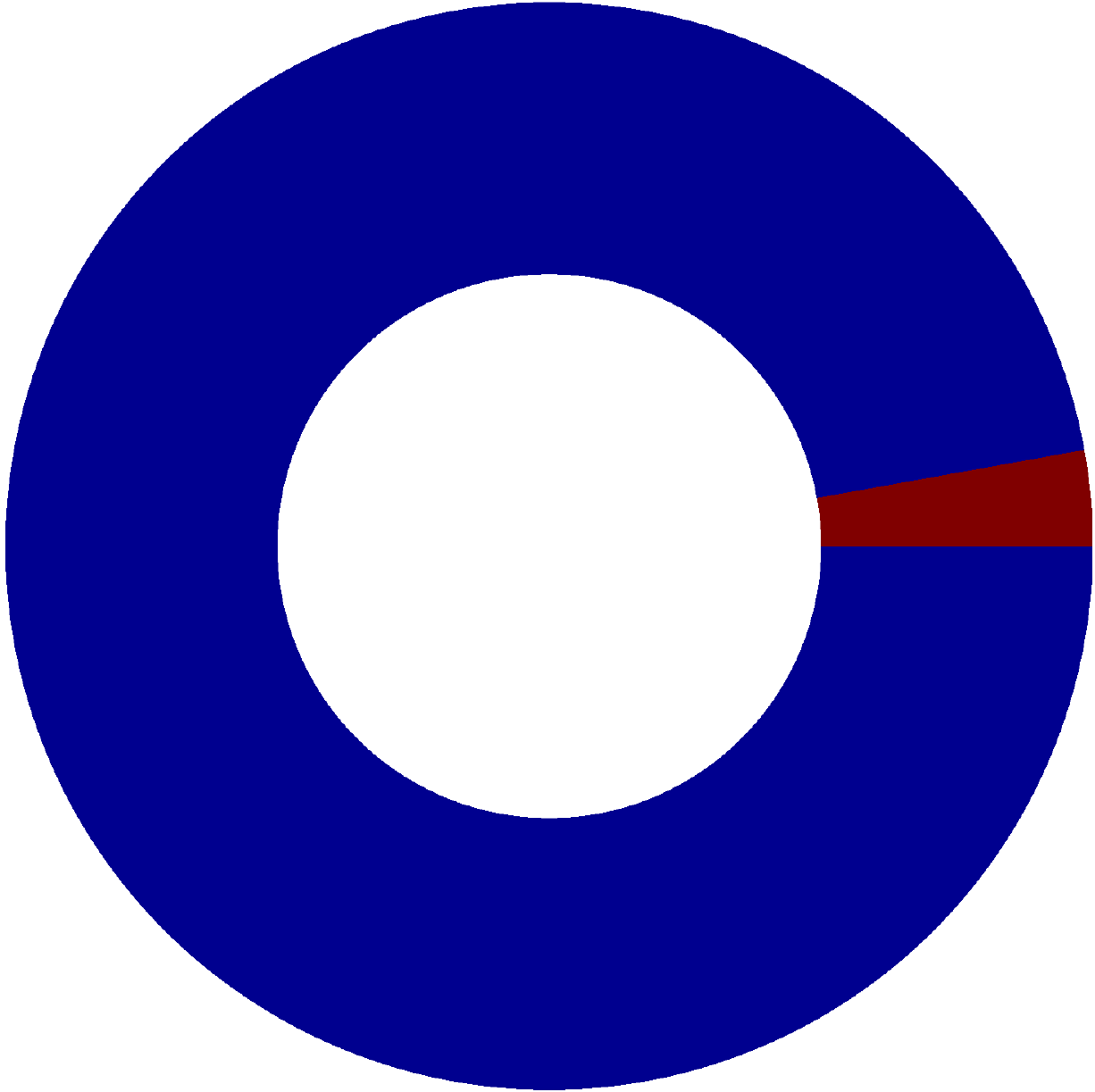}
    \end{minipage}
    &
      0.69
    & 
      0.64
     &
     1
    &
    \begin{minipage}{.2\textwidth}
      \centering
      \includegraphics[scale = 0.2]{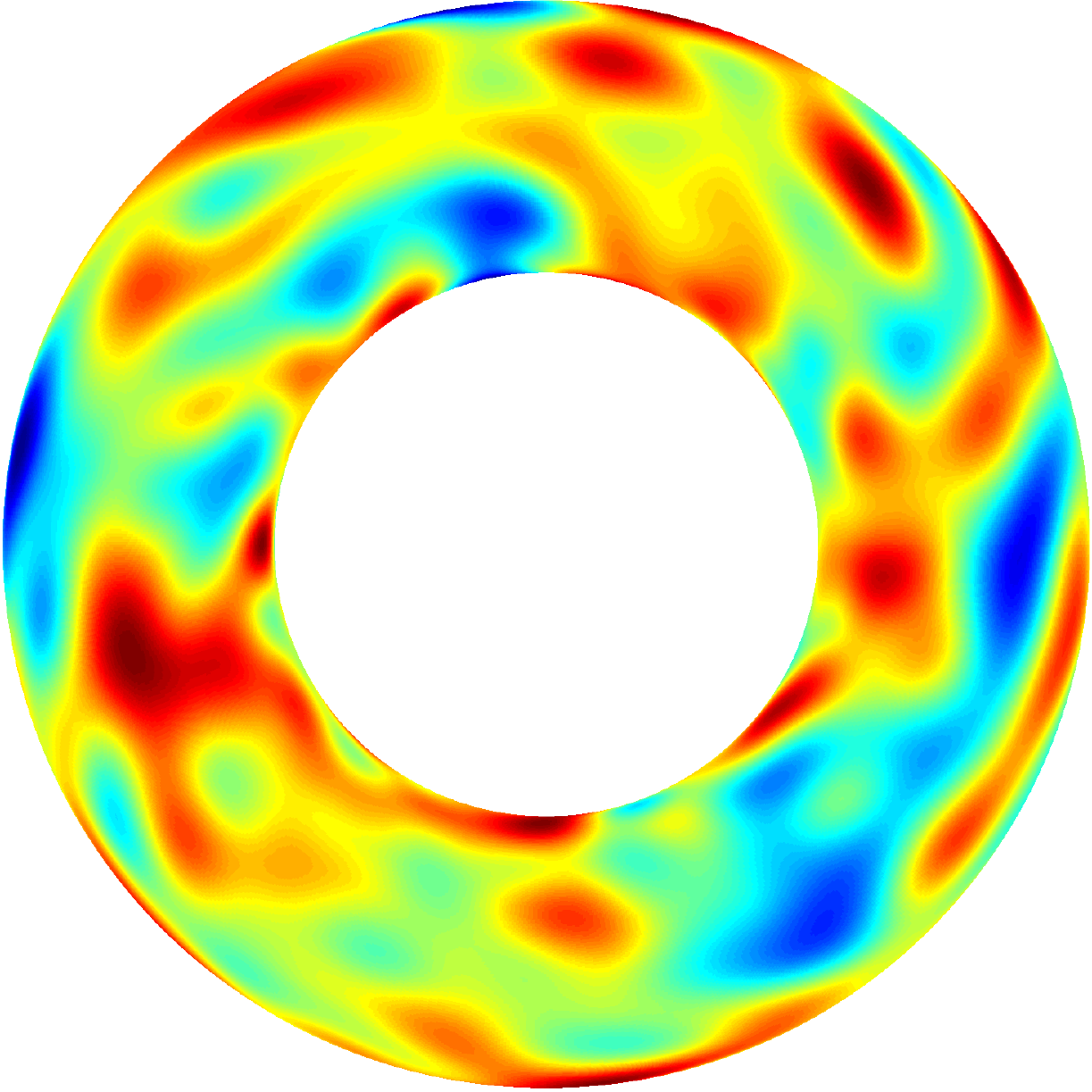}
    \end{minipage}
    &
      2.29
    & 
      0.998
     &
     1 
    \\ \hline
  \end{tabular}
\end{table}

\begin{figure}[H]
 \begin{minipage}{\textwidth}
\setcounter{subfigure}{0}
\renewcommand*{\thesubfigure}{(a-1)} 
      \hspace{0cm}\subfigure[$t = 3$ (w/ the vesicle flow)]{\scalebox{0.52}{{\includegraphics{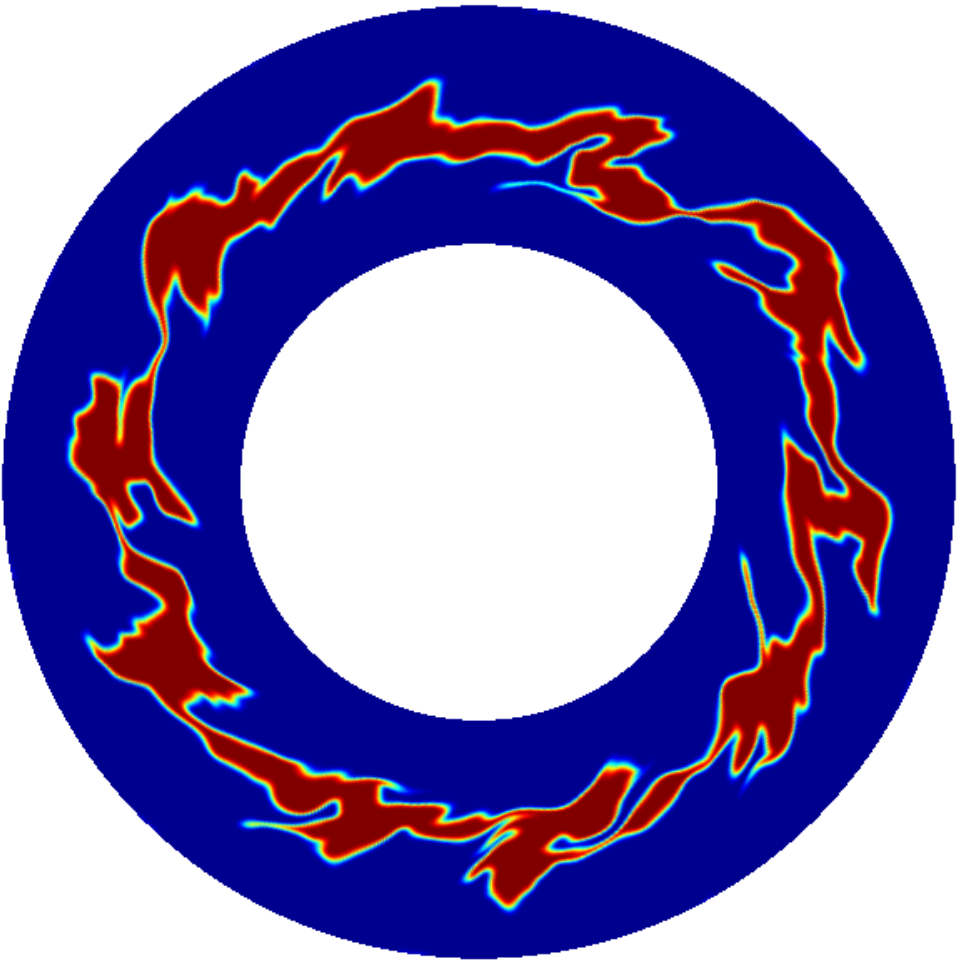}}}	
      \label{f:ConSerp1Ves}}
\setcounter{subfigure}{0}
\renewcommand*{\thesubfigure}{(a-2)} 
      \hspace{0cm}\subfigure[$t = 6$ (w/ the vesicle flow)]{\scalebox{0.52}{{\includegraphics{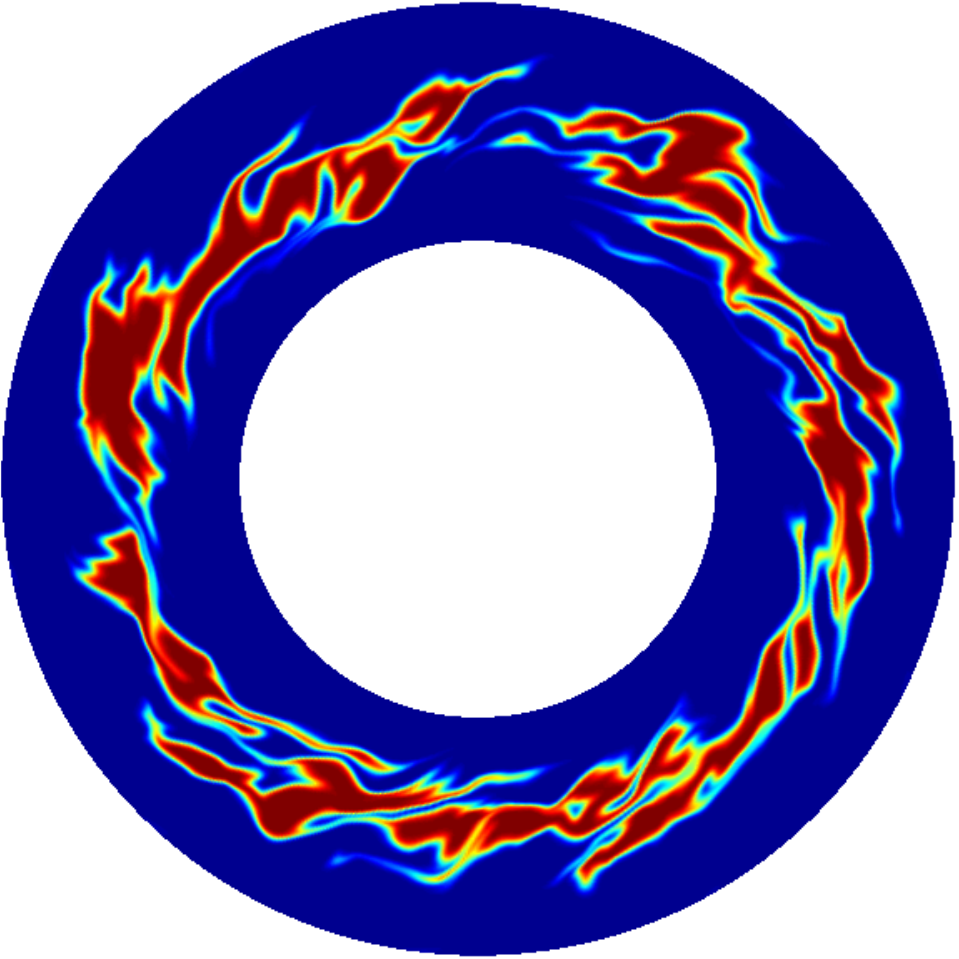}}}
      \label{f:ConSerp2Ves}}
\setcounter{subfigure}{0}
\renewcommand*{\thesubfigure}{(a-3)} 
      \hspace{0cm}\subfigure[$t = 24$ (w/ the vesicle flow)]{\scalebox{0.52}{{\includegraphics{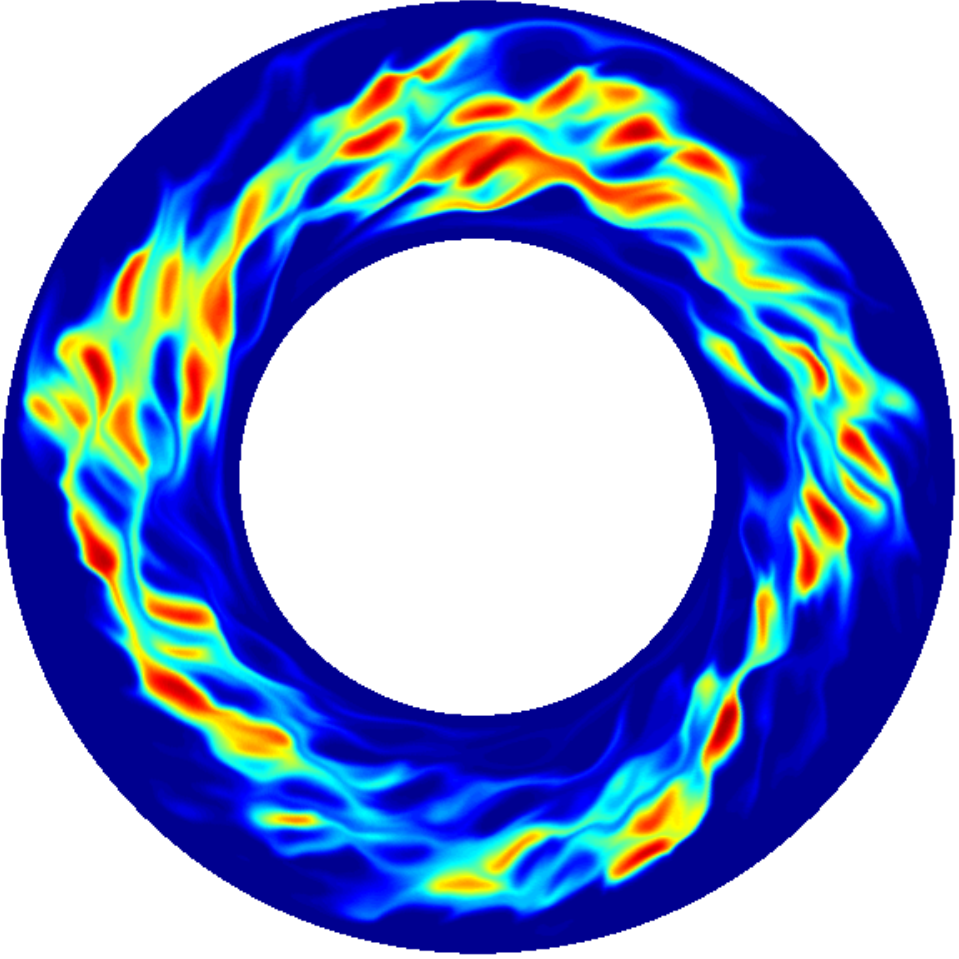}}}
      \label{f:ConSerp3Ves}}
  \end{minipage}    
  \begin{minipage}{\textwidth}    
\setcounter{subfigure}{0}
\renewcommand*{\thesubfigure}{(b-1)} 
      \hspace{0cm}\subfigure[$t = 3$ (w/ the Couette flow)]{\scalebox{0.52}{{\includegraphics{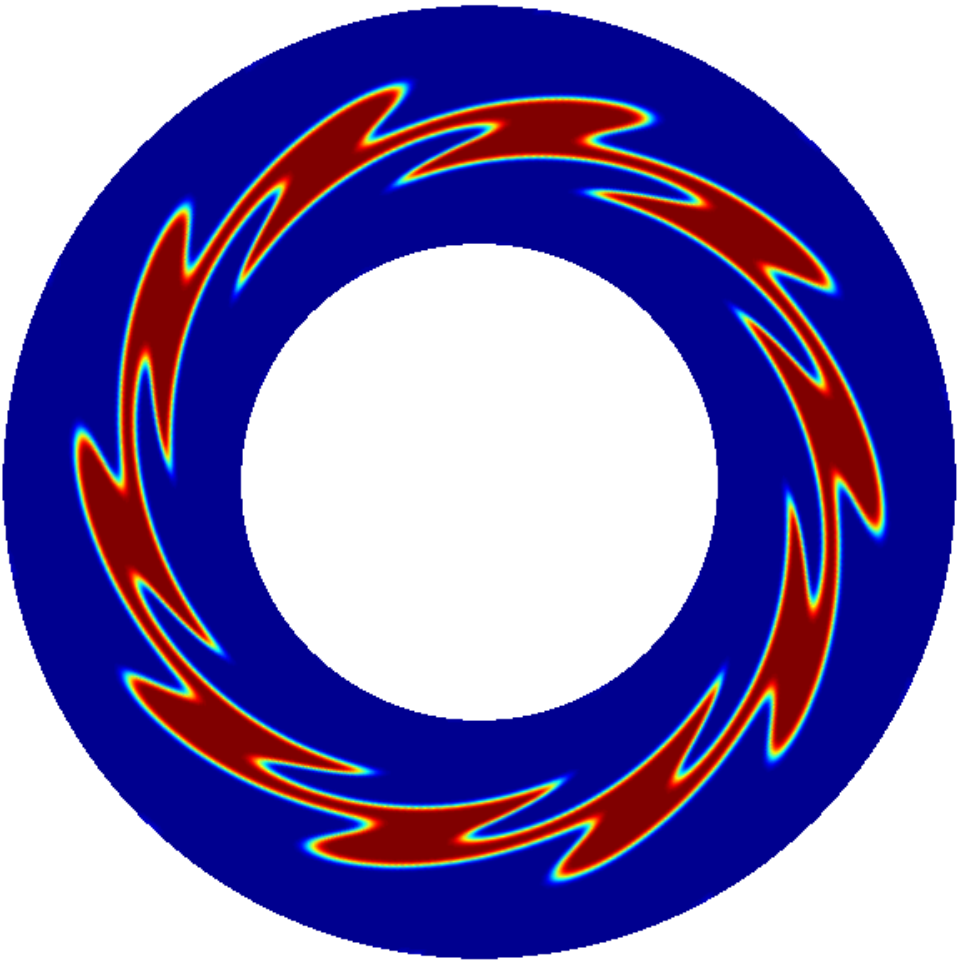}}}
      \label{f:ConSerp1Ant}}
\setcounter{subfigure}{0}
\renewcommand*{\thesubfigure}{(b-2)} 
      \hspace{0cm}\subfigure[$t=6$ (w/ the Couette flow)]{\scalebox{0.52}{{\includegraphics{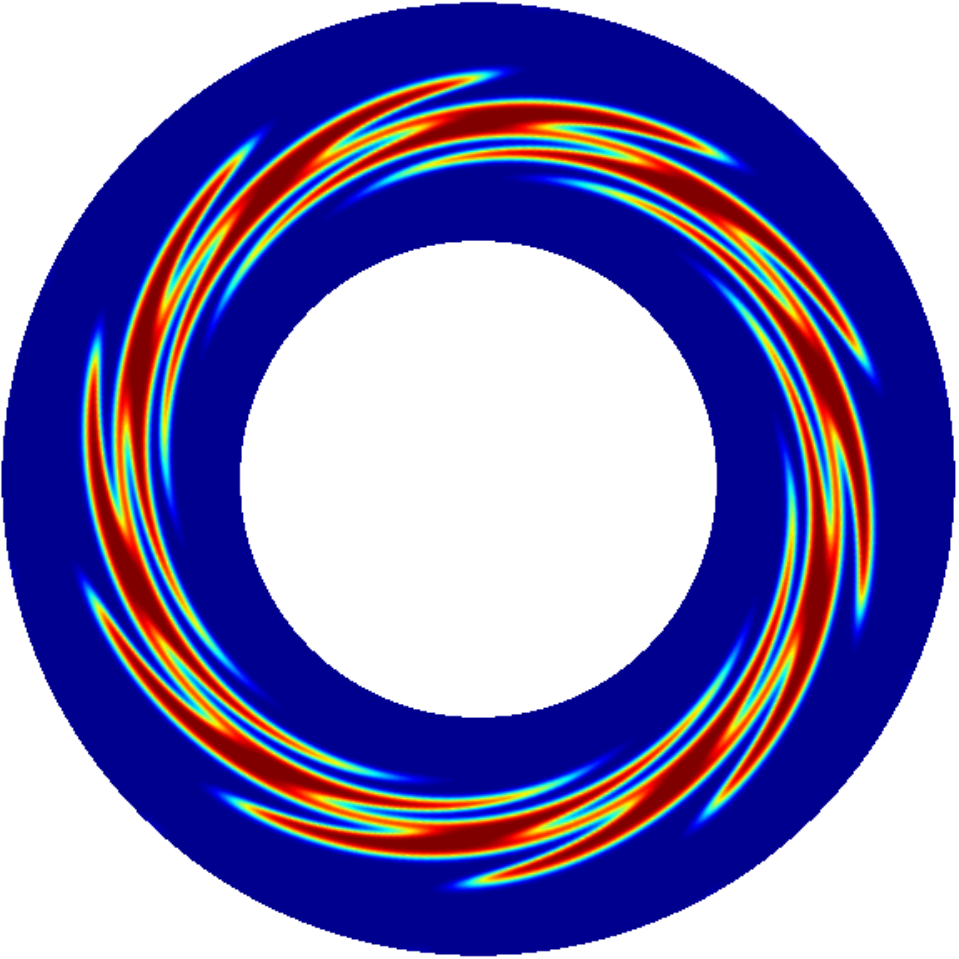}}}
      \label{f:ConSerp2Ant}}
\setcounter{subfigure}{0}
\renewcommand*{\thesubfigure}{(b-3)} 
      \hspace{0cm}\subfigure[$t = 24$ (w/ the Couette flow)]{\scalebox{0.52}{{\includegraphics{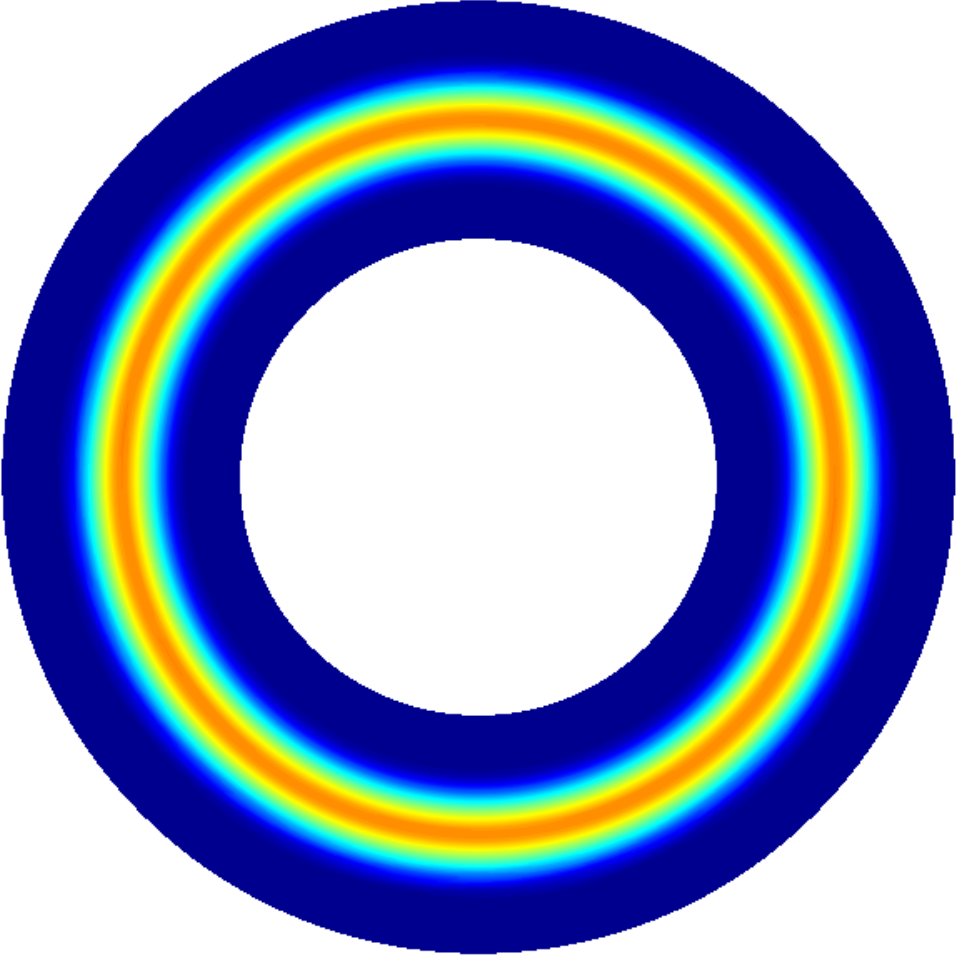}}}
      \label{f:ConSerp3Ant}} 
    \end{minipage}      
\mcaption{{We} present frames from the mixing simulation of the initial
concentration shown in \figref{f:mixEffSerp}. The frames at the first
row are from the simulation with the vesicle flow of area fraction AF =
$40\%$ and viscosity contrast VC = $1$. The ones at the second row are
from the simulation with the Couette flow. Here, both simulations have
the same Peclet number $\pec = 1e+4$. The corresponding mixing
efficiency is in \figref{f:mixEffSerp}. In particular, the mixing
efficiencies at the instances we show here are $\eta(t=3) = 0.99$,
$\eta(t=6) = 0.97$ and $\eta(t=24) = 1.08$.}{f:FramesICserp}
\end{figure}

\section{Conclusion} \label{s:conclusion}
Using an in-house integral equation solver and a pseudo-spectral
advection diffusion solver, we have studied mixing in vesicle
suspensions.  To the best of our knowledge, this is the first
study of the effect of vesicles on mixing.  Mixing measures from the literature were investigated,
and we we focused on the negative index $H^{-1}$ Sobolev norm which
quantifies mixing due to both advection and diffusion. We compare
mixing in the absence and the presence of vesicles and investigate the
effects of the Peclet number, the area fraction, and the viscosity
contrast.  The main outcomes are: 
\begin{itemize}
\item For the same average Peclet number, the presence of vesicles
suppresses mixing in most of the cases, and increasing the area
fraction suppresses it more. However, there are special initial
conditions for the transported quantity for which there is no advective
mixing in the absence of vesicles. The presence of vesicles provides
advection in those cases and hence promotes mixing.

\item For the same average Peclet number and the same area fraction,
the change in viscosity contrast does not affect mixing.

\item In order to estimate whether the presence of vesicles promotes or
suppresses mixing, we define a measure $M \propto \| \nabla \phi_0\cdot
\mathbf{v}\|_{L^1}$ where $\phi_0$ is the initial concentration field
and $\mathbf{v}$ is the default Couette velocity field. We found that
when the measure $M$ of a concentration for a passively transported
quantity approaches zero, mixing is dominated by diffusion in the
absence of vesicles.  Since the vesicle flows are more chaotic, the
presence of vesicles promotes mixing of the passively transported
quantity. The fact that the viscosity contrast doesn't have a significant impact,
suggest that these effects may hold true  for other types of suspensions.
\end{itemize}

Although here we consider only a two-dimensional cylindrical Couette flow,
similar results should hold for a planar Couette flow and Poiseuille
flow.  Additionally, mixing might present different physics in three
dimensions which we will investigate in a future study.

\bibliographystyle{plainnat} 
\bibliography{refs}

\begin{thebibliography}{49}
\providecommand{\natexlab}[1]{#1}
\providecommand{\url}[1]{\texttt{#1}}
\expandafter\ifx\csname urlstyle\endcsname\relax
  \providecommand{\doi}[1]{doi: #1}\else
  \providecommand{\doi}{doi: \begingroup \urlstyle{rm}\Url}\fi

\bibitem[Adrover et~al.(1998)Adrover, Giona, Muzzio, Cerbelli, and
  Alvarez]{Adrover-Alvarez-e98}
A.~Adrover, M.~Giona, F.J. Muzzio, S.~Cerbelli, and M.M. Alvarez.
\newblock Analytic expression for the short-time rate of growth of the
  intermaterial contact perimeter in two-dimensional chaotic flows and
  hamiltonian systems.
\newblock \emph{Physical Review Letters}, E 58:\penalty0 447--458, 1998.

\bibitem[Alvarez et~al.(1998)Alvarez, Muzzio, Cerbelli, Adrover, and
  Giona]{Alvarez-Giona-e98}
M.M. Alvarez, F.J. Muzzio, S.~Cerbelli, A.~Adrover, and M.~Giona.
\newblock Self-similar spatiotemporal structure of intermaterial boundaries in
  chaotic flows.
\newblock \emph{Physical Review Letters}, 81:\penalty0 3395--3398, 1998.

\bibitem[Annaswamy and Ghoniem(1995)]{Annaswamy-Ghoniem95}
A.~M. Annaswamy and A.~F. Ghoniem.
\newblock Active control in combustion systems.
\newblock \emph{IEEE Control Systems}, 15:\penalty0 49--63, 1995.

\bibitem[Aref and Balachandar(1986)]{Aref-Balachandar86}
H.~Aref and S.~Balachandar.
\newblock Chaotic advection in {S}tokes flows.
\newblock \emph{Physics of Fluids}, 29:\penalty0 3515--3521, 1986.

\bibitem[Ashwin et~al.(2002)Ashwin, M.Nicol, and N.Kirkby]{Ashwin-Kirkby-e02}
P.~Ashwin, M.Nicol, and N.Kirkby.
\newblock Acceleration of one-dimensional mixing by discontinuous mappings.
\newblock \emph{Physica A}, 130:\penalty0 347--363, 2002.

\bibitem[Bottausci et~al.(2004)Bottausci, Mezic, Meinhart, and
  Cardonne]{Bottausci-Cardonne-e04}
F.~Bottausci, I.~Mezic, C.~D. Meinhart, and C.~Cardonne.
\newblock Mixing in the shear superposition micromixer: three-dimensional
  analysis.
\newblock \emph{Philos. Trans. R. Soc. Lond. Ser. A}, 362:\penalty0 1001--1018,
  2004.

\bibitem[Boyd(2013)]{Boyd13}
J.~P. Boyd.
\newblock \emph{Chebyshev and {F}ourier {S}pectral {M}ethods (2nd {E}dition,
  {R}evised)}.
\newblock Dover Publications, Mineola, NY, USA, 2013.

\bibitem[Christlieb et~al.(2014)Christlieb, Guo, Morton, and
  Qiu]{Christlieb-Qiu-e14}
A.~Christlieb, W.~Guo, M.~Morton, and J-M. Qiu.
\newblock A high order time splitting method based on integral deferred
  correction for semi-{L}agrangian vlasov simulations.
\newblock \emph{Journal of Computational Physics}, 267:\penalty0 7--27, 2014.

\bibitem[Crank and Nicolson(1947)]{Crank-Nicolson1947}
John Crank and Phyllis Nicolson.
\newblock A practical method for numerical evaluation of solutions of partial
  differential equations of the heat-conduction type.
\newblock In \emph{Mathematical Proceedings of the Cambridge Philosophical
  Society}, volume~43, pages 50--67. Cambridge Univ Press, 1947.

\bibitem[D'Alessandro et~al.(1999)D'Alessandro, Dahleh, and
  Mezic]{DAlessandro-Mezic-e99}
D.~D'Alessandro, M.~Dahleh, and I.~Mezic.
\newblock Control of mixing: a maximum entropy approach.
\newblock \emph{IEEE Trans. Automatic Control}, 44:\penalty0 1852--1864, 1999.

\bibitem[Doering and Thiffeault(2006)]{Doering-Thiffeault06}
C.R. Doering and J.L. Thiffeault.
\newblock Multiscale mixing efficiencies for steady sources.
\newblock \emph{Physical Review E}, 025301, 2006.

\bibitem[Falcone and Ferretti(1998)]{Falcone-Ferretti98}
M.~Falcone and R.~Ferretti.
\newblock Convergence analysis for a class of high-order semi-{L}agrangian
  advection schemes.
\newblock \emph{SIAM J. Numer. Anal.}, 35:\penalty0 909--940, 1998.

\bibitem[Finn et~al.(2004)Finn, Cox, and Byrne]{Finn-Byrne-e04}
M.D. Finn, S.M. Cox, and H.M. Byrne.
\newblock Mixing measures for a two-dimensional chaotic {S}tokes flow.
\newblock \emph{Journal of Engineering Mathematics}, 48:\penalty0 129--155,
  2004.

\bibitem[Foures et~al.(2014)Foures, Caulfied, and Schmid]{Fourer-Schmid-e14}
D.~P.~G. Foures, C.~P. Caulfied, and P.J. Schmid.
\newblock Optimal mixing in two-dimensional plane {P}oiseuille flow at finite
  {P}eclet number.
\newblock \emph{Journal of Fluid Mechanics}, 748:\penalty0 241--277, 2014.

\bibitem[Ghigliotti et~al.(2011)Ghigliotti, Rahimian, Biros, and
  Misbah]{Ghigliotti-Misbah-e11}
G.~Ghigliotti, A.~Rahimian, G.~Biros, and C.~Misbah.
\newblock Vesicle migration and spatial organization driven by flow line
  curvature.
\newblock \emph{Physical Review Letters}, 106:028101, 2011.

\bibitem[Goldsmith and Skalak(1975)]{Goldsmith-Skalak75}
H.L. Goldsmith and R.~Skalak.
\newblock Hemodynamics.
\newblock \emph{Annu. Rev. Fluid Mech.}, 7:\penalty0 213--247, 1975.

\bibitem[Hessel et~al.(2005)Hessel, Lowe, and Schonfeld]{Hessel-Schonfeld-e05}
V.~Hessel, H.~Lowe, and F.~Schonfeld.
\newblock Micromixers - {A} review on passive and active mixing principles.
\newblock \emph{Chemical Engineering Science}, 60:\penalty0 2479--2501, 2005.

\bibitem[J.L.Thiffeault(2012)]{Thiffeault11}
J.L.Thiffeault.
\newblock Using multi scale norms to quantify mixing and transport.
\newblock \emph{Nonlinearity}, pages R1--R44, 2012.

\bibitem[Keller and Skalak(1982)]{Keller-Skalak82}
S.R. Keller and R.~Skalak.
\newblock Motion of a tank-treading ellipsoidal particle in a shear flow.
\newblock \emph{Journal of Fluid Mechanics}, 120:\penalty0 27--47, 1982.

\bibitem[Khakhar et~al.(1986)Khakhar, Rising, and Ottino]{Khakhar-Ottino86}
D.V. Khakhar, H.~Rising, and J.~M. Ottino.
\newblock Analysis of chaotic mixing in two model systems.
\newblock \emph{Journal of Fluid Mechanics}, 172:\penalty0 419--451, 1986.

\bibitem[Kraus et~al.(1996)Kraus, Wintz, Seifert, and
  Lipowsky]{Kraus-Lipowsky-e96}
M.~Kraus, W.~Wintz, U.~Seifert, and R.~Lipowsky.
\newblock Fluid vesicles in shear flow.
\newblock \emph{Physical Review Letters}, 77(17):\penalty0 3685--3688, 1996.

\bibitem[Leiken et~al.(2005)Leiken, Coulliette, Mariano, Ryan, Shay, Haller,
  and Marsden]{Leiken-Marsden-e05}
F.~Leiken, C.~Coulliette, A.~J. Mariano, E.~H. Ryan, L.K. Shay, G.~Haller, and
  J.~Marsden.
\newblock Pollution release tied to invariant manifolds: a case study for the
  coast of {F}lorida.
\newblock \emph{Physica D}, 210:\penalty0 1--20, 2005.

\bibitem[LeVeque(2007)]{Leveque07}
R.~J. LeVeque.
\newblock \emph{{F}inite {D}ifference {M}ethods for {O}rdinary and {P}artial
  {D}ifferential {E}quations}.
\newblock SIAM, Philadelphia, PA, USA, 2007.

\bibitem[Lin et~al.(2011)Lin, Thiffeault, and Doering]{Lin-Doering-e11}
Z.~Lin, J.L. Thiffeault, and C.R. Doering.
\newblock Optimal stirring strategies for passive scalar mixing.
\newblock \emph{Journal of Fluid Mechanics}, 675:\penalty0 465--476, 2011.

\bibitem[Mathew et~al.(2005)Mathew, Mezic, and Petzold]{Mathew-Petzold-e05}
G.~Mathew, I.~Mezic, and L.~Petzold.
\newblock A multiscale measure for mixing.
\newblock \emph{Physica D}, 211:\penalty0 23--46, 2005.

\bibitem[Mathew et~al.(2007)Mathew, Mezic, Grivopoulos, Vaidya, and
  Petzold]{Mathew-Petzold-e07}
G.~Mathew, I.~Mezic, S.~Grivopoulos, U.~Vaidya, and L.~Petzold.
\newblock Optimal control of mixing in {S}tokes fluid flows.
\newblock \emph{Journal of Fluid Mechanics}, 580:\penalty0 261--281, 2007.

\bibitem[Meleshko and Aref(1996)]{Meleshko-Aref96}
V.V. Meleshko and H.~Aref.
\newblock A blinking rotlet model for chaotic advection.
\newblock \emph{Physics of Fluids}, 8:\penalty0 3215--3217, 1996.

\bibitem[Misbah(2006)]{Misbah06}
C.~Misbah.
\newblock Vacillating breathing and tumbling of vesicles under shear flow.
\newblock \emph{Physical Review Letters}, 96(2), 2006.

\bibitem[Muzzio and Swanson(1991)]{Muzzio-Swanson91}
F.J. Muzzio and P.D. Swanson.
\newblock The statistics of stretching and stirring in chaotic flows.
\newblock \emph{Physics of Fluids}, A 3:\penalty0 822--834, 1991.

\bibitem[Nacev et~al.(2011)Nacev, Beni, Bruno, and Shapiro]{Nacev-Shapiro-e11}
A.~Nacev, C.~Beni, O.~Bruno, and B.~Shapiro.
\newblock The behaviors of ferro-magnetic nano-particles in and around blood
  vessels under applied magnetic fields.
\newblock \emph{J. Magn. Magn. Mater.}, 323(6):\penalty0 651--668, 2011.

\bibitem[Noguchi and Gompper(2005)]{Nogouchi-Gompper05}
H.~Noguchi and D.G. Gompper.
\newblock Shape transitions of fluid vesicles and red blood cells in capillary
  flows.
\newblock \emph{Proceedings of the National Academy of Sciences of the United
  States of America}, 102:\penalty0 14159--14164, 2005.

\bibitem[Ottino(1990)]{Ottino90}
J.~M. Ottino.
\newblock Mixing, chaotic advection, and turbulence.
\newblock \emph{Annu. Rev. Fluid Mech.}, 22:\penalty0 207--253, 1990.

\bibitem[Ottino and Wiggins(2004)]{Ottino-Wiggins04}
J.M. Ottino and S.~Wiggins.
\newblock Introduction: mixing in microfluidics.
\newblock \emph{Philos. Trans. R. Soc. Lond. Ser. A}, 362:\penalty0 923--935,
  2004.

\bibitem[Popel and Johnson(2005)]{Popel-Johnson05}
A.~S. Popel and P.~C. Johnson.
\newblock Microcirculation and {H}emorheology.
\newblock \emph{Annu. Rev. Fluid Mech.}, 37:\penalty0 43--69, 2005.

\bibitem[Pozrikidis(1992)]{Pozrikidis-92}
C.~Pozrikidis.
\newblock \emph{Boundary {I}ntegral and {S}ingularity {M}ethos for {L}inearized
  {V}iscous {F}low}.
\newblock Cambridge University Press, Cambridge, UK, 1992.

\bibitem[Quaife and Biros(2014)]{Quaife-Biros14a}
B.~Quaife and G.~Biros.
\newblock High-volume fraction simulations of two-dimensional vesicle
  suspensions.
\newblock \emph{Journal of Computational Physics}, 274:\penalty0 245--267,
  2014.

\bibitem[Quaife and Biros(2016)]{Quaife-Biros16}
B.~Quaife and G.~Biros.
\newblock Adaptive time stepping for vesicle suspensions.
\newblock \emph{Journal of Computational Physics}, 306:\penalty0 478--499,
  2016.

\bibitem[Rahimian et~al.(2010)Rahimian, Veerapaneni, and
  Biros]{Rahimian-Biros-e10}
A.~Rahimian, S.K. Veerapaneni, and G.~Biros.
\newblock Dynamic simulation of locally inextensible vesicles suspended in an
  arbitrary two-dimensional domain, a boundary integral method.
\newblock \emph{Journal of Computational Physics}, 229:\penalty0 6466--6484,
  2010.

\bibitem[Robert(1981)]{Robert81}
A.~Robert.
\newblock A stable numerical integration scheme for the primitive
  meteorological equations.
\newblock \emph{Atmosphere-Ocean}, 19:\penalty0 35--46, 1981.

\bibitem[Rothstein et~al.(1999)Rothstein, Henry, and
  Gollub]{Rothstein-Gollub-e99}
D.~Rothstein, E.~Henry, and J.P. Gollub.
\newblock Persistent patterns in transient chaotic fluid mixing.
\newblock \emph{Nature}, 401:\penalty0 770--772, 1999.

\bibitem[Strang(1968)]{Strang68}
G.~Strang.
\newblock On the construction and comparison of difference schemes.
\newblock \emph{SIAM J. Numer. Anal.}, 5(3):\penalty0 506--517, 1968.

\bibitem[Thiffeault et~al.(2004)Thiffeault, Doering, and
  Gibbon]{Thiffeault-Gibbon-e04}
J.~Thiffeault, C.~S. Doering, and J.~D. Gibbon.
\newblock A bound on mixing efficiency for the advection-diffusion equation.
\newblock \emph{Journal of Fluid Mechanics}, 521:\penalty0 105--114, 2004.

\bibitem[Torquato(2002)]{Torquato02}
S.~Torquato.
\newblock \emph{Random {H}eterogeneous {M}aterials: {M}icrostructure and
  {M}acroscopic {P}roperties}.
\newblock Springer, 2002.

\bibitem[Trefethen(2000)]{Trefethen00}
N.~L. Trefethen.
\newblock \emph{{S}pectral {M}ethods in {MATLAB}}.
\newblock SIAM, Philadelphia, PA, USA, 2000.

\bibitem[Veerapaneni et~al.(2009)Veerapaneni, Gueyffier, Zorin, and
  Biros]{Veerapaneni-Biros-e09}
S.K. Veerapaneni, D.~Gueyffier, D.~Zorin, and G.~Biros.
\newblock A boundary integral method for simulating the dynamics of
  inextensible vesicles suspended in a viscous fluid in 2d.
\newblock \emph{Journal of Computational Physics}, 228(7):\penalty0 2334--2353,
  2009.

\bibitem[Wade et~al.(2007)Wade, Khaliq, Yousuf, Vigo-Aguiar, and
  Deininger]{Wade-Deininger-e07}
B.~A. Wade, A.~Q.~M. Khaliq, M.~Yousuf, J.~Vigo-Aguiar, and R.~Deininger.
\newblock On the smoothing of the {C}rank-{N}icolson scheme and higher order
  schemes for pricing barrier options.
\newblock \emph{J. Comp. App. Math.}, 204:\penalty0 144--158, 2007.

\bibitem[Wang and Popel(1993)]{Wang-Popel93}
C.H. Wang and A.~S. Popel.
\newblock Effect of red blood cell shape on oxygen transport in capillaries.
\newblock \emph{Mathematical Biosciences}, 116:\penalty0 89--110, 1993.

\bibitem[Xiu and Karniadakis(2001)]{Xiu-Karniadakis01}
D.~Xiu and G.E. Karniadakis.
\newblock A semi-{L}agrangian high-order method for the {N}avier-{S}tokes
  equations.
\newblock \emph{Journal of Computational Physics}, 172:\penalty0 658--684,
  2001.

\bibitem[Zvan et~al.(2000)Zvan, Vetzal, and Forsyth]{Zvan-Forsyth-e00}
R.~Zvan, K.~R. Vetzal, and P.~A. Forsyth.
\newblock {PDE} methods for pricing barrier options.
\newblock \emph{J. Econ. Dynamics \& Control}, 24:\penalty0 1563--1590, 2000.

\end{thebibliography}
\biboptions{sort&compress}
\end{document}